\documentclass{article}

\usepackage{stmaryrd}
\usepackage{amssymb,amsfonts,amsmath}
\usepackage{cite,enumerate,float,indentfirst}
\usepackage{color}

\def\be{\begin{eqnarray}}
\def\ee{\end{eqnarray}}
\def\nn{\nonumber}

\def\p{\partial}

\def\Tr{{\rm Tr}\,}

\def\ph{\!\!\phantom.}

\definecolor{red}{rgb}{1,0,0}
\definecolor{orange}{rgb}{1,0.5,0}
\definecolor{violet}{rgb}{0.7,0,1}

\def\cre{\color{red}}

\def\cg{\color{green}}
\def\cb{\color{blue}}

\def\b1{(-24,18)}\def\b2{(9,29)}\def\b3{(30,0)}\def\b4{(9,-29)}\def\b5{(-24,-18)}

\def\theequation{\arabic{section}.\arabic{equation}}

\hoffset= - 1.0in         

\begin{document}

\title{\vspace{.1cm}{\LARGE {\bf Cut and join operator ring in Aristotelian tensor model
}\vspace{.5cm}}
\author{{\bf H. Itoyama$^{a,b}$},
{\bf A. Mironov$^{c,d,e,f}$},
\ {\bf A. Morozov$^{d,e,f}$}
}
\date{ }
}

\maketitle

\vspace{-6.2cm}

\begin{center}
\hfill FIAN/TD-23/17\\
\hfill IITP/TH-17/17\\
\hfill ITEP/TH-30/17\\
\hfill OCU-PHYS-472
\end{center}

\vspace{4.cm}

\begin{center}
$^a$ {\small {\it Department of Mathematics and Physics, Graduate School of Science,
Osaka City University, 3-3-138, Sugimoto, Sumiyoshi-ku, Osaka, 558-8585, Japan}}\\
$^b$ {\small {\it Osaka City University Advanced Mathematical Institute (OCAMI), 3-3-138, Sugimoto, Sumiyoshi-ku, Osaka, 558-8585, Japan}}\\
$^c$ {\small {\it I.E.Tamm Theory Department, Lebedev Physics Institute, Leninsky prospect, 53, Moscow 119991, Russia}}\\
$^d$ {\small {\it ITEP, B. Cheremushkinskaya, 25, Moscow, 117259, Russia }}\\
$^e$ {\small {\it Institute for Information Transmission Problems,  Bolshoy Karetny per. 19, build.1, Moscow 127051 Russia}}\\
$^à$ {\small {\it National Research Nuclear University MEPhI, Moscow 115409, Russia }}\\
\end{center}

\vspace{.5cm}

\begin{abstract}
Recent advancement of rainbow tensor models based on their superintegrability (manifesting itself as the existence of an explicit expression for a generic Gaussian correlator) has allowed us to {\it bypass} the long-standing problem seen as the lack of eigenvalue/determinant representation needed to establish the KP/Toda integrability.  As the mandatory next step, we discuss in this paper how to provide an adequate designation to each of the connected gauge-invariant operators that form a double coset, which is required to cleverly formulate a tree-algebra generalization of the Virasoro constraints. This problem goes beyond the enumeration problem {\it per se} tied to the permutation group, forcing us to introduce a few gauge fixing procedures to the coset. We point out that the permutation-based labeling, which has proven to be relevant for the Gaussian averages is, via interesting complexity, related to the one based on the keystone trees, whose algebra will provide the tensor counterpart of the Virasoro algebra for matrix models. Moreover, our simple analysis reveals the existence of nontrivial kernels and co-kernels for the cut operation and for the join operation respectively that prevent a straightforward construction of the non-perturbative RG-complete partition function and the identification of truly independent time variables. We demonstrate these problems by the simplest non-trivial Aristotelian {\bf RGB} model with one complex rank-3 tensor, studying its ring of gauge-invariant operators, generated by the keystone triple with the help of four operations: addition, multiplication, cut and join.
\end{abstract}

\bigskip

\bigskip

\section{Introduction}
\setcounter{equation}{0}

Tensor models \cite{tensor} begin to acquire attention that they deserve \cite{tenfirst}-\cite{tenlast}
as natural objects to study in the framework of the non-linear algebra \cite{NLA}.
In a recent series of papers \cite{IMMten1}-\cite{MMten}, we described the technique
necessary for the first step of systematic analysis of tensor models.
It turned out that the simplest problem is a complete description of the Gaussian correlators,
the problem which for many years remained unsolved in the case of matrix models,
despite a number of brilliant insights including the celebrated Harer-Zagier
formulas \cite{HZ,LZ}.
As was expected, a solution to the problem came from the synthesis of character
\cite{charcal} and Hurwitz  \cite{MMN1,KR,AMMNhur} calculi (see \cite{MMmamo}),
and it appeared
to be immediately generalizable to the tensor case \cite{Di,MMten}.
Like in the matrix model case, the simplest from this perspective is the
rectangular complex model of \cite{RCM,CMcorr,CMmod}, and, among tensor models, the easiest treatable
are rainbow models \cite{IMMten1} with the highest possible ``gauge" symmetry,
while models with restrictions on the colorings and/or reality conditions are described
by a little more complicated formulas,
with the simplest example of such complications provided by the
Hermitian matrix model (!).
Of additional interest is the subclass of starfish rainbow models \cite{IMMten1},
where the large-$N$ limit is automatically described by melonic diagrams
(these, however, will not be considered in the present paper).

As usual in quantum field theory, the study of any such model consists of
several steps:
describing the symmetries and the field content of the model,
enumeration and classification of operators,
introduction of appropriate generating functions
and evaluation of correlators/averages.
Only at the last of these steps, the action/dynamics of the model is
needed, though a clever choice of the generating functions to make can also
depend on the action and on a particular phase of the model.
Traditional analysis begins from the Gaussian phase,
and then the Ward identities are used to express the
correlation functions through the Gaussian spectral curve in {\it a functorial way}
(by the procedure known as topological recursion \cite{AMMrec}),
and transition to the non-perturbative (Dijkgraaf-Vafa) phases
goes through a deformation of the spectral curve.
This approach is successfully developed for the one-matrix
eigenvalue models
(where also integrability properties are revealed and understood),
and the present task is to extend it in two directions:
to multi-matrix and to (multi-)tensor models.
However, such extension is quite sophisticated and can hardly be made
by one simple effort.
As suggested in \cite{IMMten2}, we move by small steps,
but in a systematic way with the hope that it will be
no less straightforward for tensors than it has proven to be for
matrices.

Accordingly, the very first task is to provide an efficient enumeration of operators.
As was already mentioned,
this step is independent of the action of the model and depends just on its
field content.
The problem is purely combinatorial, but one should not underestimate its significance.
The choice of an appropriate language and notation is crucial for
the theory of tensor models, which did not advance for years,
with the main obstacle being the lack of notation like traces and determinants
(while their relevant generalizations in the character/Hurwitz calculus
are perfectly known within the context of non-linear algebra, see \cite{NLA}
and references in the last paper there).
We advocate the use of permutation-group terminology,
which was attempted for matrix models already in \cite{MMN1,KR},
but did not gain enough attention, both because of efficiency of other languages
and, as we now understand, because of the fact that its application to the Hermitian rather than
to the rainbow-like rectangular complex model (RCM) is rather clumsy.
For tensor models, however, advantages of this terminology become obvious:
it was actually used in the tensorial calculations in \cite{IMMten2},
and was made fully explicit in \cite{AMMNhur,MMmamo,MMten} and \cite{Di},
where it immediately provided {\it generic explicit expressions} for the Gaussian correlators
in arbitrary matrix and tensor models.
These formulas
can not be fully appreciated without detailed examples and explanations
of how they can be practically used.
Our consideration should now be lifted to the next level: a
systematic description of operators and their Gaussian correlators
as functions of models and index contractions
expressed in terms of the permutations and Young diagrams.

In the present text, we consider from this perspective
the simplest of the rainbow tensor models of \cite{IMMten1}:
the Aristotelian model with a single complex tensor of rank 3 and
the {\bf RGB} (red-green-blue) symmetry $U({\cre N_1})\otimes U({\cg N_2})\otimes U({\cb N_3})$.
Our main purpose is to illustrate the non-trivial interplay with the
theory of Hurwitz numbers, which, in the case of rank 3, is reduced
to the distinguished theory of the Belyi functions and Grothendick's
{\it dessins d'enfants} \cite{Bel,BHM,LZ,Amb,KR}.
We emphasize that this latter subject describes equilateral triangulations
and, thus, is often emerging in string theory studies
\cite{LevMor,DJSmit,Gopak},
but the rank-3 tensor theory is the first place where it is
practically unavoidable.
At the same time, the relation is not literal: the
operators of the Aristotelian models are
originally labeled by {\it pairs of permutations} and their variety is
not {\it fully}
reduced to {\it the admissible triples} of {Young diagrams},
i.e. to the most natural conjugacy classes.
The knowledge of the Gaussian correlators,
provided by \cite{IMMten2} and \cite{Di,MMten}
is an important tool, helping to check and validate the general
considerations.

This paper has overlaps at several pints with \cite{GR} and more recent ones \cite{KGT,GeRamg}:
some of the questions we address are close and some answers seem to be in accord.
The main difference is that we not just enumerate the gauge-invariants operators, but also reveal an additional cut and join structure in their ring\footnote{A word of precaution is necessary here: the cut and join operations here is different from the cut-and-join operator of \cite{MMN1}, since, though in the both these cases the operations are acting in the group algebra of infinite symmetric group, those from \cite{MMN1} are elements of center of the group algebra, and the operations in this paper are not.}.
These operations, along with addition and multiplication, allow one to generate all the ring starting with simplest keystone operators and provide a proper counterpart to the Virasoro constraints in the matrix model case \cite{MMVir,MMW}. In fact, versions of cut and join operations which act on graphs were considered within the context of Ward identities in tensor models earlier \cite{Virtree,GurVir}, hence, this our paper is an attempt to marry up these two structures discussed previously.

We also put more emphasis on concrete examples
and on variety of more delicate properties, which can be revealed.
In particular, we wonder if operators can be distinguished by the Gaussian correlators,
which remain independent after factoring over the coloring permutations $S_3^{color}$,
how close is the result to the set of Hurwitz admissible triples,
what is the number of independent connected operators etc.
Our answers to these questions are far from being exhaustive,
and still they are important to developing the language, which would
adequately reflect the renormalization group properties and provide an appropriate
substitute for the resolvents and for other generating functions used in the matrix model
theory.

\section{Logic and the structure of the paper
\label{tasks}}
\setcounter{equation}{0}

Since the study of tensor models is still at the very early stage
when there is no consensus even in terminology and in the main research directions,
it deserves making explicit our logic and immediate goals.
In this section, we elaborate the Introduction, introducing systematic procedures, some of which are explicitly carried out in this paper.
Schematically our plan can be described as follows:

\vspace{2.5cm}

\!\!\!\!\!\!\!\!\!\!\!\!\!\!\!\!\!\!\!\!\!\!\!\!\!\!\!\!\!\!\!\!\!

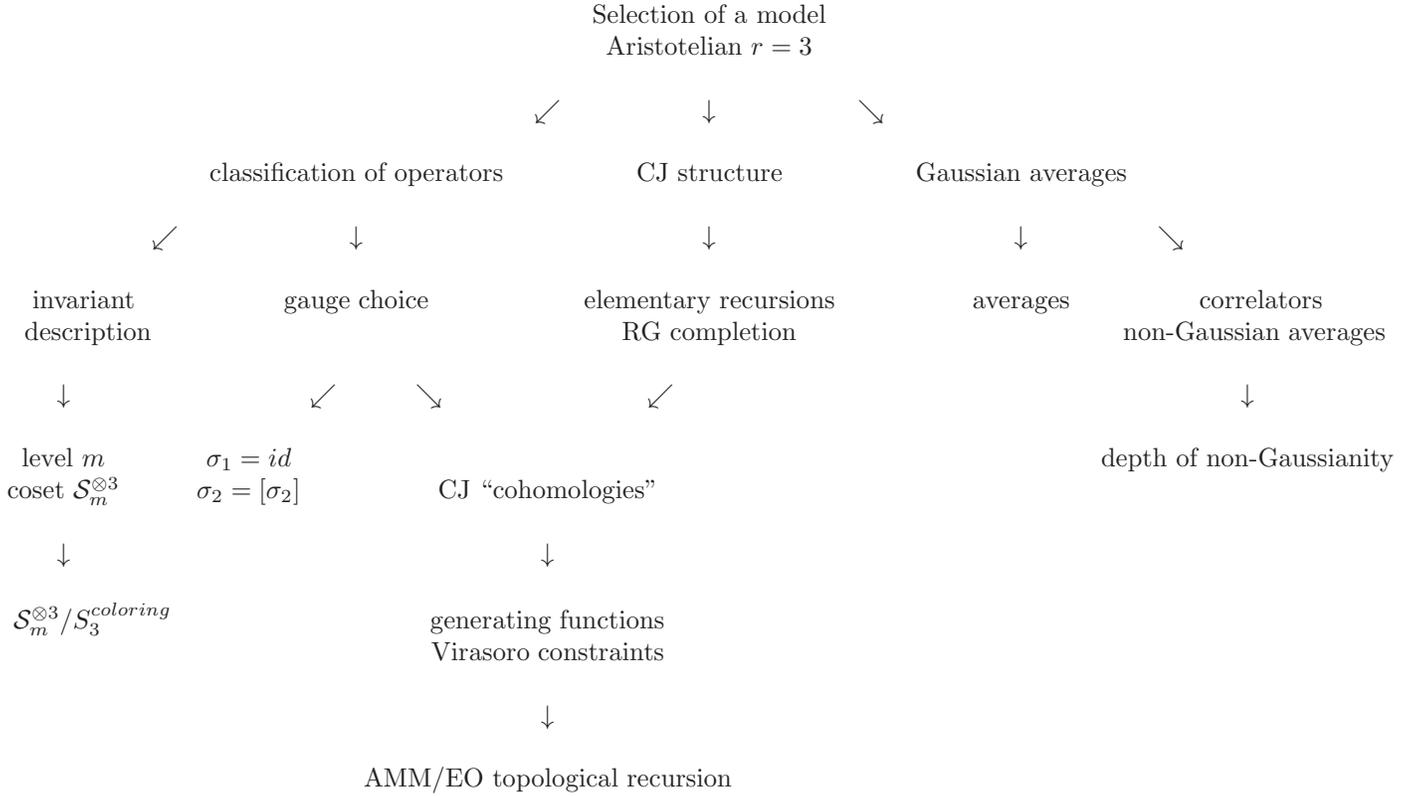
\centerline{
\begin{tabular}{ccccccccc}
&&&&Selection of a model \\ &&&&Aristotelian $r=3$ \\
\\
&&&$\swarrow$ &$\downarrow$ &$\searrow$ \\
\\
&&classification of operators&&CJ structure&&Gaussian averages\\
\\
&$\swarrow$&$\downarrow$&&$\downarrow$&&$\downarrow$&$\searrow$\!\!\!\!\\
\\
invariant  \!\!\!\!\!\!\!\!\!\!\!
 &&gauge choice&& \!\!\!\!\!elementary recursions\!\!\!\!\!
&&averages &&\!\!\!\!\!\!\!\!\!\!\!\!\!\!\!\!correlators \\
description\!\!\!\!\!\!\!\!\!\!\!
&&&&RG completion &&&&\!\!\!\!\!\!\!\!\!\!\!\!\!\!\!\!\!\!\!non-Gaussian averages \\
\\
$\downarrow$&&
\!\!\!\!\!\!\!\!\!\!\!\!\!$\swarrow$\ \ \ \ \ \ \ \ \
$\searrow$\!\!\!\!\!\!\!\!\!\!\!\!\!\!\!\!\!\!\!\!\!\!&&
\!\!\!\!\!\!\!\!\!\!\!\!\!\!\!\!\!\!\!\!\!\!$\swarrow$
&&&&\!\!\!\!\!\!\!\!\!\!\!\!\!\!\!\!\!\!\!\!\!\!$\downarrow$\\
\\
level $m$&$\sigma_1=id$
\!\!\!\!\!\!\!\!\!\!\!\!\!\!\!\!\!\!\!\!\!\!\!\!\!\!\!\!\!\!\!\!\!\!\!\!\!\!\!\!
&&\!\!\!\!\!\!\!\!\!\!\!\!\!\!\!\!\!\!\!\!\!\!\!\!\!\!\!\!\!\!\!\!\!\!\!\!\!\!\!\!\!
\!\!\!\!\!\!\!\!\!\!\!\!\!\!\!\!\!\!\!\!\!\!\!\!\!\!\!\!\!\!\!\!\!
&&&&&\!\!\!\!\!\!\!\!\!\!\!\!\!\!\!\!\!\!\!\!\!\!depth of non-Gaussianity
\\
coset ${\cal S}_m^{\otimes 3}$
&$\sigma_2=[\sigma_2]$
\!\!\!\!\!\!\!\!\!\!\!\!\!\!\!\!\!\!\!\!\!\!\!\!\!\!\!\!\!\!\!\!\!\!\!\!\!\!\!\!
&& \!\!\!\!\!\!\!\!\!\!\!\!\!\!\!\!\!\!\!\!\!\!\!\!\!\!\!\!\!\!\!\!\!\!\!\!\!
CJ ``cohomologies"
\!\!\!\!\!\!\!\!\!\!\!\!\!\!\!\!\!\!\!\!\!\!\!\!\!\!\!\!\!\!\!\!\!\!\!\!\! \\
\\
$\downarrow$ &&& \!\!\!\!\!\!\!\!\!\!\!\!\!\!\!\!\!\!\!\!\!\!\!\!\!\!\!\!\!\!\!\!\!
$\downarrow$
\!\!\!\!\!\!\!\!\!\!\!\!\!\!\!\!\!\!\!\!\!\!\!\!\!\!\!\!\!\!\!\!\!
 \\
\\
${\cal S}_m^{\otimes 3}/S_3^{coloring}$\!\!\!\!\!\!\!\!\!\!\!\!\!  &&&
\!\!\!\!\!\!\!\!\!\!\!\!\!\!\!\!\!\!\!\!\!\!\!\!\!\!\!\!\!\!\!\!\!
generating functions
\!\!\!\!\!\!\!\!\!\!\!\!\!\!\!\!\!\!\!\!\!\!\!\!\!\!\!\!\!\!\!\!\!
\\
&&& \!\!\!\!\!\!\!\!\!\!\!\!\!\!\!\!\!\!\!\!\!\!\!\!\!\!\!\!\!\!\!\!\!
Virasoro constraints
\!\!\!\!\!\!\!\!\!\!\!\!\!\!\!\!\!\!\!\!\!\!\!\!\!\!\!\!\!\!\!\!\!\\
\\
&&& \!\!\!\!\!\!\!\!\!\!\!\!\!\!\!\!\!\!\!\!\!\!\!\!\!\!\!\!\!\!\!\!\!
$\downarrow$
\!\!\!\!\!\!\!\!\!\!\!\!\!\!\!\!\!\!\!\!\!\!\!\!\!\!\!\!\!\!\!\!\!\\
\\
&&& \!\!\!\!\!\!\!\!\!\!\!\!\!\!\!\!\!\!\!\!\!\!\!\!\!\!\!\!\!\!\!\!\!\!\!\!\!\!\!\!\!\!\!
AMM/EO topological recursion
\!\!\!\!\!\!\!\!\!\!\!\!\!\!\!\!\!\!\!\!\!\!\!\!\!\!\!\!\!\!\!\!\!\!\!\!\!\!\!\!\!\!\!
\end{tabular}
}

\bigskip

\bigskip

The very first step is specification of a model, and,
in this paper, it will be the simplest one of all:
the rank-three Aristotelian model, which was also the choice in
\cite{IMMten2}   and \cite{KGT,GeRamg}.
Actually, chosen at this stage is the (``gauge") symmetry and the field content,
but some work is still needed before the choice of dynamics (Lagrangian)
can be discussed.
Given the field content, there are two immediate directions to follow:
one can ask what are the ``local" operators
and how they ``communicate" at the perturbative level.
The second question involves the Gaussian averaging
and the Feynman diagram technique.
This leads us to consider the Gaussian averages of the local operators and
their Gaussian correlators, the latter are actually a step towards the
perturbative consideration of non-Gaussian Lagrangians.
A distinguished step in this consideration is inserting
a single propagator: according to the Wick theorem, calculation of arbitrary
Gaussian correlators is multiple applications (iterations) of this
elementary operation.
Inserting a propagator can join two disconnected operators and
can cut one connected operator into two disconnected ones.
Thus, this operation introduces a peculiar {\it cut and join} (CJ)
structure in the operator ring, and this is actually the one which
stands behind the renormalization group (RG) properties and the celebrated
Virasoro-type constraints which further lead to the AMM/EO topological recursion.
Despite that they underlie the RG structure in {\it all} models of quantum field theory,
the particular formulation of Virasoro type constraints and their technical efficiency
strongly depend on the clever choice of the generating functions for the
RG-complete set of local operators: in matrix models, these are just the ordinary
resolvents, but they are clearly not just so simple in tensor models.
We refer the reader to an introductory discussion of the issue in \cite{IMMten2},
and, in the present paper, we will not reach the level of generating functions.
Our goal in this paper is more modest: it is to prepare the necessary ingredients,
namely, to discuss classification/enumeration of gauge-invariant operators and the
CJ structure on this set.
This step is already highly non-trivial, and it is by no means fully performed in the
present paper, we rather formulate problems and provide enlightening examples
of how they can be dealt with.
The crucial point is that needed is not just {\it some} classification procedure,
but the one which is relevant for the deep study of tensor models, their dynamics,
integrability and {\it superintegrability}. (For this last issue, see the last section.)

If this were not the case,
the set of independent gauge-invariant operators in the Aristotelian model
would be easy to characterize: for the operators made from $m$ pairs of the rank-$r$
tensor fields $M$ and $\bar M$,
it is the double coset
${\cal S}^{r}_m = S_m\backslash S_m^{\otimes r}/S_m$
where $S_m$ is the symmetric group, consisting of permutations of $m$ elements.
The textbook symmetric group calculus \cite{SG} says that the size of this set, i.e.
the number of linearly independent gauge-invariant operators at level $m$
(linear generators of the operator ring), is 
\be
\Big|\Big|{\cal S}^{r}_m\Big|\Big| = \sum_{\Delta\vdash m} z_\Delta^{r-2}
\label{sizecoset}
\ee
where the sum goes over all Young diagrams (conjugacy classes) $\Delta$ with lines $\delta_1\ge\delta_2\ge\ldots$ of the size $m=\sum_i\delta_i$,
and $z_\Delta = \prod_i i^{k_i}k_i!$ is the number of conjugations
which leave the permutation with $k_i$ cycles (lines of the Young diagram $\Delta$) of length $i$ intact,
and the plethystic logarithm can be used to extract the number of  independent
{\it connected} operators (multiplicative generators of the ring),
these simple formulas are well known in tensor model theory \cite{GR,KGT,GeRamg}.
However, this powerful invariant technique is nearly inapplicable for any further
considerations, even for asking appropriate ``physical" questions.
One of the ways out is to abandon the invariant formalism and proceed in concrete
gauges, somehow fixing some of the $r$ permutations in the conjugacy classes
${\cal S}^{r}_m$.
For the Aristotelian model {\it per se}, i.e. for $r=3$, a possible gauge choice
is $\sigma_1=id$ and $\sigma_2$ identified with its Young diagram $[\sigma_2]$,
i.e. is reduced just to a set of numbers of cycles of different lengths, this is
what has led to the formalism of ``red-green cycles", efficiently used in \cite{IMMten2}.
It allows us to enumerate the operators in terms of simple pictures
and it also provides an acceptably simple description of the CJ structure.
It is important that these cut and join operations connect only levels adjacent to level $m$,
which makes their description at a given $m$ a {\it finite} problem.
It also splits the size $||{\cal S}_m^r||$ into finer and informative structure,
characterizing the number of independent $\sigma_3$ when the two Young diagrams
$[\sigma_2]$ and $[\sigma_3]$ are fixed, which measures the deviation (degeneracy)
of the operator classification problem from the better studied Hurwitz calculus
of \cite{MMN1}\footnote{There is a confusion in terminology: the calculus of coverings directly related to tensor model calculus,
is somewhat different from the symmetric group calculus of \cite{MMN1}
based on the Burnside-Frobenius formula \cite{FroD,LZ} $\sum_R d_R^{2-2g} \prod_i \varphi_R(\Delta_i)$
which is nowadays associated with ``Hurwitz $\tau$-functions": they coincide only for
simple ramification points, i.e. for the Young diagrams of the type $\Delta=[2^k,1^l]$.
This is reflected in the fact that, beyond this intersection domain, the
Hurwitz $\tau$-function are not of the KP/Toda type
(and this is exactly what makes them mysterious and so interesting)
unlike what one expects \cite{MMten} for the truly combinatorial partition functions
of tensor models.
}.
The main drawback of this formalism (gauge choice) is that it breaks the global ``symmetry"
$S_r^{coloring}$, which is very important for decreasing the number of independent
operators: for $r=3$, the red-green symmetry is easily seen, while the
red-blue and green-blue ones are more difficult to see.
Consideration of this formalism, its various realizations and applications
to the study of CJ structure  will be one of the main topics of the present paper.
What we actually do is the level-after-level analysis of the two stories:
the operator set ${\cal S}_m^3$ and the CJ action on it, for $m=2,3,4,5$,
with the most interesting things starting to happen at the $m=5$ level\footnote{
Since  Paolo Ruffini and Niels Henrik Abel it is known that $m=5$ is the threshold for symmetric group theory
to become really interesting, though insolvability of $S_m$ for $m\geq 5$ {\it per se},
which explains Abel's impossibility theorem of solving degree $m$ equations in radicals,
still needs to find its place in
the tensor model story.}.

An immediate benefit of this naive classification of gauge-invariant operators
is the possibility of extracting the primary dynamical information by
looking at their Gaussian averages: like matrix models \cite{MMmamo},
the rainbow tensor models are also {\it superintegrable}, and all the Gaussian
correlators are immediately and explicitly calculable (expressed through the
finite sums of symmetric group characters) \cite{MMten,Di}.
It is of course very appealing to use this extraordinary strong result in the
study of operator classification (Hilbert and Fock spaces) and of the CJ structures on it.
The problem, however, is that, starting from $m=5$, the Gaussian averages do {\it not}
fully distinguish the gauge-invariant operators: there are some operators which are different,
but have the same averages.
Since the operators are {\it different}, this degeneracy is of course lifted
in the non-Gaussian case, but any such pair of degenerate operators is separated
at its own level (depth) of non-Gaussianity, and this provides an additional
``depth" structure on the set ${\cal S}_m^r$.

However, if one wants to address the truly interesting dynamical questions, the
Gaussian averages and correlators are not enough.
The powerful approach to study of non-Gaussian phases in matrix models
is obtained through the Virasoro-like constraints, which are applicable in any backgrounds.
They are the Ward identities associated with the change of integration variable (quantum field)
$\delta M = \frac{\p}{\p \bar M} {\cal K}_{\Sigma_0}$ with a gauge-invariant operator
${\cal K}_{\Sigma_0}(M,\bar M)$
and are fully expressed through the CJ structure of the operator ring.
If the theory has an action $-\mu\Tr M\bar M + \sum_\Sigma t_\Sigma {\cal K}_\Sigma$,
then, the Ward identities are
\be\label{vir}
\mu|\Sigma_0|\cdot \Big<{\cal K}_{\Sigma_0}\Big> = \sum_\Sigma t_\Sigma \cdot\Big<\{{\cal K}_\Sigma, {\cal K}_{\Sigma_0}\}\Big> +
\Big<\Delta{\cal K}_{\Sigma_0}\Big>
\ee
where $|\Sigma|$ denotes the degree of $\Sigma$.
One can rewrite this relation in more details introducing the structure constants:
\be
\{{\cal K}_{\Sigma'},{\cal K}_{\Sigma''}\} =
\sum_{\Sigma'''} \gamma_{\Sigma',\Sigma''}^{\Sigma'''} {\cal K}_{\Sigma'''}
\label{join}
\ee
and
\be
\Delta{\cal K}_\Sigma = \sum_{\Sigma',\Sigma''}
\Delta_\Sigma^{\Sigma',\Sigma''} {\cal K}_{\Sigma'}{\cal K}_{\Sigma''}
\label{cut}
\ee
The cut operation $\Delta$ is at most quadratic when acting on connected operators,
because inserting the propagator can cut any
connected operator into two disconnected parts at most.
On the disconnected
operators, $\Delta$ acts with the help of $\{\,\}$:
\be
\Delta({\cal K}_\Sigma\cdot {\cal K}_{\Sigma'}) =
\Delta{\cal K}_\Sigma\cdot{\cal K}_{\Sigma'} + {\cal K}_\Sigma\cdot\Delta{\cal K}_{\Sigma'}
+  \{{\cal K}_\Sigma,{\cal K}_{\Sigma'}\} + \{{\cal K}_{\Sigma'},{\cal K}_{\Sigma}\}
\ee
With the help of (\ref{join}) and (\ref{cut}),
the Ward identities for the partition function $Z\{t\}$ acquire the familiar form
\be
{\cal L}_\Sigma {\cal Z}\{t\} = 0
\ee
with
\be
{\cal L}_\Sigma = -\mu \cdot |\Sigma| \frac{\p}{\p t_\Sigma}
+ \sum_{\Sigma',\Sigma'''} \gamma_{\Sigma,\Sigma'}^{\Sigma'''}\cdot
t_{\Sigma'}\frac{\p}{\p t_{\Sigma'''}}
+ \Delta_\Sigma^{\Sigma',\Sigma'''}\cdot \frac{\p^2}{\p t_{\Sigma'}\p t_{\Sigma'''}}
\ee
What we need for an efficient formalism is an appropriate description of $\gamma$ and $\Delta$.
The problem is that they are governed by somewhat different structures.

What stands behind $\gamma$ is just a Lie algebra, and a appropriate labeling $\Sigma$
of operators should properly take into account this algebraic structure.
In general, this algebra is just that of {\it rooted trees} \cite{Virtree,GurVir,IMMten2},
which, in turn, appear after one specifies a set of keystone operators (see \cite{IMMten2}
for details).
Namely, one can form operators as a sequence of action of $\{,\}$ on keystones,
implying the labeling like
$$
{\cal K}_{\Big[[AB]\big[[AA]B\big]\Big]} =
\Big\{ \{{\cal K}_A,{\cal K}_B\},\big\{\big\{{\cal K}_A,{\cal K}_A\},{\cal K}_B\big\}\Big\}
$$
(this example is for the case of two keystones ${\cal K}_A$ and ${\cal K}_B$).
This is a very clear labeling, but definitely different from the one natural for the
coset ${\cal S}_m^r$ within the framework of symmetric group theory.
An interplay between the two is an important and challenging problem.
Moreover, the tree labeling is incomplete: the join operation has a huge cokernel in ${\cal S}_m^r$,
and there are many operators that are not tree descendants of the keystones.
Still, they can be produced by the cut operation $\Delta$ and then should be included
into an RG-complete non-perturbative partition function
as a new {\it secondary keystone} operators.

The most important feature of $\Delta$ is its {\it degeneracy},
which means that the partition function actually depends on less number of variables then one
would think.
The problem could be seen even at the matrix model level.
Imagine that we have made an erroneous choice of the operator set and included
into the action the terms like $t_{k,l}\Tr M^k\Tr M^l$.
Clearly, many of them will be mapped into the same operators by the action of $\Delta$,
and the result of this will be that the partition function satisfies some
``trivial" relations such as
$$
\frac{\p {\cal Z}}{\p t_{k,l}} = \frac{\p^2 {\cal Z}}{\p t_k\p t_l}
$$
which reduce it to a function of $t_k$'s only.
In this matrix model example, this simply means that one should include only
{\it connected} operators into the action.
Remarkably, even this restriction is already non-trivial for tensor models:
description of the ``connected subset" of ${\cal S}_m^r$ is not fully straightforward
in symmetric group language.
But, in fact, even for the connected operators the operation $\Delta$ remains degenerate:
this is simply because the number of different gauge-invariant operators increases
with level, unlike in the matrix model case, where there is just one operator
$\Tr (M\bar M)^m$ at each level $m$.
This means that the actual set of independent time variables is much smaller
than all elements of $({\cal S}_m^{r})_{conn}$,
and one needs to look for an appropriate set of bases from this point of view.

At the same time, images of the operations $\{\,\}$ and $\Delta$ at a given level $m$
are essentially different: not all of the operators which $\Delta$ creates from
the  operators of level $m+1$ are multilinear combinations of those produced by
$\{\,\}$ from the lower levels: in this sense, the set of operators is bigger than
generated by $\{\,\}$, and this is contrary to what happens in matrix models (at $r=2$).
In particular, the original (primary) keystone operators need to be complemented
by secondary keystones arising from $\Delta$-images of the join-descendants of the
original ones.

The next problem is that the labeling that is relevant for the Lie algebra structure and
underlies the join operation $\{\,\}$, is not immediately consistent with the
one separating non-degenerate subspace for the cut operation $\Delta$,
and, as we already said,
{\it neither} is provided by the symmetric group theory, at least, naively.

This completes our brief survey of the problems that we see when approaching
the tensor model theory and which motivate our study in the present paper.
Hopefully, these comments would help the reader to get through these examples
and extract technical lessons which, at the next stage, can be used in attacking
the generating function problem and appropriate formulation of
Virasoro-like
(actually, the Bogoliubov-Zimmermann rooted-tree algebra \cite{CK,GMSel,Virtree,GurVir,RA,ART,IMMten2})
relations
and associated version of the AMM/EO topological recursion.

Throughout the paper, we use in examples various concrete gauge-invariant operators from levels $m=1,2,3,4,5$. These operators are all listed in Appendix A, hence, all notation of operators can be found there. In Appendix B, we collected tables describing various numbers of operators and their association with permutations.

\section{Models, operators and Gaussian averages}
\setcounter{equation}{0}

The models referred to as rainbow models are the ones with $|I|$ complex tensors $M^I_{a_1\ldots a_r}$ of the rank $r$,
with $I\in \{1,2,\ldots,|I|\}$
and with the ``gauge" symmetry ${\cal U}=U(N_1)\otimes\ldots\otimes U(N_r)$.
The simplest among these are
\begin{itemize}
\item The Aristotelian model with a single tensor, $|I|=1$: it includes the vector model
at $r=1$, the rectangular matrix model (RCM)
for $r=2$, {\bf the Aristotelian (red-green-blue) model of \cite{IMMten2} {\it per se}
at $r=3$}, and many more models with arbitrary $r>3$
\item The $AB$ model with $|I|=2$, i.e. with the two tensors of rank $r$ named $A$ and $B$:
it includes the peculiar two-matrix model at $r=2$
\item The 3-tensor $ABC$ models with $|I|=3$: at the matrix model level of $r=2$,
the interesting chiral keystone  operator
(see \cite{IMMten2}) $\Tr ABC$ appears
\item The tetrahedron ($ABCD$) model with $|I|=4$:
the interesting chiral keystone
operator (tetrahedron vertex) appears at $r=3$
\item The starfish models \cite{IMMten1} with interesting starfish keystone operators
at $r=|I|-1$
\item $\ldots$
\end{itemize}

\noindent
Boldfaced in this list is the Aristotelian model, which we will actually focus on in the
present paper.

\bigskip

The operators of interest are invariants of ${\cal U}$, made by contraction of all indices
of $m$ tensors $M$ and $m$ complex conjugates of tensors $\bar M$.
We call $m$ the ``level" of the operator,
especially simple being invariants at level $1$:
\be
\Tr M^I\bar M_J = \sum_{a_1=1}^{N_1}\ldots \sum_{a_r = 1}^{N_r} M_{a_1\ldots a_r}^I
\bar M_J^{a_1\ldots a_r}
\ee
In the case
of bilinear operators, we may denote the obvious contraction of all indices
by ``Tr", though in general the notion of trace has no direct meaning for tensors.
These operators at level one are used in the definition of Gaussian averages,
when the action is given by the bilinear kinetic term
\be
\sum_I \Tr M^I\bar M_I
\label{quadact}
\ee
We do not consider the space-time dependence, because it adds nothing
new to the combinatorial aspect of the story, which is our main interest in this paper.
In what follows, we also put $|I|=1$, i.e. consider a single complex tensor.

At level $m=2$, the  gauge-invariant operators are made from
$$
M_{a_1\ldots a_r}M_{a'_1\ldots a'_r}\bar M^{b_1\ldots b_r}\bar M^{b'_1\ldots b'_r}
$$
by the contraction of each pair $b_i,b'_i$ with the corresponding  $a_i,a'_i$
(one can not contract $b_i$ and $a_j$  with $i\neq j$ because of the huge
symmetry of the rainbow models).
There are two possibilities for each $i$: one can put $b_i=a_i, b'_i=a'_i$ or put
$b_i=a'_i, b'_i=a_i$, i.e. a total of $2^r$ choices labeled by $r$ permutations
from the symmetric group $S_2$ (of permutations of two elements).
To write down a formula, we need to change the notation, from $'$ to numeric superscript,
taking values $1$ and $2$, say, $a_i=a_i^1$, $a_i'=a_i^2$,
and the $2^r$ gauge-invariant operators at level $2$ are
\be
{\cal K}_{\vec\sigma\in S_2^{\otimes r}} =
M_{a_1^1\ldots a_r^1}M_{a_1^2\ldots a_r^2} \,
\bar M^{a_1^{\sigma_1(1)}\ldots\ a_r^{\sigma_r(1)}}
\bar M^{a_1^{\sigma_1(2)}\ldots\ a_r^{\sigma_r(2)}  }
\ee
At level $m$, the gauge-invariant operators are labeled by $r$ permutations
from the symmetric group $S_m$:
\be
{\cal K}_{\sigma_1\otimes\ldots\otimes\sigma_r}^{(m)} =
\prod_{p=1}^m M_{a_1^p\ldots a_r^p} \bar M^{a_1^{\sigma_1(p)}\ldots\ a_r^{\sigma_r(p)}}
\label{gop}
\ee
In fact, one can now permute the $m$ tensors $M$ or the $m$ tensors $\bar M$
as a whole, i.e. multiply all the permutations $\sigma_i$ by two common permutations,
from the right and from the left sides, which factorizes
$S_m^{\otimes r}$ by $S_m$ both from the left and from the right sides
and provides the double coset \cite{GR,IMMten2,Di,MMten,KGT}
\be
{\cal S}_m^r = S_m\backslash S_m^{\otimes r}/S_m
\label{coset}
\ee
An explicit description/parameterization of this coset
can begin from violating the ``symmetry" $S_r$ between different
indices (i.e. colorings), for example, by always putting $\sigma_1=id$
(this ``symmetry" is, in any case, violated by the difference between $N_i$
in the gauge groups $U(N_i)$).
This leaves us with $(m!)^{r-1}$ classes of operators at level $m$,
in particular, with $m!$ for the complex matrix model ($r=2$)
and with $(m!)^2$ for the Aristotelian model with the tensor of rank $r=3$.
These are still not the minimal classes:
the remaining freedom is the common {\it conjugation}.
For complex matrix model, this means that the gauge-invariant operators are
enumerated by conjugacy classes in $S_m$, which are labeled by Young diagrams $\mu$:
\be
r=2:& \ \ \ \ &
{\cal K}_\mu = \prod_{k=1}^{l(\mu)} \Tr (M\bar M)^{\mu_k},
\ \ \ \ \ \ \ \mu=\{\mu_1\geq\mu_2\geq\ldots\geq \mu_{l(\mu)}>0\},
\ \ \ \ \ \ \ \ |\mu|=\sum_{k=1}^{l(\mu)} \mu_k=m \ \ \ \ \ \
\ee
For the Aristotelian model, we get two permutations $\sigma_2\otimes \sigma_3$
modulo common conjugation:
\be
r=3: &  \ \ \ &
{\cal K}_{<\sigma_2\otimes\sigma_3>} = \sum_{\{\vec a,\vec b,\vec c\}}
\left(
\prod_{p=1}^m M_{a_pb_pc_p}\bar M^{a_pb_{\sigma_2(p)}c_{\sigma_3(p)}}
\right)
\nn\\
&&\!\!\!\!\!\!\!\!\!\!\!\!
\sigma_2\otimes\sigma_3^{\pm 1}\cong (\sigma\otimes\sigma)\circ
(\sigma_2\otimes\sigma_3^{\pm 1})\circ(\sigma^{-1}\otimes\sigma^{-1}),
\ \ \ \ \ {\rm i.e.} \ \ \ \
\left\{\begin{array}{c}
\sigma_2 \cong \sigma\circ\sigma_2\circ\sigma^{-1} \\
\sigma_3 \cong \sigma\circ\sigma_3\circ\sigma^{-1} \\
\sigma_2\circ\sigma_3^{\pm 1} \cong \sigma\circ\sigma_2\circ\sigma_3^{\pm 1}\circ\sigma^{-1}
\end{array}\right.
\label{conjAr}
\ee
The first task in any study of tensor models is to describe these conjugacy classes.
This was partly done in  \cite{GR,IMMten2,Di,MMten,KGT,GeRamg},
but, as explained in sec.\ref{tasks} above, much more details are actually needed
and different relevant classification schemes should be somehow matched.
As was also mentioned, the useful tool
(though of a limited capacity),
which allows one to illustrate general arguments by explicit formulas,
is use of the Gaussian averages,
which are integrals with the quadratic action (\ref{quadact})
and are defined by the Wick theorem:
\be
\left<\!\!\!\left<
\prod_{p=1}^m M_{a_{ 1}^p\ldots a_{r}^p} \prod_{p=1}^m \bar M^{b_{1}^p\ldots b_{r}^p}
\right>\!\!\!\right> \ =
\sum_{\gamma\in S_m} \left( \prod_{p=1}^m\prod_{i=1}^r \,
\delta_{a_{i}^p}^{b_i^{\gamma(p)}} \right)
\label{Wth}
\ee
Hereafter, we normalize the averages so that $\Big<1\Big>=1$.
The Gaussian average of an arbitrary operator at (\ref{gop})
is, therefore, known in full generality \cite{MMten}, see also \cite{Di} and \cite{KGT}:
\be
\Big<\!\!\Big<
{\cal K}^{(m)}_{\underbrace{\sigma_1\otimes\ldots\otimes \sigma_r}_{ \vec\sigma}}
\Big>\!\!\Big>\
= \sum_{\gamma\in S_m} \prod_{s=1}^r N_s^{\#(\gamma\circ\sigma_s)}
=\sum_{\vec R\,\vdash m}\left( \prod_{s=1}^r D_{_{R_s}}(N_s) \psi_{\vec R}(\vec\sigma)\right)
\ee
where $D_{_R}(N)=\chi_{_R}\{p_k=N\}$ is the dimension of representation $R$ of $sl(N)$,
given by the hook formula, and
\be
\psi_{\vec R}(\vec\sigma) = \sum_{\gamma\in S_m}
\left(\prod_{s=1}^r \psi_{_{R_s}}(\gamma\circ\sigma_s)\right)
\ee
The sums in these formulas are finite and run over $r$ Young diagrams $R_1,\ldots,R_r$,
of the size $m$ each,
and over $2^m$ permutations from the symmetric group $S_m$.
The symmetric group characters $\psi_{_R}(\sigma)$ depend only on the conjugacy class
of the permutation $\sigma$, i.e. on the associated Young diagram $[\sigma]$
and are easily available in MAPLE and Mathematica. The degeneracy, the size of the
conjugacy class $[\sigma]$ is equal to $\frac{m!}{z_{[\sigma]}}$.

As noted already in \cite{IMMten2}, some of the Gaussian averages are actually factorized.
This usually happens, when the Young diagram $[\sigma_s]$ has single-box lines
and follows from the factorization of the sum over cycles
\be
\sum_{\gamma\in S_m} N^{\#(\gamma)} = N(N+1)\ldots (N+m-1) = \frac{\Gamma(N+m)}{\Gamma(N)}
\ee
and its generalizations such as
\be
\sum_{\gamma\in S_m} N_1^{\#(\gamma\circ\sigma_{m-k})} N_2^{\#(\gamma)} =
\frac{\Gamma(N_1N_2+2k)}{\Gamma(N_1N_2+k)} \sum_{\gamma\in S_{m-k}} N_1^{\#(\gamma\circ\hat\sigma_{m-k})}N_2^{\#(\gamma)}
\ee
where $\sigma_{m-k}\in S_m$ is a permutation that contains $k$ unit cycles and $\hat\sigma_{m-k}\in S_{m-k}$ is the same permutation with the unit cycles dropped off.
The explanation of these formulas is that
each permutation from $S_{m+1}$ is a composition of a permutation
from $S_m$ and an additional permutation given by the length 2 cycle: $(i,m+1)$.
Then, for $i=m+1$, we have $\#(id\circ \sigma) = \#(\sigma)+1$, while for all other
$i=1,\ldots,m$ the number of cycles remains intact: $\#\Big((i,m+1)\circ \sigma\Big)
= \#\Big(\sigma\Big)$ for all $\sigma\in S_m$.
This means that for $\sigma\in S_m$, i.e. for $\sigma\otimes (m+1)\in S_{m+1}$
\be
\sum_{\gamma\in S_{m+1}} N^{\#\big(\gamma\circ\left(\sigma\otimes (m+1)\right)\big)} =
(N+m)\sum_{\gamma\in S_m} N^{\#(\gamma\circ\sigma)}
\ee
and this provides the necessary factorization when the cycle of unit length is added.
Generalization from $N$ to $N_1,\ldots,N_r$ is straightforward.

As a first illustration for the structure of operator set and for its Gaussian averages,
we list them for the few simplest levels $m$.
For $m=1$, there is one diagram and one correlator equal to $N_1N_2N_3$.
For $m=2$, there are one connected correlator with the  average $N_1N_2N_3(N_1N_2+N_3)$
(it is one correlator modulo permutations of colourings)
and one disconnected correlator, coming from the previous level: its average is $N_1N_2N_3(N_1N_2N_3+1)$, etc:
$$
\begin{array}{|c|c|c|c|}
\hline
&&&\\
m=1 & 1 & 1 & {\cal K}_1={\cal K}_{id,id,id}\\
&&&\\
\hline
&&&\\
m=2& 1 &N_1N_2N_3(N_1N_2+N_3) &{\cal K}_{\cre 2}={\cal K}_{id,(12),(12)}\\
&&&\\
&1& N_1N_2N_3(N_1N_2N_3+1) & {\cal K}_1^2={\cal K}_{id,id,id} \\
&&&\\
\hline
&&&\\
m=3& 3&N_1N_2N_3(3N_1N_2N_3+N_1^2+N_2^2+N_3^2) &{\cal K}_{W} = {\cal K}_{id,(123),(132)}\\
&&N_1N_2N_3(N_2^2N_3^2+3N_1N_2N_3+N_1^2+1)\ &{\cal K}_{\cre 3} = {\cal K}_{id,(123),(123)} \\
&& N_1N_2N_3(N_1N_2N_3^2+N_1^2N_3+N_2^2N_3+2N_1N_2+N_3)\
&{\cal K}_{{\cg 2},{\cb 2}} = {\cal K}_{id,(12),(13)}\\
&&&\\
&2&N_1N_2N_3(N_1N_2+N_3)(N_1N_2N_3+2)\
&{\cal K}_{\cre 2}{\cal K}_1 = {\cal K}_{id,(12),(12)} \\
&& N_1N_2N_3(N_1N_2N_3+1)(N_1N_2N_3+2)\ &{\cal K}_1^3={\cal K}_{id,id,id}\\
&&&\\
\hline
\end{array}
$$
where $(N_1N_2+N_3)(N_1N_2N_3+2)$ at level 3 comes
from the product of connected correlators from $S_1$ and $S_2$
and $(N_1N_2N_3+1)(N_1N_2N_3+2)$
comes from the product of three connected correlators from $S_1$.
An extended and detailed version of this table can be found in sec.\ref{exalevels},
see, in particular, (\ref{ops3}) and (\ref{ops3G}).

\section{Cut and join operations and Virasoro-like recursions: a primer
\label{reminderCJ} }
\setcounter{equation}{0}

For the rectangular complex matrix model (RCM),
the elements of the coset
${\cal S}^2_m = S_m\backslash S_m\otimes S_m/S_m$, i.e. operators
\be
{\cal K}_\mu^{^{RCM}} = \prod_{i=1}^{l_\mu} {\cal K}_{m_i}^{^{RCM}}
\ee
forming the linear basis of the operator ring, are labeled by Young diagrams
$\mu=\{m_1\geq m_2\geq \ldots m_{l_\mu}>0\}$ of the size $m$,
and all connected (non-factorizable) operators
\be
{\cal K}_m^{^{RCM}}=\Tr (M\bar M)^m = \sum_{\vec a,\vec b}
\left(\prod_{p=1}^m M_{a_pb_p}\bar M^{a_pb_{p+1}}\right)=\sum_{\vec a,\vec b}
\left(\prod_{p=1}^m M_{a_pb_p}\bar M^{a_pb_{\sigma(p)}}\right)
\ee
with $\sigma(p) = (123\ldots m)$ being the longest cycle, are represented by polygons (red-green cycles) of the size $2m$: we depict one example $m=3$,

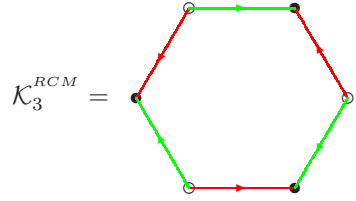
\begin{picture}(300,80)(-220,-40)
\put(-47,-2){\mbox{${\cal K}_{3}^{^{RCM}}=  $}}
\put(0,0){\circle*{4}}
\put(20,34){\circle{4}}
\put(60,34){\circle*{4}}
\put(80,0){\circle{4}}
\put(20,-34){\circle{4}}
\put(60,-34){\circle*{4}}
{\cre
\qbezier(0,0)(10,17)(20,34)\put(10,17){\vector(-1,-2){2}}
\qbezier(80,0)(70,17)(60,34)\put(70,17){\vector(-1,2){2}}
\qbezier(20,-34)(40,-34)(60,-34)\put(40,-34){\vector(1,0){2}}
}
\put(0,0){{\cg
\qbezier(0,0)(10,-17)(20,-34)\put(10,-17){\vector(-1,2){2}}
\qbezier(80,0)(70,-17)(60,-34)\put(70,-17){\vector(-1,-2){2}}
\qbezier(20,34)(40,34)(60,34)\put(40,34){\vector(1,0){2}}
}}
\end{picture}

\noindent
The index is just the Dedekind function
\be
\eta_{_{RCM}}(q) = \prod_{m=1}^\infty \frac{1}{1-q^m} = {\rm PE}\left(\frac{q}{1-q}\right),
\ \ \ \ \ \ \ \ \ \ \ \
\eta^{\rm conn}_{_{RCM}}(q) = \frac{q}{1-q}
\ee
where we have denoted by PE plethystic exponential (the Euler transform) \cite{Little}.
Finding this function for a more complicated tensor model is a less trivial
exercise.
Even if that is solved, however, it does not provide enough information for building a
reasonable generating function: some deeper structures on the ring must be
revealed for this.

The first two important operations on the ring of gauge-invariant operators are
the cut
\be
\Delta{\cal K}
= \Tr\frac{\p^2\,{\cal K}}{\p M\p\bar M}
= \sum_{a_1,\ldots a_r}
\frac{\partial^2 {\cal K}}{\partial M_{a_1\ldots a_r}\partial\bar M^{a_1\ldots a_r}}
\ee
and join
\be
\{{\cal K},{\cal K}'\} = \sum_{a_1,\ldots a_r} \frac{\p {\cal K}}{\p M_{a_1\ldots a_r}}
\cdot\frac{\p {\cal K}'}{\p\bar M^{a_1\ldots a_r}}
\ee
In fact, there are many different possibilities of choosing the cut and join operations, with different properties. Our choice in this paper serve as an archetypical example and enjoys additional interesting structures. Note that this join operation is definitely not like a Poisson bracket, it is neither associative, {\it nor} antisymmetric,
in fact, it is also not always symmetric, though non-symmetric examples first appear at level $m=5$:
e.g. the join operation involving the black-white asymmetric operator ${\cal K}_{XVIII}$.
They necessarily appear in the description of Virasoro-like recursions for the averages
\cite{UFN3,Virtree,GurVir,IMMten2}:
as we explained in the previous section, if the action $S$ is some combination of gauge-invariant operators, then the
averages (functional integrals) satisfy the Ward identities,
following from invariance under the shift of integration variables
$\delta \bar M = \nabla_M {\cal K}$:
\be
\Big< \{{\cal K},S\} \Big>_S = \hbar\, \Big<\Delta{\cal K}\Big>_S
\label{recS}
\ee
or, more generally,
\be
 \Big< \{{\cal K},S\}\cdot{\cal K}' \Big>_S =
 \Big<\{{\cal K},{\cal K}'\}\Big>_S +
\hbar\, \Big<\Delta{\cal K}\cdot{\cal K}'\Big>_S
\ee
where we restored for a moment the Plank constant $\hbar$ in order to emphasize that the cut operation comes from the variation of measure in the path integral.
For Gaussian averages, when $S = {\cal K}_1=\Tr M\bar M$,
the l.h.s. of (\ref{recS}) reduces to rescaling of ${\cal K}$,
which just multiplies the operator by its degree in $M$:
\be
{\rm deg}_{\cal K}\cdot \Big<\!\!\Big<{\cal K}\Big>\!\!\Big>
= \hbar\, \Big<\!\!\Big<\Delta{\cal K}\Big>\!\!\Big>
\ee
Iteration of this formula gives for the operator of degree ${\rm deg}_{\cal K}^{(m)} = m$
\be
\Big<\!\Big<{\cal K}^{(m)}\Big>\!\Big>
= \frac{\hbar^m}{m!} \,\Big<\!\Big<\Delta^m {\cal K}^{(m)}\Big>\!\Big>\,,
\ee
where $\Delta^m {\cal K}^{(m)}$ is just a number:
this expression is nothing but the Wick theorem for the Gaussian correlators.
In what follows, we omit $\hbar$, which counts the degree (grading)
and can be easily restored.

Let us choose a set of ``keystone" operators \cite{IMMten2}, which is a subset of the whole graded ring ${\cal R}$ of gauge-invariant operators. This subset
generates a sub-ring ${\cal R}_{\{,\};\Delta}$ by application of the addition, multiplication,
cut and join operations
(i.e. is not just a set of multiplicative generators of the ring) and
introduces this way an additional structure in the operator ring:
all other (non-keystone) operators can be represented as ``descendants" of the keystone ones,
and what matters is their ``depth", the number of times the
cut and join operators are applied to produce them from the keystone ones.
If the operator belongs to the sub-ring generated only by the join operation, this operator is of the ``tree" type,
otherwise, it is of the ``loop" type.
This structure is at the operator level, and does not depend on the choice
of the action and manifests itself in all the averages, not being limited to the Gaussian ones.
For the matrix RCM, the action of the introduced cut and join operations is
\be
\Delta{\cal K}_m^{^{RCM}} = m(N_1+N_2){\cal K}_{m-1}^{^{RCM}}
+ m\sum_{k=1}^{m-2} {\cal K}_k^{^{RCM}}{\cal K}_{m-k-1}^{^{RCM}}\nn\\
\{{\cal K}_m^{^{RCM}},{\cal K}_n^{^{RCM}}\} = mn\,{\cal K}_{m+n-1}^{^{RCM}}
\ee
i.e. it closes on the ring, and the keystone operator ${\cal K}_2^{^{RCM}}$
serves just as a multiplicative generator.
Already at the Aristotelian model, the situation changes: while
the keystone set is provided by the
three operators ${\cal K}_{\cre 2}$, ${\cal K}_{\cg 2}$, ${\cal K}_{\cb 2}$,
the ring itself is far more complicated:
keystones are not longer its multiplicative generators,
they generate the ring only if cut and join operators are added.

Convenient for study of the cut and join operations is
the following pictorial representation of operators
${\cal K}^{(m)}_{{\cg \sigma_2},{\cb \sigma_3}}$:

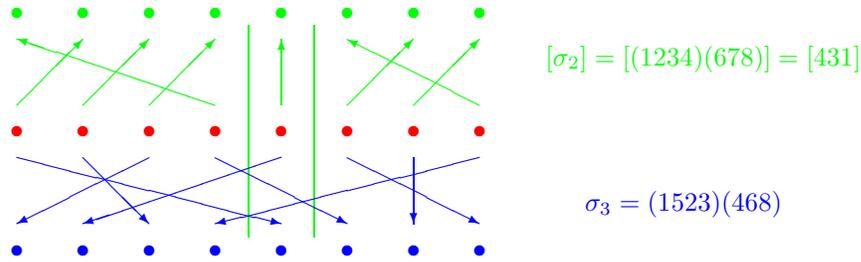
\begin{picture}(400,120)(-100,-60)

\put(0,0){{\cre
\put(0,0){\circle*{4}}
\put(25,0){\circle*{4}}
\put(50,0){\circle*{4}}
\put(75,0){\circle*{4}}
\put(100,0){\circle*{4}}
\put(125,0){\circle*{4}}
\put(150,0){\circle*{4}}
\put(175,0){\circle*{4}}
}}

\put(0,0){{\cg
\put(0,45){\circle*{4}}
\put(25,45){\circle*{4}}
\put(50,45){\circle*{4}}
\put(75,45){\circle*{4}}
\put(100,45){\circle*{4}}
\put(125,45){\circle*{4}}
\put(150,45){\circle*{4}}
\put(175,45){\circle*{4}}
\put(0,10){\vector(1,1){25}}
\put(25,10){\vector(1,1){25}}
\put(50,10){\vector(1,1){25}}
\put(75,10){\vector(-3,1){75}}
\put(87.5,40){\line(0,-1){80}}
\put(100,10){\vector(0,1){25}}
\put(112.5,40){\line(0,-1){80}}
\put(125,10){\vector(1,1){25}}
\put(150,10){\vector(1,1){25}}
\put(175,10){\vector(-2,1){50}}
\put(200,25){\mbox{$[\sigma_2]=[(1234)(678)]=[431]$}}
}}

\put(0,0){{\cb
\put(0,-45){\circle*{4}}
\put(25,-45){\circle*{4}}
\put(50,-45){\circle*{4}}
\put(75,-45){\circle*{4}}
\put(100,-45){\circle*{4}}
\put(125,-45){\circle*{4}}
\put(150,-45){\circle*{4}}
\put(175,-45){\circle*{4}}
\put(0,-10){\vector(4,-1){100}}
\put(25,-10){\vector(1,-1){25}}
\put(50,-10){\vector(-2,-1){50}}
\put(75,-10){\vector(2,-1){50}}
\put(100,-10){\vector(-3,-1){75}}
\put(125,-10){\vector(2,-1){50}}
\put(150,-10){\vector(0,-1){25}}
\put(175,-10){\vector(-4,-1){100}}
\put(215,-30){\mbox{$\sigma_3=(1523)(468)$}}
}}

\end{picture}

\noindent
The green permutation ${\cg\sigma_2}$ is in the canonical form
of the Young diagram, and the blue one ${\cb\sigma_3}$ is just a permutation
(defined modulo conjugations that leave ${\cg\sigma_2}$ intact).
The operator is connected, if the vertical green lines do not cut the blue
permutation ${\cb\sigma_3}$
into independent pieces (collections of cycles).
Here is an example of disconnected operator:
${\cal K}^{(8)}_{{\cg (1234)(678)},{\cb(1253)(68)}} =
{\cal K}^{(5)}_{{\cg (1234) },{\cb(1253) }}\cdot
{\cal K}^{(3)}_{{\cg (123)},{\cb (13)}} $

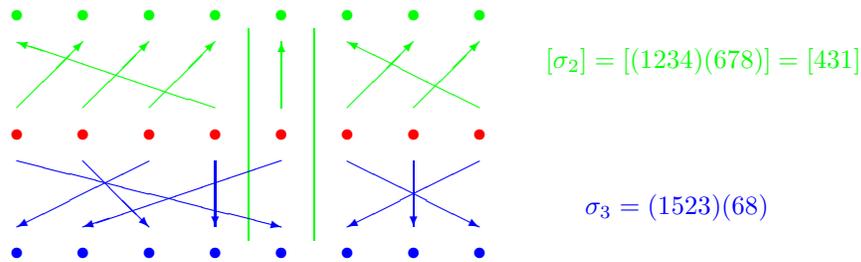
\begin{picture}(400,120)(-100,-60)

\put(0,0){{\cre
\put(0,0){\circle*{4}}
\put(25,0){\circle*{4}}
\put(50,0){\circle*{4}}
\put(75,0){\circle*{4}}
\put(100,0){\circle*{4}}
\put(125,0){\circle*{4}}
\put(150,0){\circle*{4}}
\put(175,0){\circle*{4}}
}}

\put(0,0){{\cg
\put(0,45){\circle*{4}}
\put(25,45){\circle*{4}}
\put(50,45){\circle*{4}}
\put(75,45){\circle*{4}}
\put(100,45){\circle*{4}}
\put(125,45){\circle*{4}}
\put(150,45){\circle*{4}}
\put(175,45){\circle*{4}}
\put(0,10){\vector(1,1){25}}
\put(25,10){\vector(1,1){25}}
\put(50,10){\vector(1,1){25}}
\put(75,10){\vector(-3,1){75}}
\put(87.5,40){\line(0,-1){80}}
\put(100,10){\vector(0,1){25}}
\put(112.5,40){\line(0,-1){80}}
\put(125,10){\vector(1,1){25}}
\put(150,10){\vector(1,1){25}}
\put(175,10){\vector(-2,1){50}}
\put(200,25){\mbox{$[\sigma_2]=[(1234)(678)]=[431]$}}
}}

\put(0,0){{\cb
\put(0,-45){\circle*{4}}
\put(25,-45){\circle*{4}}
\put(50,-45){\circle*{4}}
\put(75,-45){\circle*{4}}
\put(100,-45){\circle*{4}}
\put(125,-45){\circle*{4}}
\put(150,-45){\circle*{4}}
\put(175,-45){\circle*{4}}
\put(0,-10){\vector(4,-1){100}}
\put(25,-10){\vector(1,-1){25}}
\put(50,-10){\vector(-2,-1){50}}
\put(75,-10){\vector(0,-1){25}}
\put(100,-10){\vector(-3,-1){75}}
\put(125,-10){\vector(2,-1){50}}
\put(150,-10){\vector(0,-1){25}}
\put(175,-10){\vector(-2,-1){50}}
\put(215,-30){\mbox{$\sigma_3=(1523)(68)$}}
}}

\end{picture}

The action of the cut operation
$\Delta = \frac{\partial^2}{\partial M_{{\cre i}{\cg j}{\cb k}}
\partial \bar M^{{\cre i}{\cg j}{\cb k}}}$
produces a double sum over indices $p,q=1,\ldots m$ which label the $M$ and $\bar M$
tensors respectively:
\be
\Delta {\cal K}^{(m)}_{{\cg \sigma_2},{\cb \sigma_3}}
= \sum_{p,q}  {\cal K}^{(m-1)}_{{\cg \sigma_2^{(p,q)}}{\cb \sigma_3^{(p,q)}}}
\ee
where $\sigma^{(p,q)}\in S_{m-1}$ are best described pictorially. Consider an example of $p=3$ and $q=7$:

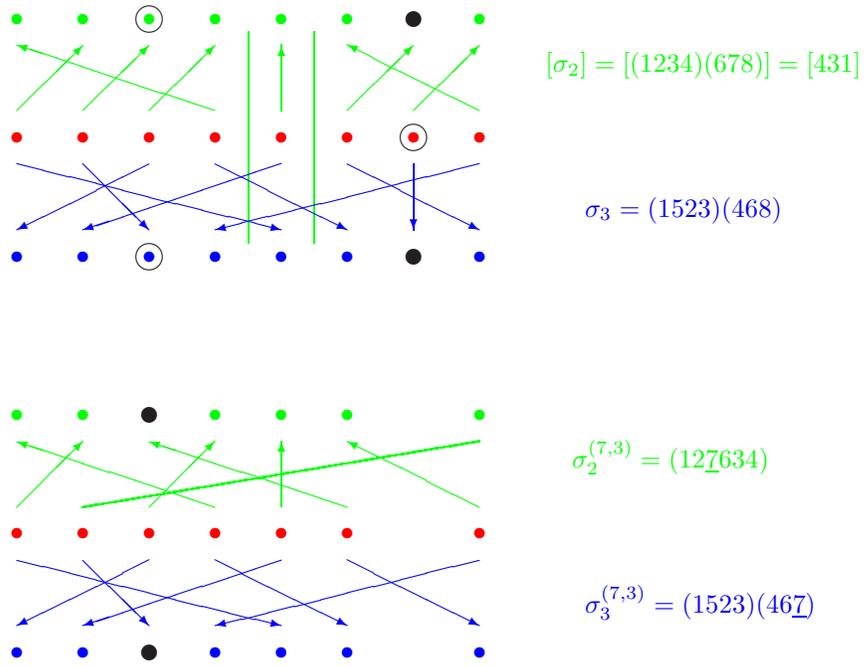
\begin{picture}(400,270)(-100,-210)

\put(0,0){{\cre
\put(0,0){\circle*{4}}
\put(25,0){\circle*{4}}
\put(50,0){\circle*{4}}
\put(75,0){\circle*{4}}
\put(100,0){\circle*{4}}
\put(125,0){\circle*{4}}
\put(150,0){\circle*{4}}
\put(175,0){\circle*{4}}
}}

\put(0,0){{\cg
\put(0,45){\circle*{4}}
\put(25,45){\circle*{4}}
\put(50,45){\circle*{4}}
\put(75,45){\circle*{4}}
\put(100,45){\circle*{4}}
\put(125,45){\circle*{4}}
\put(150,45){\circle*{4}}
\put(175,45){\circle*{4}}
\put(0,10){\vector(1,1){25}}
\put(25,10){\vector(1,1){25}}
\put(50,10){\vector(1,1){25}}
\put(75,10){\vector(-3,1){75}}
\put(87.5,40){\line(0,-1){80}}
\put(100,10){\vector(0,1){25}}
\put(112.5,40){\line(0,-1){80}}
\put(125,10){\vector(1,1){25}}
\put(150,10){\vector(1,1){25}}
\put(175,10){\vector(-2,1){50}}
\put(200,25){\mbox{$[\sigma_2]=[(1234)(678)]=[431]$}}
}}

\put(0,0){{\cb
\put(0,-45){\circle*{4}}
\put(25,-45){\circle*{4}}
\put(50,-45){\circle*{4}}
\put(75,-45){\circle*{4}}
\put(100,-45){\circle*{4}}
\put(125,-45){\circle*{4}}
\put(150,-45){\circle*{4}}
\put(175,-45){\circle*{4}}
\put(0,-10){\vector(4,-1){100}}
\put(25,-10){\vector(1,-1){25}}
\put(50,-10){\vector(-2,-1){50}}
\put(75,-10){\vector(2,-1){50}}
\put(100,-10){\vector(-3,-1){75}}
\put(125,-10){\vector(2,-1){50}}
\put(150,-10){\vector(0,-1){25}}
\put(175,-10){\vector(-4,-1){100}}
\put(215,-30){\mbox{$\sigma_3=(1523)(468)$}}
}}

\put(150,0){\circle{10}}
\put(50,45){\circle{10}}
\put(50,-45){\circle{10}}
\put(150,45){\circle*{6}}
\put(150,-45){\circle*{6}}

\put(0,-150){
\put(0,0){{\cre
\put(0,0){\circle*{4}}
\put(25,0){\circle*{4}}
\put(50,0){\circle*{4}}
\put(75,0){\circle*{4}}
\put(100,0){\circle*{4}}
\put(125,0){\circle*{4}}
\put(175,0){\circle*{4}}
}}

\put(0,0){{\cg
\put(0,45){\circle*{4}}
\put(25,45){\circle*{4}}
\put(50,45){\circle*{4}}
\put(75,45){\circle*{4}}
\put(100,45){\circle*{4}}
\put(125,45){\circle*{4}}
\put(175,45){\circle*{4}}
\put(0,10){\vector(1,1){25}}
\qbezier(25,10)(100,22.5)(175,35)
\put(50,10){\vector(1,1){25}}
\put(75,10){\vector(-3,1){75}}
\put(100,10){\vector(0,1){25}}
\put(125,10){\vector(-3,1){75}}
\put(175,10){\vector(-2,1){50}}
\put(210,25){\mbox{$\sigma_2^{(7,3)}=(12\underline{7}634) $}}
}}

\put(0,0){{\cb
\put(0,-45){\circle*{4}}
\put(25,-45){\circle*{4}}
\put(50,-45){\circle*{4}}
\put(75,-45){\circle*{4}}
\put(100,-45){\circle*{4}}
\put(125,-45){\circle*{4}}
\put(175,-45){\circle*{4}}
\put(0,-10){\vector(4,-1){100}}
\put(25,-10){\vector(1,-1){25}}
\put(50,-10){\vector(-2,-1){50}}
\put(75,-10){\vector(2,-1){50}}
\put(100,-10){\vector(-3,-1){75}}
\put(125,-10){\vector(2,-1){50}}
\put(175,-10){\vector(-4,-1){100}}
\put(215,-30){\mbox{$\sigma_3^{(7,3)}=(1523)(46\underline{7})$}}
}}

\put(50,45){\circle*{6}}
\put(50,-45){\circle*{6}}
}

\end{picture}

\noindent
The black circles move here from $q$ to $p$,
together with the arrows which pointed at them,
while the arrows which pointed to the white circles become pointing to the images of the white circle in the middle.
Underlined are the elements which are re-numbered (shifted by unity) when
switching from permutation from $S_m$ to those from $S_{m-1}$.
If $p=q$ then one puts an extra factor  ${\cre N_1}$,
while in the case, where two white circles were connected by an arrow, one puts
factors ${\cg N_2}$ or ${\cb N_3}$.

\bigskip

{\bf Example:} $\Delta{\cal K}^{(3)}_{{\cre 2},{\cg 2}}=
\Delta {\cal K}^{(3)}_{{\cg(123)},{\cb (12)}}$

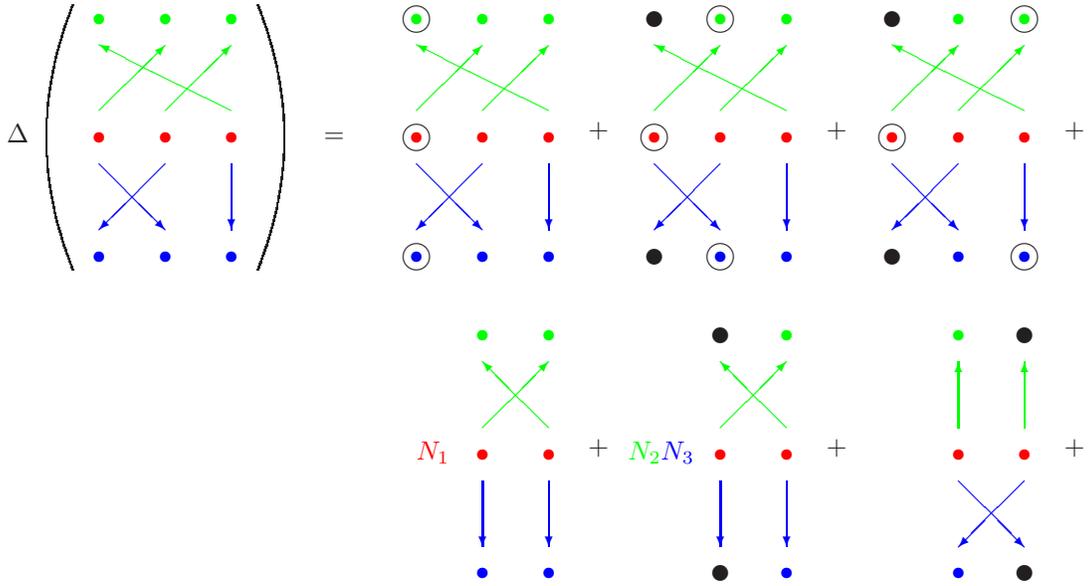
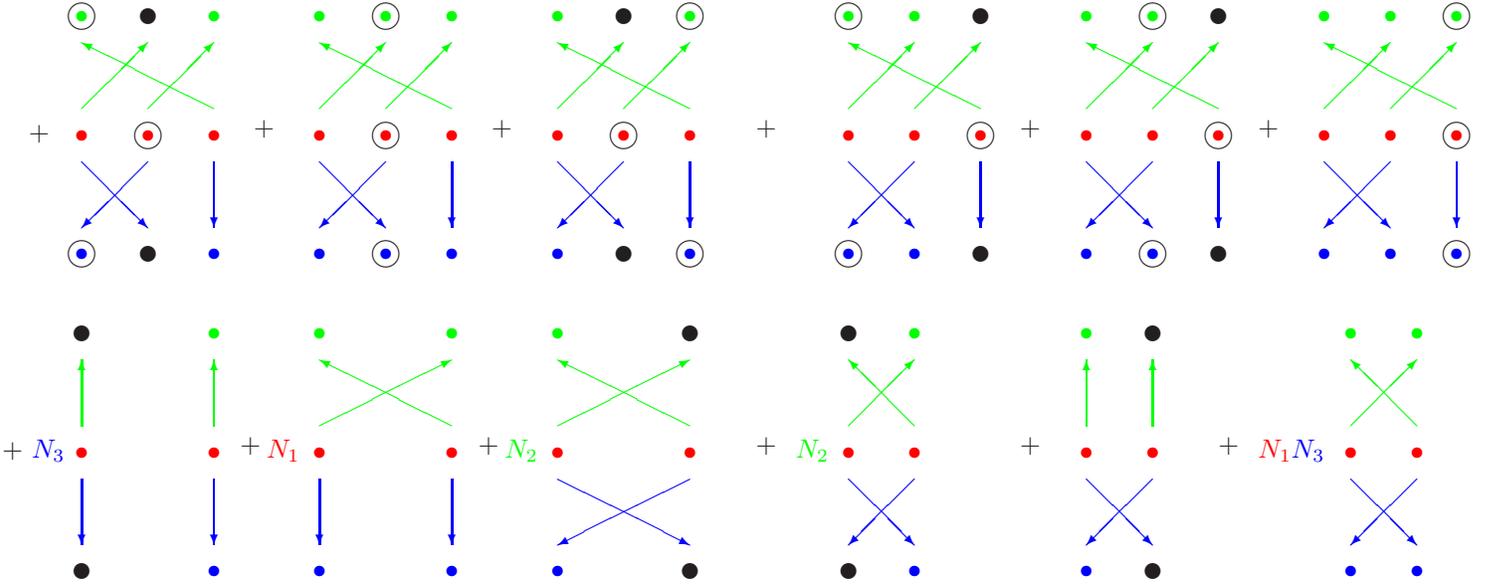
\begin{picture}(400,500)(-60,-440)

\put(0,0){
\put(-35,-2){\mbox{$\Delta$}}
\qbezier(-10,-50)(-30,0)(-10,50)
\qbezier(60,-50)(80,0)(60,50)
\put(85,-2){\mbox{$=$}}
\put(0,0){{\cre
\put(0,0){\circle*{4}}
\put(25,0){\circle*{4}}
\put(50,0){\circle*{4}}
}}
\put(0,0){{\cg
\put(0,45){\circle*{4}}
\put(25,45){\circle*{4}}
\put(50,45){\circle*{4}}
\put(0,10){\vector(1,1){25}}
\put(25,10){\vector(1,1){25}}
\put(50,10){\vector(-2,1){50}}
}}
\put(0,0){{\cb
\put(0,-45){\circle*{4}}
\put(25,-45){\circle*{4}}
\put(50,-45){\circle*{4}}
\put(0,-10){\vector(1,-1){25}}
\put(25,-10){\vector(-1,-1){25}}
\put(50,-10){\vector(0,-1){25}}
}}

}

\put(120,0){
\put(0,0){{\cre
\put(0,0){\circle*{4}}
\put(25,0){\circle*{4}}
\put(50,0){\circle*{4}}
}}
\put(0,0){{\cg
\put(0,45){\circle*{4}}
\put(25,45){\circle*{4}}
\put(50,45){\circle*{4}}
\put(0,10){\vector(1,1){25}}
\put(25,10){\vector(1,1){25}}
\put(50,10){\vector(-2,1){50}}
}}
\put(0,0){{\cb
\put(0,-45){\circle*{4}}
\put(25,-45){\circle*{4}}
\put(50,-45){\circle*{4}}
\put(0,-10){\vector(1,-1){25}}
\put(25,-10){\vector(-1,-1){25}}
\put(50,-10){\vector(0,-1){25}}
}}
\put(0,0){\circle{10}}
\put(0,45){\circle{10}}
\put(0,-45){\circle{10}}
\put(65,0){\mbox{$+$}}
}

\put(210,0){
\put(0,0){{\cre
\put(0,0){\circle*{4}}
\put(25,0){\circle*{4}}
\put(50,0){\circle*{4}}
}}
\put(0,0){{\cg
\put(0,45){\circle*{4}}
\put(25,45){\circle*{4}}
\put(50,45){\circle*{4}}
\put(0,10){\vector(1,1){25}}
\put(25,10){\vector(1,1){25}}
\put(50,10){\vector(-2,1){50}}
}}
\put(0,0){{\cb
\put(0,-45){\circle*{4}}
\put(25,-45){\circle*{4}}
\put(50,-45){\circle*{4}}
\put(0,-10){\vector(1,-1){25}}
\put(25,-10){\vector(-1,-1){25}}
\put(50,-10){\vector(0,-1){25}}
}}
\put(0,0){\circle{10}} \put(0,45){\circle*{6}} \put(0,-45){\circle*{6}}
\put(25,45){\circle{10}}
\put(25,-45){\circle{10}}
\put(65,0){\mbox{$+$}}
}

\put(300,0){
\put(0,0){{\cre
\put(0,0){\circle*{4}}
\put(25,0){\circle*{4}}
\put(50,0){\circle*{4}}
}}
\put(0,0){{\cg
\put(0,45){\circle*{4}}
\put(25,45){\circle*{4}}
\put(50,45){\circle*{4}}
\put(0,10){\vector(1,1){25}}
\put(25,10){\vector(1,1){25}}
\put(50,10){\vector(-2,1){50}}
}}
\put(0,0){{\cb
\put(0,-45){\circle*{4}}
\put(25,-45){\circle*{4}}
\put(50,-45){\circle*{4}}
\put(0,-10){\vector(1,-1){25}}
\put(25,-10){\vector(-1,-1){25}}
\put(50,-10){\vector(0,-1){25}}
}}
\put(0,0){\circle{10}} \put(0,45){\circle*{6}} \put(0,-45){\circle*{6}}
\put(50,45){\circle{10}}
\put(50,-45){\circle{10}}
\put(65,0){\mbox{$+$}}
}
\put(120,-120){
\put(0,-2){\mbox{${\cre N_1}  $}}
\put(0,0){{\cre
\put(25,0){\circle*{4}}
\put(50,0){\circle*{4}}
}}
\put(0,0){{\cg
\put(25,45){\circle*{4}}
\put(50,45){\circle*{4}}
\put(25,10){\vector(1,1){25}}
\put(50,10){\vector(-1,1){25}}
}}
\put(0,0){{\cb
\put(25,-45){\circle*{4}}
\put(50,-45){\circle*{4}}
\put(25,-10){\vector(0,-1){25}}
\put(50,-10){\vector(0,-1){25}}
}}
\put(65,0){\mbox{$+$}}
}

\put(210,-120){
\put(-10,-2){\mbox{${\cg N_2}{\cb N_3} $}}
\put(0,0){{\cre
\put(25,0){\circle*{4}}
\put(50,0){\circle*{4}}
}}
\put(0,0){{\cg
\put(25,45){\circle*{4}}
\put(50,45){\circle*{4}}
\put(25,10){\vector(1,1){25}}
\put(50,10){\vector(-1,1){25}}
}}
\put(0,0){{\cb
\put(25,-45){\circle*{4}}
\put(50,-45){\circle*{4}}
\put(25,-10){\vector(0,-1){25}}
\put(50,-10){\vector(0,-1){25}}
}}
\put(25,45){\circle*{6}} \put(25,-45){\circle*{6}}
\put(65,0){\mbox{$+$}}
}

\put(300,-120){
\put(0,0){{\cre
\put(25,0){\circle*{4}}
\put(50,0){\circle*{4}}
}}
\put(0,0){{\cg
\put(25,45){\circle*{4}}
\put(50,45){\circle*{4}}
\put(25,10){\vector(0,1){25}}
\put(50,10){\vector(0,1){25}}
}}
\put(0,0){{\cb
\put(25,-45){\circle*{4}}
\put(50,-45){\circle*{4}}
\put(25,-10){\vector(1,-1){25}}
\put(50,-10){\vector(-1,-1){25}}
}}
\put(50,45){\circle*{6}} \put(50,-45){\circle*{6}}
\put(65,0){\mbox{$+$}}
}
\put(-100,-260){
\put(-20,-2){\mbox{$+$}}
\put(0,0){{\cre
\put(0,0){\circle*{4}}
\put(25,0){\circle*{4}}
\put(50,0){\circle*{4}}
}}
\put(0,0){{\cg
\put(0,45){\circle*{4}}
\put(25,45){\circle*{4}}
\put(50,45){\circle*{4}}
\put(0,10){\vector(1,1){25}}
\put(25,10){\vector(1,1){25}}
\put(50,10){\vector(-2,1){50}}
}}
\put(0,0){{\cb
\put(0,-45){\circle*{4}}
\put(25,-45){\circle*{4}}
\put(50,-45){\circle*{4}}
\put(0,-10){\vector(1,-1){25}}
\put(25,-10){\vector(-1,-1){25}}
\put(50,-10){\vector(0,-1){25}}
}}
\put(25,0){\circle{10}}
\put(25,45){\circle*{6}}\put(25,-45){\circle*{6}}
\put(0,45){\circle{10}}
\put(0,-45){\circle{10}}
\put(65,0){\mbox{$+$}}
}

\put(-10,-260){
\put(0,0){{\cre
\put(0,0){\circle*{4}}
\put(25,0){\circle*{4}}
\put(50,0){\circle*{4}}
}}
\put(0,0){{\cg
\put(0,45){\circle*{4}}
\put(25,45){\circle*{4}}
\put(50,45){\circle*{4}}
\put(0,10){\vector(1,1){25}}
\put(25,10){\vector(1,1){25}}
\put(50,10){\vector(-2,1){50}}
}}
\put(0,0){{\cb
\put(0,-45){\circle*{4}}
\put(25,-45){\circle*{4}}
\put(50,-45){\circle*{4}}
\put(0,-10){\vector(1,-1){25}}
\put(25,-10){\vector(-1,-1){25}}
\put(50,-10){\vector(0,-1){25}}
}}
\put(25,0){\circle{10}}
\put(25,45){\circle{10}}
\put(25,-45){\circle{10}}
\put(65,0){\mbox{$+$}}
}

\put(80,-260){
\put(0,0){{\cre
\put(0,0){\circle*{4}}
\put(25,0){\circle*{4}}
\put(50,0){\circle*{4}}
}}
\put(0,0){{\cg
\put(0,45){\circle*{4}}
\put(25,45){\circle*{4}}
\put(50,45){\circle*{4}}
\put(0,10){\vector(1,1){25}}
\put(25,10){\vector(1,1){25}}
\put(50,10){\vector(-2,1){50}}
}}
\put(0,0){{\cb
\put(0,-45){\circle*{4}}
\put(25,-45){\circle*{4}}
\put(50,-45){\circle*{4}}
\put(0,-10){\vector(1,-1){25}}
\put(25,-10){\vector(-1,-1){25}}
\put(50,-10){\vector(0,-1){25}}
}}
\put(25,0){\circle{10}}
\put(25,45){\circle*{6}} \put(25,-45){\circle*{6}}
\put(50,45){\circle{10}}
\put(50,-45){\circle{10}}
\put(75,0){\mbox{$+$}}
}

\put(190,-260){
\put(0,0){{\cre
\put(0,0){\circle*{4}}
\put(25,0){\circle*{4}}
\put(50,0){\circle*{4}}
}}
\put(0,0){{\cg
\put(0,45){\circle*{4}}
\put(25,45){\circle*{4}}
\put(50,45){\circle*{4}}
\put(0,10){\vector(1,1){25}}
\put(25,10){\vector(1,1){25}}
\put(50,10){\vector(-2,1){50}}
}}
\put(0,0){{\cb
\put(0,-45){\circle*{4}}
\put(25,-45){\circle*{4}}
\put(50,-45){\circle*{4}}
\put(0,-10){\vector(1,-1){25}}
\put(25,-10){\vector(-1,-1){25}}
\put(50,-10){\vector(0,-1){25}}
}}
\put(50,0){\circle{10}}
\put(50,45){\circle*{6}} \put(50,-45){\circle*{6}}
\put(0,45){\circle{10}}
\put(0,-45){\circle{10}}
\put(65,0){\mbox{$+$}}
}

\put(280,-260){
\put(0,0){{\cre
\put(0,0){\circle*{4}}
\put(25,0){\circle*{4}}
\put(50,0){\circle*{4}}
}}
\put(0,0){{\cg
\put(0,45){\circle*{4}}
\put(25,45){\circle*{4}}
\put(50,45){\circle*{4}}
\put(0,10){\vector(1,1){25}}
\put(25,10){\vector(1,1){25}}
\put(50,10){\vector(-2,1){50}}
}}
\put(0,0){{\cb
\put(0,-45){\circle*{4}}
\put(25,-45){\circle*{4}}
\put(50,-45){\circle*{4}}
\put(0,-10){\vector(1,-1){25}}
\put(25,-10){\vector(-1,-1){25}}
\put(50,-10){\vector(0,-1){25}}
}}
\put(50,0){\circle{10}}
\put(50,45){\circle*{6}} \put(50,-45){\circle*{6}}
\put(25,45){\circle{10}}
\put(25,-45){\circle{10}}
\put(65,0){\mbox{$+$}}
}

\put(370,-260){
\put(0,0){{\cre
\put(0,0){\circle*{4}}
\put(25,0){\circle*{4}}
\put(50,0){\circle*{4}}
}}
\put(0,0){{\cg
\put(0,45){\circle*{4}}
\put(25,45){\circle*{4}}
\put(50,45){\circle*{4}}
\put(0,10){\vector(1,1){25}}
\put(25,10){\vector(1,1){25}}
\put(50,10){\vector(-2,1){50}}
}}
\put(0,0){{\cb
\put(0,-45){\circle*{4}}
\put(25,-45){\circle*{4}}
\put(50,-45){\circle*{4}}
\put(0,-10){\vector(1,-1){25}}
\put(25,-10){\vector(-1,-1){25}}
\put(50,-10){\vector(0,-1){25}}
}}
\put(50,0){\circle{10}}
\put(50,45){\circle{10}}
\put(50,-45){\circle{10}}
}

\put(-100,-380){
\put(-30,-2){\mbox{$+\  {\cb N_3}$}}
\put(0,0){{\cre
\put(0,0){\circle*{4}}
\put(50,0){\circle*{4}}
}}
\put(0,0){{\cg
\put(0,45){\circle*{4}}
\put(50,45){\circle*{4}}
\put(0,10){\vector(0,1){25}}
\put(50,10){\vector(0,1){25}}
}}
\put(0,0){{\cb
\put(0,-45){\circle*{4}}
\put(50,-45){\circle*{4}}
\put(0,-10){\vector(0,-1){25}}
\put(50,-10){\vector(0,-1){25}}
}}
\put(0,45){\circle*{6}}\put(0,-45){\circle*{6}}
\put(60,0){\mbox{$+$}}
}

\put(-10,-380){
\put(-20,-2){\mbox{${\cre N_1}$}}
\put(0,0){{\cre
\put(0,0){\circle*{4}}
\put(50,0){\circle*{4}}
}}
\put(0,0){{\cg
\put(0,45){\circle*{4}}
\put(50,45){\circle*{4}}
\put(0,10){\vector(2,1){50}}
\put(50,10){\vector(-2,1){50}}
}}
\put(0,0){{\cb
\put(0,-45){\circle*{4}}
\put(50,-45){\circle*{4}}
\put(0,-10){\vector(0,-1){25}}
\put(50,-10){\vector(0,-1){25}}
}}
\put(60,0){\mbox{$+$}}
}

\put(80,-380){
\put(-20,-2){\mbox{${\cg N_2}$}}
\put(0,0){{\cre
\put(0,0){\circle*{4}}
\put(50,0){\circle*{4}}
}}
\put(0,0){{\cg
\put(0,45){\circle*{4}}
\put(50,45){\circle*{4}}
\put(0,10){\vector(2,1){50}}
\put(50,10){\vector(-2,1){50}}
}}
\put(0,0){{\cb
\put(0,-45){\circle*{4}}
\put(50,-45){\circle*{4}}
\put(0,-10){\vector(2,-1){50}}
\put(50,-10){\vector(-2,-1){50}}
}}
\put(50,45){\circle*{6}} \put(50,-45){\circle*{6}}
\put(75,0){\mbox{$+$}}
}

\put(190,-380){
\put(-20,-2){\mbox{${\cg N_2}$}}
\put(0,0){{\cre
\put(0,0){\circle*{4}}
\put(25,0){\circle*{4}}
}}
\put(0,0){{\cg
\put(0,45){\circle*{4}}
\put(25,45){\circle*{4}}
\put(0,10){\vector(1,1){25}}
\put(25,10){\vector(-1,1){25}}
}}
\put(0,0){{\cb
\put(0,-45){\circle*{4}}
\put(25,-45){\circle*{4}}
\put(0,-10){\vector(1,-1){25}}
\put(25,-10){\vector(-1,-1){25}}
}}
\put(0,45){\circle*{6}} \put(0,-45){\circle*{6}}
\put(65,0){\mbox{$+$}}
}

\put(280,-380){
\put(0,0){{\cre
\put(0,0){\circle*{4}}
\put(25,0){\circle*{4}}
}}
\put(0,0){{\cg
\put(0,45){\circle*{4}}
\put(25,45){\circle*{4}}
\put(0,10){\vector(0,1){25}}
\put(25,10){\vector(0,1){25}}
}}
\put(0,0){{\cb
\put(0,-45){\circle*{4}}
\put(25,-45){\circle*{4}}
\put(0,-10){\vector(1,-1){25}}
\put(25,-10){\vector(-1,-1){25}}
}}
\put(25,45){\circle*{6}} \put(25,-45){\circle*{6}}
\put(50,0){\mbox{$+$}}
}

\put(380,-380){
\put(-35,-2){\mbox{${\cre N_1}{\cb N_3} $}}
\put(0,0){{\cre
\put(0,0){\circle*{4}}
\put(25,0){\circle*{4}}
}}
\put(0,0){{\cg
\put(0,45){\circle*{4}}
\put(25,45){\circle*{4}}
\put(0,10){\vector(1,1){25}}
\put(25,10){\vector(-1,1){25}}
}}
\put(0,0){{\cb
\put(0,-45){\circle*{4}}
\put(25,-45){\circle*{4}}
\put(0,-10){\vector(1,-1){25}}
\put(25,-10){\vector(-1,-1){25}}
}}
}

\end{picture}

\noindent
We read off from the pictures in the lower lines
\be
\Delta{\cal K}_{{\cre 2},{\cg 2}}=
\Delta {\cal K}^{(3)}_{{\cg(123)},{\cb (12)}}
= (2{\cre N_1}+{\cg N_2}{\cb N_3}) {\cal K}^{(2)}_{{\cg (12)},{\cb ()}}
+ 2 {\cal K}^{(2)}_{{\cg ()},{\cb (12)}}
+ {\cb N_3} {\cal K}^{(2)}_{{\cg ()},{\cb ()}}
+(2{\cg N_2}+{\cre N_1}{\cb N_3}) {\cal K}^{(2)}_{{\cg (12)},{\cb (12)}}
= \nn \\
= (2{\cg N_2}+{\cre N_1}{\cb N_3}) {\cal K}_{\cre 2}
+ (2{\cre N_1}+{\cg N_2}{\cb N_3}) {\cal K} _{{\cg 2} }
+ 2 {\cal K}_{ {\cb 2}}
+ {\cb N_3} {\cal K}_1^2
\ee
which is in full accordance with (\ref{Deltafrom3level}) below.

\bigskip

The join operation $\{\ ,\ \}$ can be described in exactly the same way,
with two operators drawn one after another, so that the operation maps two
pairs of permutation $\{(\sigma_2\otimes \sigma_3),(\sigma_2'\otimes \sigma_3')\}
\in S_{m_1+m_2}^{\otimes 2}$ to a sum
of permutations in $S_{m_1+m_2-1}^{\otimes 2}$.

\bigskip

{\bf Kernels and cokernels.}

In the tensor model, there is no clear notion of genus expansion.
Moreover, there is no obvious characteristic of non-planarity even in application
to operators themselves. (We remind that the diagram technique for description of
gauge-invariant operators and of Feynman rules for their correlators
are distinct: in the former case, the propagators are colored lines,
while in the latter case, they are $r$-colored tubes).
In the case of operators, however, the cut-and-join structure can provide
a characteristic of this type.
For a given set of (primary) keystone operators, one can consider a {\it join} pyramid, a sub-ring ${\cal R}_{\{,\}}$
obtained by multiple action of the join operation $\{\,\}$.
In variance with matrix models, however, there are connected gauge-invariant operators of the loop type
which do not belong to the join pyramid, but which arise in the sub-ring ${\cal R}_{\{,\};\Delta}$. These operators have to be added to the {\it full} pyramid as new, secondary keystone operators.
Thus, for an operator at level $m$, we actually have
\be
{\cal K}^{(m)} = \Delta^p \Big(\{{\cal K}^{(i)},{\cal K}^{(j)}\}\Big)
\ee
where ${\cal K}^{(i)}$ and ${\cal K}^{(j)}$ already belong to the join pyramid
(and $i+j=m+p+1$).
The minimal value of $p$ for the given operator is referred to as degree of the secondary operator.
The first such operator in the Aristotelian model of degree one is
\be
{\cal K}_{3W} \in \Delta\Big({\cal K}_{{\cre 2},{\cb 2},{\cg 2}}\Big)
\in \Delta\Big(\{ {\cal K}_{{\cre 2},{\cg 2}},{\cal K}_{\cb 2}\}\Big)
\in \Delta\left( \Big\{ \{{\cal K}_{\cre 2},{\cal K}_{\cg 2}\},{\cal K}_{\cb 2}\Big\}\right)
\ee

\bigskip

Since $\Delta$ maps all the operators at level $m$ to those at level $m-1$,
the number of which is much less, $\Delta$ inevitably has a huge  kernel.
In fact, since the multiplication by ${\cal K}_1$ converts all gauge-invariant operators at level
$m-1$ into independent disconnected operators at level $m$,
the dimension of
the kernel of $\Delta$ is equal to the dimension of the space of {\it all connected} operators at level $m$,
plus those disconnected which do not possess a ${\cal K}_1$ factor:
\be
{\rm Ker}(\Delta) \cong {\cal R}_m/{\cal R}_{m-1}
\label{kerDelta}
\ee
where ${\cal R}_m$ denotes the grading $m$ part of the ring ${\cal R}$.
More precisely, the basis in the kernel is labeled by
connected and above-mentioned disconnected operators,
but the basis vectors are their linear combinations
with the ${\cal K}_1$-multiples.

In fact, from $\Delta({\cal K}_1{\cal K}^{(m)}) =
{\cal K}_1\Delta({\cal K}^{(m)}) + (2m+1){\cal K}^{(m)}$,
it follows that
\be
\Delta \circ {\cal K}_1 + {\cal K}_1\circ \Delta = -(2m+1)
\ee
i.e. that ${\cal K}_1$ is a kind of operator inverse of $\Delta$
(modulo grading twist).
Similarly to the case of hierarchy of anomalies \cite{UFN1},
existence and simplicity of such operator can be a clue to the structure of
the Virasoro-like constraints.

Very schematically, the CJ structure can be represented as a collection of pyramides,
with secondary ones separated from the main (parent) join pyramid by a
distance, which is the degree of secondary keystone operators:

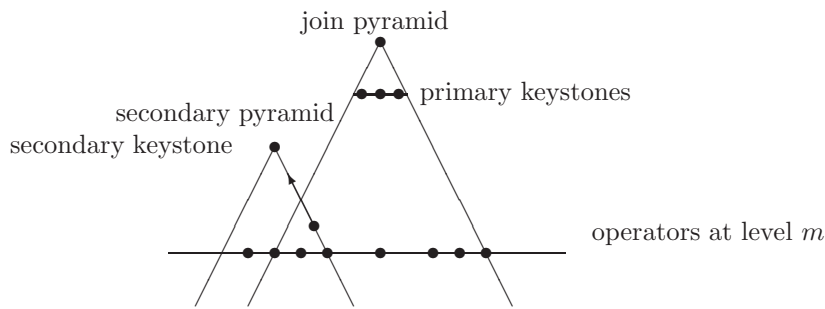
\begin{picture}(200,150)(-200,-130)

\put(0,0){\circle*{4}}
\put(0,0){\line(1,-2){50}} \put(-30,5){\mbox{join pyramid}}
\put(0,0){\line(-1,-2){50}}
\put(-7,-20){\circle*{4}}\put(0,-20){\circle*{4}}\put(7,-20){\circle*{4}}
\put(-10,-20){\line(1,0){20}} \put(15,-22){\mbox{primary keystones}}
\put(-100,-30){\mbox{secondary pyramid}}
\put(-40,-40){\line(1,-2){30}} \put(-40,-40){\circle*{4}}
\put(-40,-40){\line(-1,-2){30}} \put(-140,-42){\mbox{secondary keystone}}
\put(-25,-70){\vector(-1,2){10}} \put(-25,-70){\circle*{4}}
\put(-80,-80){\line(1,0){150}}
\put(80,-75){\mbox{operators at level $m$}}
\put(-50,-80){\circle*{4}}\put(-40,-80){\circle*{4}}\put(-30,-80){\circle*{4}}
\put(-20,-80){\circle*{4}}\put(0,-80){\circle*{4}}\put(20,-80){\circle*{4}}
\put(30,-80){\circle*{4}}\put(40,-80){\circle*{4}}

\end{picture}

The upgoing arrow shows the action of operation $\Delta$, which can be done multiple of times.

\bigskip

In \cite{GeRamg}, there is an attempt made to describe the ring ${\cal R}$ of all gauge-invariant operators in an invariant way. However, we need more: an invariant description of the full pyramid.
Ideally, we need this description in the language of symmetric groups,
but in this paper we make just a first step:
compiling tables of the CJ structure constants
in particular examples at lower levels $m$. See sec.\ref{exalevels} below.

\section{Counting diagrams in the Aristotelian model}
\setcounter{equation}{0}

Before describing the CJ-structure of the operator ring,
we need a description of the operator ring itself.
While we will proceed to level-by-level analysis in sec.\ref{exalevels}, it deserves to present
some general results on the numbers related to the cosets ${\cal S}_m^r$, which is the purpose of the present section.
This simple issue is a direct application of the Hurwitz (symmetric group)
calculus {\it a la} \cite{MMN1},
and it has already been partly presented in \cite{GR,IMMten2,Di,KGT,GeRamg}. We simplify some of the previous presentations.
We begin from a short summary for the particular case of $r=3$ and then go into
more lengthy comments on these formulas.

Our main goals here are to evaluate  at the given level $m$ the numbers of: (i) gauge-invariant operators (i.e. the dimension of ${\cal S}_m^r$); (ii) connected gauge-invariant operators; (iii) gauge-invariant operators with a given number of red-green cycles (for $r=3$, i.e. for the Aristotelian model); (iv) orbits of the colour permutation group acting on gauge-invariant operators (i.e. gauge-invariant operators symmetrized over colours). We proceed in two different gauges, which we also discuss in detail.

\subsection{Index functions and number of connected operators}

The gauge-invariant operators form a graded ring, and the number of
independent operators
at each level $m$ is defined by the index function
\be
\eta(q) = 1 + \sum_{m=1}^\infty \#_m q^m = {\rm PE}\Big(\eta^{\rm conn}(q)\Big)
= {\rm PE}\left(\sum_{m=1}^\infty \#_m^{\rm conn}q^m\right)
= \prod_{m=1}^\infty\frac{1}{(1-q^m)^{\#^{\rm conn}_m}}
\label{indexfunc}
\ee
The numbers $\#^{\rm conn}_m$ can be read off from the plethystic logarithm
\be\label{PL}
\hbox{PLog}\left(\eta(q)\right)=\sum_{k=1}^\infty \#^{\rm conn}_kq^k=\sum_{m=1}^\infty{\mu(m)\over m}\log\eta(q^m)
\ee
where $\mu(m)$ is the M\"obius function
\be
\mu(m)=\left\{\begin{array}{lr}
0& m \hbox{ has at least one repeated prime factor}\\
1& m=1\\
(-1)^n& m \hbox{ is a product of $n$ distinct primes}
\end{array}\right.
\ee

The
index functions (\ref{indexfunc}),
counting the numbers of all and of connected gauge-invariants at different levels
can be read from (\ref{sizecoset}),
and for $r=3$ they are
\be\label{etaAr}
\eta_{_{\rm Arist}}(q) = 1 + q+4q^2+11q^3 + 43q^4+161q^5 + 901 q^6 +
5579 q^7 +43206 q^8 + \ldots =
\nn \\
= \frac{1}{(1-q)(1-q^2)^{^3}(1-q^3)^{^7}(1-q^4)^{^{26}}(1-q^5)^{^{97}}(1-q^6)^{^{624}}
(1-q^7)^{^{4163}}\ldots}
\!\!\!\!\!\!\!\!\!\!\!
 \\ \nn \\ \nn \\
\eta^{\rm conn}_{_{\rm Arist}}(q) = q+3q^2+7q^3+26q^4+97q^5+624q^6 +
4163 q^7 + \ldots
 \ee
These are sequences A110143 and  A057005 from \cite{OEIS} respectively,
the latter is known to enumerate also the unlabeled dessins d'enfants with $m$ edges).
A more economic classification is provided by the orbits of the $S_r^{coloring}$ group,
which  permutes the colorings $N_i$, and thus relates the operators. This consideration applies to any kind of their averages, not limited to Gaussian.
For $r=3$, the corresponding index functions are \cite{GR}
\be\label{symgf}
\zeta_{_{\rm Arist}}(q) = 1+q+2q^2+5q^3+15q^4+44q^5 + 199q^6+ 1069q^7+ \ldots =  \nn\\
=\frac{1}{(1-q)(1-q^2)(1-q^3)^{^3}(1-q^4)^{^{9}}(1-q^5)^{^{26}}(1-q^6)^{^{139}}
(1-q^7)^{^{814}}\ldots }
\!\!\!\!\!\!\!\!\!\!
 \\ \nn \\ \nn \\
\tilde\zeta_{_{\rm Arist}}^{\rm conn}(q) = q+q^2+3q^3+ 9 q^4+26q^5+139q^6+814q^7+ \ldots\nn
\ee
However, ${\cal K}_{\cre 2}^2$ should be distinguished
from ${\cal K}_{\cg 2}{\cal K}_{\cb 2}$,
and this is not done in the above $\tilde\zeta_{_{\rm Arist}}^{\rm conn}(q)$: we need
$\zeta_{_{\rm Arist}}^{\rm conn}(q) = sym$(PLog)
rather than $\tilde\zeta_{_{\rm Arist}}^{\rm conn}(q)=\hbox{PLog}(sym)$.
The proper index function has been first calculated in \cite{GR}:
\be
\zeta_{_{\rm Arist}}^{\rm conn}(q) = q+q^2+3q^3+ 8 q^4 +24q^5+72 q^6+ \ldots
\ee
Anyhow, the averages of such disconnected operators, even Gaussian, are not factorized
beyond the planar limit.

\subsection{On gauge choices for ${\cal S}^3_m$}

While one can analyze the coset spaces in invariant terms,
still sometimes it is useful to use more economic descriptions,
which break the symmetries, but which allow one to visualize the patterns better:
invariant descriptions are typically multi-dimensional, while visualizations
require projections into lower dimensions.
In the tensor model story, different ``projections" begin from different
ways to represent/draw the operators:
we already encountered two absolutely different types of pictures in sec.\ref{reminderCJ}
above, and we will encounter more in sec.\ref{exalevels} below.
In this subsection, we briefly review description of these ``projections"
in terms of the gauge-fixing procedure, the most familiar one in quantum field theory language,
and we do this for the case of $r=3$, to avoid overloading with details.

\subsubsection{The double coset}

As we already know, the space of operators is a double coset
${\cal S}^r_m = S_m\backslash S_m^{\otimes r}/S_m$,
where the original $S_m^{\otimes r}$ is just a set of $r$ permutations
$\sigma_1,\ldots,\sigma_r$ from the symmetric group $S_m$, where $m$
is the ``level", the number of fields $M$ in the gauge-invariant operator
(the number of the conjugate fields $\bar M$ is also $m$).
For $r=3$, we often call the three permutations ``red" ${\cre \sigma_1}$,
``green" ${\cg \sigma_2}$ and ``blue" ${\cb \sigma_3}$.
The coset is obtained through the equivalence relation of the triple by common left and right multiplications:
\be
({\cre \sigma_1},{\cg \sigma_2},{\cb \sigma_3}) \ \cong \
(h_L\circ{\cre \sigma_1}\circ h_R,\ h_L\circ{\cg \sigma_2}\circ h_R,\
h_L\circ {\cb \sigma_3}\circ h_R)
\ee

\subsubsection{{\bf RG}-gauge: enumeration by red-green cycles
\label{rgcycles}}

If we eliminate $\sigma_1\longrightarrow id$
by left multiplication, selecting $h_L=h_R^{-1}\circ \sigma_1^{-1}$, it converts
\be\label{gtr}
\sigma_1\otimes\sigma_2\otimes\sigma_3 \longrightarrow
id\otimes h_R^{-1}\circ \sigma_1^{-1}\circ\sigma_2 \circ h_R
\otimes h_R^{-1}\circ \sigma_1^{-1}\circ\sigma_3\circ h_R
\ee
In this formula, $h_R$ is still unfixed, hence, there is a gauge transformation remaining which preserves the gauge condition $\sigma_1=1$.
This can be enough to bring $\sigma_2$ in $id\otimes\sigma_2\otimes\sigma_3$
into its ``canonical" form $\sigma_2^{can} =(1,\ldots,m_1)(m_1+1,\ldots, m_1+m_2)\ldots$,
with $m_1\geq m_2 \geq\ldots$, namely, into the lexicographically ordered elements in the cycles
described by the Young diagram $[\sigma_2]=\{m_1\geq m_2\geq \ldots\}$.
Invariant under the conjugation is also the class (Young diagram) $[\sigma_3^{-1}\circ\sigma_2]$.

We call the gauge $\sigma_1=id$, $\sigma_2 = \sigma_2^{can}$ {\bf RG}-gauge
(which is derived from ``red-green"; it should not be mixed with RG for ``renormalization group", hence, boldface).
It corresponds to representing the gauge-invariant operators by red-green cycles
of lengths $m_1,m_2, \ldots$ with vertices, connected by blue contractions.
This enumeration was proposed in \cite{IMMten2}.
Different sets of red-green cycles are labeled by $\#_m$ Young diagrams
$\{m_1\geq m_2\geq \ldots \geq m_l>0\}$  of the size $m=m_1+\ldots+ m_l = \sum k\nu_k$.
They are identified with cycles in the permutation ${\cg \sigma_2}$ in the gauge
${\cre \sigma_1}=id$.
The maximal number of blue contractions for each set is $m!$. The actual number is, however,
much smaller.
For example, for the cycles of unit length, the blue contractions form from them new blue cycles
(with unit red-green cycles playing the role of vertices),
and the total number of these is again $\#_m\leq m!$.

\subsubsection{Orbits of $S_3^{coloring}$}

Permutations from the global symmetry group $S_3^{coloring}$ such as
$$
\sigma_1\otimes\sigma_2\otimes\sigma_3 \longrightarrow \sigma_2\otimes\sigma_1\otimes\sigma_3
$$
or
$$
\sigma_1\otimes\sigma_2\otimes\sigma_3 \longrightarrow \sigma_3\otimes\sigma_2\otimes\sigma_1
$$
in the {\bf RG} gauge $\sigma_1=id$ can be described as follows:
$$
id\otimes\sigma_2\otimes\sigma_3 \longrightarrow
id\otimes\sigma_2^{-1}\otimes\sigma_2^{-1}\sigma_3
$$
and
$$
id\otimes\sigma_2\otimes\sigma_3 \longrightarrow
id\otimes\sigma_3^{-1}\circ\sigma_2\otimes\sigma_3^{-1}
$$
since the {\bf RG}-gauge is given by the transformation is $h_L=h_R^{-1}\circ \sigma_1^{-1}$, i.e., in this gauge, $\tilde\sigma_2=h_R^{-1}\circ \sigma_1^{-1}\circ\sigma_2 \circ h_R$, $\tilde\sigma_3=h_R^{-1}\circ \sigma_1^{-1}\circ\sigma_3\circ h_R$ ( see (\ref{gtr})) so that $(\tilde\sigma_2,\tilde\sigma_3)\to (\tilde\sigma_2^{-1},\tilde\sigma_2^{-1}\tilde\sigma_3)$ under permuting $\sigma_1\leftrightarrow\sigma_2$ and $(\tilde\sigma_2,\tilde\sigma_3)\to (\tilde\sigma_3^{-1}\tilde\sigma_2,\tilde\sigma_3^{-1})$ under permuting $\sigma_1\leftrightarrow\sigma_3$.

\subsubsection{Hurwitz gauge
\label{HuGa}}

Instead of fixing $\sigma_1=1$, which breaks the global symmetry $S_3^{coloring}$ badly,
one can impose other conditions, for example:
\be
\sigma_1\circ\sigma_2\circ\ldots\circ \sigma_r = id
\label{HuGn}
\ee
One could call this gauge Hurwitz, because the Hurwitz numbers ${\cal N}^H([\sigma_1],\ldots,[\sigma_r])$
count the number of solutions to this equation for fixed conjugacy classes (Young diagrams)
$[\sigma_1]$,\ldots,$[\sigma_r]$.

The problem, however, is that this description is inconvenient for our purposes:
it fails to distinguish the gauge-invariant operators in any nice way.
For example, the admissible triple of permutations ${\cre (123)},{\cg(123)},{\cb (123)}$
corresponds to exactly the same operator ${\cal K}_1^3$ as another triple
${\cre [\,]},{\cg [,]},{\cb [,]}$.
Of course, the coset ${\cal S}_m^r$ can be described in any gauge, including (\ref{HuGn}).

Another issue with this gauge is that (\ref{HuGn}) gives rise to a sophisticated non-linear
equation for $h_L$.
The way out suggested is that, instead of (\ref{HuGn}), we consider as the Hurwitz gauge
the alternating product
\be
\sigma_1\circ\sigma_2^{-1}\circ\sigma_3\circ\sigma_4^{-1}\circ\ldots\circ \sigma_r^{(-1)^{r+1}} = id
\label{HuG}
\ee
where the l.h.s. is now multiplicatively transformed by the left and right multiplications,
so this true Hurwitz gauge fixes $h_L$.
In exchange, the $S_r^{coloring}$ symmetry acts non-trivially,
not just permutating $\sigma_i$, as one expects from
(\ref{HuGn}).

\subsection{Size of the coset}

\subsubsection{Symmetric Schur polynomials and the $z_\Delta$ factors}

If we proceed to exact results for the numbers of operators, then all of the answers will be
expressed through the single quantity
\be\label{zfactor}
\boxed{
z_\Delta = \prod_{i=1} i^{k_i} \cdot k_i!},
\ee
where the set of non-negative integers $\{k_i\}$ parameterizes the Young diagram
$\Delta = [\ldots, 4^{k_4},3^{k_3},2^{k_2},1^{k_1}]$.
This symmetry factor is used to define the symmetric Schur polynomials
\be
{\rm Schur}_m\{p\} = \sum_{\Delta\vdash m} \frac{p_\Delta}{z_\Delta}
\ee
where $p_\Delta = \prod_i p_i^{k_i}$ is a monomial made from the time variables
$p_i = it_i$, and $\Delta\vdash m$ means that size of the Young diagram is $m$,
\be
|\Delta| = \sum_i ik_i = m
\ee
The symmetric Schur polynomials have a simple generating function
\be
\sum_{m=0}^\infty t^m\cdot {\rm Schur}_m\{p\} = \exp\left(\sum_k \frac{t^kp_k}{k}\right)
\ee
The more general Cauchy formula
\be
\exp\left(\sum_k \frac{t^kp_k\bar p_k}{k}\right) =
\sum_R t^{|R|}\cdot {\rm Schur}_R\{p\}\cdot {\rm Schur}_R\{\bar p\}
\ee
involves Schur polynomials for arbitrary Young diagrams $R$,
not only for the symmetric ones, $R=[m]$, but we do not need them in the present paper.

\subsubsection{Estimates of the coset size}

Permutations are characterized by their cycle structure: by $k_i$ we denote
the number of cycles of length $i$. They are divided
into conjugacy classes, labeled by Young diagrams $\Delta$.
The class $\Delta$ contains $||\Delta||= \frac{|\Delta|!}{z_\Delta}$
permutations, each left invariant under $z_\Delta$ conjugations.
Other conjugations from $S_m$, with $m=|\Delta|$ convert one of the permutations
in the class into another in the same class.

Accordingly,
the size of our double coset is  restricted by the two inequalities:
\be
\frac{(m!)^{r-1}}{m!} \leq ||{\cal S}_m^r|| \leq (m!)^{r-1}
\label{ineqsforsize}
\ee
Indeed,
\be
\sum_\Delta z_\Delta^{r-2} \geq z_{[1]^m}^{r-2} = (m!)^{r-2}
\ee
and
\be
\sum_\Delta z_\Delta^{r-2} \leq (m!)^{r-2} \left(\sum_\Delta 1\right)\leq (m!)^{r-1}
\ee
since, for each $\Delta$, $z_\Delta\leq m!$ (and only $z_{[1^m]}=m!$ saturates this inequality),
and the number of Young diagrams, defined by the coefficient of $q^m$
in the expansion of $\prod_{n=1}^\infty (1-q^n)^{-1}$ is smaller than the corresponding
coefficient in the expansion of $(1-q)^{-m}$, which is equal to $\frac{\Gamma(2m)}{m!\cdot \Gamma(m)} \leq m!$.

Note that the inequalities (\ref{ineqsforsize}) are sometimes saturated: the right one, at $m=2$,
and the left one, at $m\to\infty$.

\subsubsection{The lemma that is not Burnside's}

In order to obtain precise formulas for the size of the coset, we need the general expression:
it is given by the celebrated formula
often attributed to W. Burnside, but, in fact, discovered by A.-L. Cauchy and later re-discovered by F.G. Frobenius.
Afterwards, the lemma is often named ``the lemma that is not Burnside's" to emphasize that it is only one
of the many important and original claims made by W. Burnside in his very important book \cite{Burns}.

If a finite group $H$ acts on a set $X$, $x\longrightarrow h\circ x$,
then the number of orbits is equal to
\be
||X/H|| = \frac{1}{||H||} \sum_{h\in H}\sum_{x\in X}  \delta(h\circ x, x)
\label{cosizegen}
\ee
For example, if the group does not act at all: $h\circ x = x$ for all $g$ and $x$,
then $||X/H||  = \frac{1}{||H||}\cdot ||H||\cdot ||X|| = ||X||$.
On the contrary, if the group acts with no fixed points, i.e. $h\circ x = x$ implies $h=id$,
then, $||X/H|| = \frac{1}{||H||} \cdot ||X|| = \frac{||X||}{||H||}$.

In general, (\ref{cosizegen}) is proved by the following chain of relations:
\be
||X/H|| = \sum_{{\rm orbits\ of\ }H} 1
= \sum_{{\rm orbits}}\left(\sum_{x\in {\rm orbit}}\frac{1}{||{\rm orbit}_x||}\right)
= \sum_{x\in X}  \frac{1}{||{\rm orbit}_x||} = \nn \\
= \frac{1}{||H||} \sum_{x\in X} ||{\it stabilizer\ subgroup\ w.r.t.} \ x||
= \frac{1}{||H||} \sum_{h\in H} \sum_{x\in X} \delta(h\circ x,x)
\ee

\subsection{The number of gauge-invariant operators: $||{\cal S}_m^r||$}

If we fix the gauge $\sigma_1=id$, then
${\cal S}_m^r$ is just a set of equivalence classes under simultaneous conjugations
$\sigma_j \longrightarrow h \sigma_j h^{-1}$ for $h,\sigma_j \in S_m$ and $j=2,\ldots,r$.
Eq.(\ref{cosizegen}) then implies that the size
\be
||{\cal S}_m^r|| = \frac{1}{m!} \sum_{h\in S_m} \sum_{\sigma_2,\ldots,\sigma_r\in S_m}
\prod_{j=2}^r \delta( \sigma_j^{-1} h\sigma_j,h)
= \frac{1}{m!} \sum_{h\in S_m}
\left(\sum_{\sigma\in S_m} \delta( \sigma^{-1} h\sigma,h)\right)^{r-1}\,,
\ee
where we have substituted all $\delta(h\sigma_j h^{-1},\sigma_j)$ with the equivalent conditions
$\delta( \sigma_j^{-1} h\sigma_j,h)$.
Now, if $h$ is a permutation of the type $[h]=\Delta$, then there are exactly $z_\Delta$
conjugations from $S_m$ which leave it intact.
As a consequence, the number of different permutations of the type $[h]=\Delta$ is
$||\Delta|| = \frac{||S_m||}{z_\Delta} = \frac{m!}{z_\Delta}$.
This means that $\sum_{\sigma\in S_m} \delta( \sigma^{-1} h\sigma,h) = z_{[h]}$,
and, therefore,
\be
||{\cal S}_m^r|| =  \frac{1}{m!} \sum_{h\in S_m}  z_{[h]}^{r-1} =
\frac{1}{m!} \sum_{\Delta\vdash m} ||\Delta|| \cdot z_\Delta^{r-1} =
\sum_{\Delta\vdash m} z_\Delta^{r-2}
\label{cosize'}
\ee
This is exactly eq.(\ref{sizecoset}).
In particular, for $r=2$, where the operators are
$\prod_p \Tr (M\bar M)^{m_p}$ with $m =\sum_p {m_p}$, we obtain $||{\cal S}_m^r|| = \sum_\Delta 1 =$
the number of Young diagrams of size $m$.
Similarly, for $r=1$, i.e., for the vector model, where there
is just one operator $(M_i\bar M^i)^m$ at each level, $\sum_{\Delta\vdash m} z_\Delta^{-1} = 1$.

\subsection{The number of connected operators: index function $\eta(q)$}

The simple form of the formula (\ref{cosize'}) (or (\ref{sizecoset})) implies that the index function
$\eta(q)$ is actually factorized:
\be
\eta_r(q) = \sum_\Delta z_\Delta^{r-2} q^{|\Delta|} =
\sum_{\{k_i\}}  \prod_i (i^{r-2}q^i)^{k_i} (k_i!)^{r-2} =
\prod_i \hat\eta_r(s_i)
\ee
where
\be
\hat\eta_r(s) =  \sum_k  (k!)^{r-2}s^k
\ee
and
$s_i = i^{r-2}q^i$.
The functions $\hat\eta_1(s) = e^s$ and $\hat\eta_2(s) = \frac{1}{1-s}$ are elementary,
and they lead to the obvious index formulas
\be
\eta_1(q) = \prod_{i=1}^\infty \hat\eta_1 (q^i/i) = \exp\left(\sum_{i=1} \frac{q^i}{i}\right)=
\frac{1}{1-q}
\ee
and
\be
\eta_2(q) = \prod_{i=1}^\infty \hat\eta_2 (q^i) = \frac{1}{\prod_{i=1}^\infty (1-q^i)}
\ee
for the vector and matrix models respectively. The latter counts the number of Young diagrams
of different sizes.
For the Aristotelian model with $r=3$, we obtain the textbook example of divergent
and Borel-summable series
\be
\hat \eta_3(s) = \sum_k k!\cdot s^k = \int_0^\infty \frac{e^{-x}\,dx}{1-sx} +
const \cdot \frac{e^{-1/s}}{s}
\ee
for which we have had a chance to discuss in \cite{IMMcb}.
This function is well-known for its non-perturbative ambiguity, which, however, is not seen
at the level of the formal series in $q$. As soon as we are interested in the coefficients of the perturbative series, we neglect the non-perturbative term so that
\be
\hat \eta_3(s) ={e^{-{1\over s}}\over s}\hbox{Ei}\Big({1\over s}\Big)
\ee
where Ei$(x)$ is the exponential integral \cite{Grad}, and
\be
\eta_{Arist}(q) = \eta_3(q) = \prod_{i=1}^\infty \hat\eta_3(iq^i)
\ee
The number of connected operators $\#^{\rm conn}_k$ is then calculated by taking the plethystic logarithm (\ref{PL})
\be
\sum_{k=1}^\infty \#^{\rm conn}_kq^k=\sum_{k,i=1}^\infty{\mu(k)\over k}\log\hat\eta_3(iq^{ik})
\ee
The expansion of the logarithm of $\hat\eta_3(s)$ is given by
\be
\log\hat\eta_3(s)=\sum_{k=1}^\infty \nu_k{s^k\over k}
\ee
where the numbers $\nu_k$ is sequence A003319 from \cite{OEIS}. Now one immediately obtains
(\ref{etaAr}). Generalizations to higher $r$ are straightforward.

\subsection{The number of operators with fixed number of the red-green cycles ({\bf RG} gauge)\label{RGcycles}}

Let us evaluate the number of operators with fixed number of the red-green cycles. We start from the Aristotelian model
at $r=3$. In the {\bf RG} gauge, we fix not only ${\cre\sigma_1}=id$, but also the conjugacy class of $[{\cg \sigma_2}]$,
and enumerate the conjugacy classes of ${\cb \sigma_3}$ under the condition that the conjugations
preserve the selected ${\cg \sigma_2}$.
According to (\ref{cosizegen}), the number of such orbits,
which depends only on the conjugacy class of ${\cg\sigma_2}$, is
\be\label{55}
{\cal N}^{[{\cg \sigma_2}]} = \frac{1}{z_{[{\cg \sigma_2}]}}
\sum_{\sigma_3,h} \delta(h\circ \sigma_3\circ h^{-1},\sigma_3)\cdot
\delta(h\circ \sigma_2\circ h^{-1},\sigma_2)
= \frac{1}{z_{[{\cg \sigma_2}]}}
\sum_{h} z_h \cdot
\delta(h\circ \sigma_2\circ h^{-1},\sigma_2)
\ee
where we explicitly summed over $\sigma_3$. We define $z_{h}^R$, the number of permutations $\sigma \in R$
which commute with $h$, $\sigma^{-1}h\sigma=h$. Like $z_h$, these $z_h^R$ depend only on the conjugacy class $[h]$ of $h$. The table of the lowest $z_h^{R}$ can be found in Appendix B1.

Now we can further take in (\ref{55}) an average over all $\sigma_2$ of the given type
$[\sigma_2]$, which provides a factor
$z_h^{[\sigma_2]}$:
\be
{\cal N}^{[{\cg \sigma_2}]} = \frac{1}{||{\cg \sigma_2}||\cdot z_{[{\cg \sigma_2}]}}
\sum_{h} z_h\cdot z_h^{[{\cg \sigma_2}]}
= \frac{1}{||{\cg \sigma_2}||\cdot z_{[{\cg \sigma_2}]}}
\sum_\Delta ||\Delta||\cdot z_\Delta\cdot z_\Delta^{[{\cg \sigma_2}]}
= \sum_\Delta z_\Delta^{[{\cg \sigma_2}]}
\ee
The last equality comes from $||\Delta||\cdot z_\Delta   =
||{\cg \sigma_2}||\cdot z_{[{\cg \sigma_2}]} = m!$.
These ${\cal N}^{[{\cg \sigma_2}]}$ are exactly the numbers
in the penultimate line of the table (\ref{tabzhR}), and the index function is now
\be
\eta_{Arist}^{(f)}(p)=1+p_1+\Big(2p_{[2]}+2p_{[11]}\Big)+\Big(4p_{[3]}+4p_{[21]}+3p_{[111]}\Big)+\Big(10p_{[4]}+10p_{[31]}+8p_{[22]}+10p_{[211]}+\nn\\+5p_{[1111]}\Big)
+\Big(28p_{[5]}+34p_{[41]}+26p_{[32]}+26p_{[311]}+22p_{[221]}+18p_{[2111]}+7p_{[11111]}\Big)+\ldots
\ee
This index function is immediately reduced to (\ref{etaAr}) on the subspace $p_{R}=q^{|R|}$, i.e. upon summation of all the coefficients at a given level (which are marked by parentheses in the formula). If one realizes the time variables $p_k$ through the Miwa variables: $p_k=\sum_ix_i^k$, this corresponds to leaving just one $x_i$: $x_1=q$. This case corresponds to the Schur functions non-zero only when the Young diagram contains just one line (symmetric representation).

For generic $r$, the summand contains an additional factor $z_\Delta^{r-3}$:
then the number of operators for the pattern of red-green circles
specified by $[{\cg \sigma_2}]$ is
\be\label{55p}
{\cal N}_r^{[{\cg \sigma_2}]} =
\sum_\Delta z_{\Delta}^{[{\cg \sigma_2}]}z_\Delta^{r-3}
\ee
In particular,
\be
\sum_{R\vdash m} {\cal N}_r^{R}  = {\cal N}_r
\ee
and
\be
{\cal N}_r^{[1^m]} = {\cal N}_{r-1}
\ee
since there is just one permutation in the class $[1^m]$.

\subsection{Symmetrizing operators in colours: the number of orbits of $S_r^{coloring}$}

The last number that we are going to discuss in this paper is how many essentially different gauge-invariant operators are there, i.e. how many after identification of those differing only by permutations of colours. To put it differently, we ask how many orbits of the permutation group $S_r^{coloring}$ there are. Here we can again use (\ref{cosizegen}).

\subsubsection{A toy example: $S_2^{coloring}$}

As a warm-up example, consider just a set $X=Y^{\otimes 2}$.
What is the number of symmetric pairs?
The symmetry group $S_2$ consists of two transformations
$(y_1,y_2) \longrightarrow (y_1,y_2)$ and  $(y_1,y_2) \longrightarrow (y_2,y_1)$,
thus according to (\ref{cosizegen}),
\be
{\rm Sym}(Y^{\otimes 2}) = \frac{1}{2}\left\{ \delta\Big((y_1,y_2),\,(y_1,y_2)\Big)
+ \delta\Big((y_2,y_1),\,(y_1,y_2)\Big)\right\}
= \frac{1}{2}\Big(|Y|^2 + |Y|\Big) = \frac{|Y|(|Y|+1)}{2}
\ee
Similarly,
\be
{\rm Sym}(Y^{\otimes 3}) = \frac{1}{3!}\left\{\delta\Big((y_1,y_2,y_3),\,(y_1,y_2,y_3)\Big)
+ 3\delta\Big((y_2,y_1,y_3),\,(y_1,y_2,y_3)\Big) + 2\delta\Big((y_2,y_3,y_1),\,(y_1,y_2,y_3)\Big)
\right\} =\nn \\
= \frac{1}{6}\Big(|Y|^3+3|Y|^2+2|Y|\Big) = \frac{|Y|(|Y|+1)(|Y|+2)}{6}
\ee

Coming closer to our problem, if we symmetrize ${\cal S}_m^r$ w.r.t. the $S_2^{coloring}$
permutations of ${\cg \sigma_2}$ and ${\cb \sigma_3}$, i.e., just over two colors, green and blue,
\be
{\rm Sym}_{{\cg g},{\cb b}}({\cal S}_m^3) =
\frac{1}{2!m!} \sum_{\sigma_2,\sigma_3,h\in S_m}\Big\{
\delta(h^{-1}\sigma_2 h,\sigma_2)\delta(h^{-1}\sigma_3 h,\sigma_3) +
\delta(h^{-1}\sigma_3 h,\sigma_2)\delta(h^{-1}\sigma_2 h,\sigma_3)\Big\} = \nn \\
= \frac{1}{2! m!} \sum_\Delta ||\Delta|| \cdot\Big(z_\Delta^2 + z_{\Delta^2}\Big)
= \frac{1}{2} \sum_\Delta  \Big(z_\Delta + \frac{z_{\Delta^2}}{z_\Delta}\Big)
\ee
In the second term of the sum, after the substitution $\sigma_2 = h^{-1}\sigma_3 h$ into
$\sigma_3=h^{-1}\sigma_2 h = h^{-2} \sigma_3 h^2$, i.e. $h^2 = \sigma_3 h^2 \sigma_3^{-1}$,
we need the number of conjugations (by $\sigma_3$)
which leave invariant $h^2$, not $h$, and this number is  $z_{\Delta^2}$.
At the lowest levels (for $\xi$'s, see examples in (\ref{exaxis}) below):
\be
\begin{array}{|c||c|c|c||c|c|c|c|}
\hline
m & \Delta & z_\Delta &  z_{\Delta^2}&\ldots & \xi^{(2)}_{\Delta^2} & \xi^{(3)}_{\Delta^3}&\ldots \\
\hline
2 & [2] & 2 & z_{[11]}=2 && \xi^{(2)}_{[11]} = 2  & \xi^{(3)}_{[2]} =1& \\
& [11] & 2 & z_{[11]} =2 && \xi^{(2)}_{[11]} = 2 & \xi^{(3)}_{[11]} = 1& \\
\hline
3 & [3] & 3 & z_{[3]}=3 &&  \xi^{(2)}_{[3]} = 1  & \xi^{(3)}_{[111]} = 3&  \\
& [21] & 2 & z_{[111]}=6 &&\xi^{(2)}_{[111]} = 4  & \xi^{(3)}_{[21]} = 1&\\
& [111] & 6 & z_{[111]} = 6 && \xi^{(2)}_{[111]} = 4  & \xi^{(3)}_{[111]} = 3&\\
\hline
\end{array}
\ee
and we get
\be
{\rm Sym}_{{\cg g},{\cb b}}({\cal S}_2^3)
= \frac{1}{2} \Big( 2 + \frac{2}{2} + 2 + \frac{2}{2}\Big) = 3\,,
\ee
which is the case: in the set ${\cal K}_{\cre 2},{\cal K}_{\cg 2},{\cal K}_{\cb 2},{\cal K}_1^2$,
the two operators in the middle get identified by symmetrization, reducing the
total number from $4$ to $3$.
Similarly\
\be
{\rm Sym}_{{\cg g},{\cb b}}({\cal S}_3^3)
= \frac{1}{2} \Big( 3 + \frac{3}{3} + 2 + \frac{6}{2}+ 6 + \frac{6}{6}\Big) = 8\,,
\ee
which is also true.

\subsubsection{Aristotelian model: $S_3^{coloring}$}

For symmetrization over $S_3^{coloring}$, the problem is a little more complicated:
\be
{\rm Sym}_{{\cre r},{\cg g},{\cb b}}({\cal S}_m^3) =
\frac{1}{3!(m!)^2}\sum_{h_L,h_R,\sigma_{1,2,3} \in S_m}
\Big\{ \delta(h_L\sigma_1 h_R,\sigma_1)\cdot \delta(h_L\sigma_2 h_R,\sigma_2)
\cdot \delta(h_L\sigma_3 h_R,\sigma_3) + \nn \\
+ 3\cdot \delta(h_L\sigma_1 h_R,\sigma_1)\cdot \delta(h_L\sigma_2 h_R,\sigma_3)
\cdot \delta(h_L\sigma_3 h_R,\sigma_2) +
2\cdot \delta(h_L\sigma_1 h_R,\sigma_2)\cdot \delta(h_L\sigma_2 h_R,\sigma_3)
\cdot \delta(h_L\sigma_3 h_R,\sigma_1)\Big\}
\ee
In the first two terms of this sum $h_L = \sigma_1 h_R^{-1}\sigma_1^{-1}$ and
we obtain the problem of conjugation of $\tilde\sigma_i=\sigma_1^{-1}\sigma_i$ by $h_R$,
which we already solved.
In the last term, $h_L = \sigma_2 h_R^{-1}\sigma_1^{-1}$ and the other two conditions are
\be
\sigma_2 h_R^{-1}\sigma_1^{-1} \sigma_2 h_R = \sigma_3 \nn \\
 \sigma_2 h_R^{-1}\sigma_1^{-1} \sigma_3 h_R = \sigma_1
\ee
i.e.
\be
\tilde\sigma_2 = h_R \tilde\sigma_2^{-1} \tilde\sigma_3 h_R^{-1} \nn \\
\tilde \sigma_3 = h_R \tilde\sigma_2^{-1}  h_R^{-1}
\ee
and substituting $\tilde\sigma_3$ from the second equation into the first one, we obtain
\be
\tilde\sigma_2 = h_R\tilde\sigma_2^{-1} h_R \tilde\sigma_2^{-1}  h_R^{-2}
\ \Longrightarrow  \   \sigma^{3} = h_R^3
\ee
where $\sigma = h_R\tilde\sigma_2^{-1}$.
Thus, what we need here is the number of $\sigma$ with a given cube, which itself
is a cube, which we denote by $\xi^{(3)}_{[h_R^3]}$
(this number actually depends only on the conjugacy class of $h_R^3$, not of $h_R$).
  In all the three cases, the sum over a common factor in $\sigma_{1,2,3}$ gives just $m!$, and
the answer is
\be
{\rm Sym}_{{\cre r},{\cg g},{\cb b}}({\cal S}_m^3) = \frac{1}{3!\cdot m!}
\sum_\Delta ||\Delta||\cdot\Big(z_\Delta^2 + 3z_{\Delta^2} + 2\xi^{(3)}_{\Delta^3} \Big)
= \frac{1}{6}\sum_\Delta \left(z_\Delta + \frac{3 z_{\Delta^3}}{z_\Delta}
+\frac{2\xi^{(3)}_{\Delta^3}}{z_\Delta}\right) =
\ee
\be
= \left\{
\begin{array}{cccc|cc}
m=2: && \frac{1}{6}\Big( (2+2)+3\cdot(2/2+2/2)+2\cdot(1/2+1/2)\Big) = 2 &&&
{\cal K}_1^2,{\cal K_2} \\
&&&&&\\
m=3: &&  \frac{1}{6}\Big( (3+2+6) + 3\cdot(3/3+6/2+6/6)+2\cdot(3/3+1/2+3/6)\Big) = 5
&&& {\cal K}_1^3, {\cal K}_2{\cal K}_1, {\cal K}_3, {\cal K}_{2,2}, {\cal K}_{3W} \\
&&&&& \\
\ldots
\end{array}
\right.
\nn
\ee
The corresponding generation function is given by (\ref{symgf}).

\subsubsection{Generic $r$}

In general we need the quantities of two types:
\be
\# \sigma: \ \sigma h^r = h^r \sigma  \ \ {\rm this\ is} \ \  z_{[h^r]}
\ee
and
\be
\# \sigma: \ \sigma^r = h \ \ {\rm we\ denote\ it } \ \xi^{(r)}_{[h]}
\ee
What is needed to find these quantities is that
$h^r$ consists of the same cycles as $h$ does, if their length is coprime with $r$.
The cycle of the length $l$, which has a biggest common divisor $l_r=BCD(l,r)$,
gets split into $l/l_r$ cycles of length $l_r$, for example:
\be
(1234)^2 = (13)(24) \nn \\
(12)^6 = () \nn \\
(123456789)^6 = (174)(285)(396)
\ee
Then, say,
\be
[\ldots,4^{k_4},3^{k_3},2^{k_2},1^{k_1}]^2 = [\ldots,4^{2k_8},3^{k_3+2k_6},2^{2k_4},1^{k_1+2k_2}]
\ee
while $\xi^{(r)}$ is obtained from the re-expansion of the symmetric Schur polynomial
\be
{\rm Schur}_m\{p\} = \sum_{\Delta\vdash m} \frac{p_\Delta}{z_\Delta} \nn \\
{\rm Schur}_m\{p_{rk} = p_k^r\} = \sum_{\Delta\vdash m} \xi_\Delta^{(r)} \frac{p_\Delta}{z_\Delta}
\ee
For example,
\be
{\rm Schur}_3\{p_{2k}=p_k^2\} =
\left.\frac{p_3}{3}+\frac{p_2p_1}{2}+\frac{p_1^3}{6}\right|_{p_2=p_1^2}
= \frac{p_3}{3} + \frac{2p_1^3}{3} \ \ \Longrightarrow \ \
\xi^{(2)}_{[3]} = 1, \ \xi^{(2)}_{[21]} = 0, \ \xi^{(2)}_{[111]} = 4 \nn \\
{\rm Schur}_3\{p_{3k}=p_k^3\} = \left.\frac{p_3}{3}+\frac{p_2p_1}{2}+\frac{p_1^3}{6}\right|_{p_3=p_1^3}
= \frac{p_2p_1}{2} + \frac{p_1^3}{2}    \ \ \Longrightarrow \ \
\xi^{(3)}_{[3]} = 0, \ \xi^{(3)}_{[21]} = 1, \ \xi^{(3)}_{[111]} = 3
\label{exaxis}
\ee
Let us put it differently: define that the Young diagram $\Delta_1$ is $r$-larger than the Young diagram $\Delta_2$, $\Delta_1\succ_r\Delta_2$, if $\Delta_2$ is obtained from $\Delta_1$ (different from $\Delta_2$) by replacing all lines of lengths $r\lambda_i$ with $r$ lines of lengths $\lambda_i$ for each $i$. For instance,
$[31]\succ_3[1111]$, $[91]\succ_3 [3331]$, $[841]\succ_4[222211111]$, etc. Then,
\be\label{xird}
\xi^{(r)}_{\Delta}=
\left\{
\begin{array}{lrl}
0&\hbox{if:}& \hbox{1) there are no diagrams $r$-larger than $\Delta$;}\\
&&\hbox{2) there is at least one line with length multiple $r$}\\
&&\\
1&\hbox{if:}&\hbox{1) there are no diagrams $r$-larger than $\Delta$;}\\
&&\hbox{2) there are no lines with lengths multiple $r$}\\
&&\\
z_\Delta\cdot\sum_{\Delta'\succ_r\Delta}\displaystyle{1\over z_{\Delta'}}&&\hbox{otherwise}
\end{array}
\right.
\ee
where the sum goes over all $r$-larger diagrams and does not include the diagram $\Delta$ itself if there are diagrams $r$-smaller than $\Delta$. Otherwise, the sum includes also $\Delta$.

{\bf Examples:}
\be
& \xi^{(r)}_{[1^r]} \ \stackrel{(\ref{xird})}{=}\
z_{[1^r]} \left(\frac{1}{z_{[1^r]}} +\frac{1}{z_{[r]}}\right)
\ \stackrel{(\ref{zfactor})}{=}\  1+\frac{r!}{r} = 1+(r-1)! \nn \\
{\rm for} \   r<s<2r: & \xi^{(r)}_{[1^{s}]} \ \stackrel{(\ref{xird})}{=}\
z_{[1^s]}\left(\frac{1}{z_{[1^s]}} + \frac{1}{z_{[r,1^{s-r}]}}\right)
\ \stackrel{(\ref{zfactor})}{=}\
1 + \frac{s!}{r\cdot (s-r)!} \nn \\
{\rm for} \   2r\leq s<3r: & \xi^{(r)}_{[1^{s}]} \ \stackrel{(\ref{xird})}{=}\
z_{[1^s]}\left(\frac{1}{z_{[1^s]}} + \frac{1}{z_{[r,1^{s-r}]}}
+ \frac{1}{z_{[2r,1^{s-2r}]}}\right) \ \stackrel{(\ref{zfactor})}{=}\
1 + \frac{s!}{r\cdot (s-r)!}  + \frac{s!}{2r^2\cdot (s-2r)!} \nn \\
\ldots
\ee
From the last formula we can get, say, $\xi^{(4)}_{[1^{8}]}=1681$
or $\xi^{(4)}_{[1^{9}]}=12097$, what is true.

Likewise one obtains
\be
\xi^{(4)}_{[3,1^6]} \ \stackrel{(\ref{xird})}{=}\
z_{[3,1^6]}\left(\frac{1}{z_{[3,1^6]}} + \frac{1}{z_{[4311]}}\right)
\ \stackrel{(\ref{zfactor})}{=}\
1+ \frac{3\cdot 6!}{4\cdot 3\cdot 2!} = 91
\ee
and so on.

At last, as an example of (\ref{xird}) when the sum does not include the diagram itself, consider
\be
\xi^{(r)}_{[r^r]}={z_{[r^r]}\over z_{[r\cdot r]}}={r^r r!\over r^2}=r^{r-2}r!
\ee
where $[r^r]$ denotes the diagram consisting of $r$ lines with the same length $r$. For instance, $\xi^{(3)}_{[333]}=18$, which is, indeed, the case.

\section{Hurwitz gauges and Hurwitz numbers}
\setcounter{equation}{0}

In this section, we are going to discuss two issues: 1) how to calculate the number of operators in the Hurwitz gauge, which is associated with the Hurwitz numbers; and 2) how to calculate the numbers of operators in the {\it true} Hurwitz gauge.

\subsection{Calculations in the Hurwitz gauge}

The Hurwitz gauge is naturally related to the Hurwitz numbers. These latter, in particular, are simply related with the structure constants of the ring associated with the center of group algebra of the symmetric group. One can construct the numbers of gauge-invariant operators both from these structure constants and from the structure constants of the dual ring, which are nothing but the Clebsh-Gordan coefficients associated with representations of the symmetric group. Here we discuss all these issues.

\subsubsection{Hurwitz numbers and Clebsch-Gordon coefficients}

\paragraph{The Hurwitz numbers} are related to counting the ramified coverings of the connected Riemann surface. These numbers are
numbers of the permutations of given types, i.e. belong to given conjugacy classes, $\sigma_i\in \Delta_i$,
and satisfy the condition $\sigma_1\circ\ldots\circ\sigma_r=id$, which is nothing but the Hurwitz gauge, s.\ref{HuGa}.
On the other hand, these numbers enumerate operators in the Hurwitz gauge and fixed numbers of the red, green and blue cycles (in fact, the answer is symmetric w.r.t. permuting these latter numbers) and are given by the Burnside-Frobenius formula \cite{FroD,LZ}
\be
{\cal N}^H(\Delta_1,\ldots,\Delta_r) =
m! \sum_{R\,\vdash m} d_R^2 \varphi_R(\Delta_1)\ldots \varphi_R(\Delta_r)
\label{Fro}
\ee
where the characters of symmetric group $S_m$, $\psi_R(\Delta)$ are related with $\varphi_R(\Delta)$ via $\psi_R(\Delta) = d_R z_\Delta \varphi_R(\Delta)$. They
can be also considered as the coefficients of the Schur functions
\be
{\rm Schur}_R\{p\} = \sum_{\Delta\,\vdash m} d_R\varphi_R(\Delta)p_\Delta
\label{Schurvarphi}
\ee
In these formulas, all the Young diagrams are of the size $m$
(for more general situation see \cite{MMN1}), and
$d_R={\rm Schur}(p_k=\delta_{k,1})$,
i.e. is the coefficient in front of $p_1^{|R|}=p_1^m$,
so that $\varphi_R([1^{|R|}]) = 1$.

\begin{flushright}\parbox{15.8cm}{{\bf Example:}
Choose all three conjugacy classes to be one longest cycle for $r=3$ and group $S_3$. Then, there are only two possible solutions to the equation
$\sigma_1\circ\sigma_2\circ\sigma_3=id$: $\sigma_1=\sigma_2\sigma_3=(123)$ and $\sigma_1=\sigma_2\sigma_3=(132)$, i.e. ${\cal N}^H([3],[3],[3])=2$. Indeed, using the Burnside-Frobenius formula, one obtains
\be
{\cal N}^H([3],[3],[3]) =
3!\cdot\left(
\frac{2^3}{6^2} + \frac{(-1)^3}{3^2} + \frac{2^3}{6^2}
\right)
=3!\,{9\over 3^3}=2
\ee
Similarly, in the case of all three conjugacy classes being $[21]$, one immediately realizes that $\sigma_1\circ\sigma_2\circ\sigma_3=id$ has no solutions. It agrees with
\be
{\cal N}^H([21],[21],[21]) =
3!\cdot \left(
\frac{3^3}{6^2} + \frac{(-3)^3}{6^2}
\right)
= 0
\ee
\hrule width 5cm
}
\end{flushright}

\paragraph{The orthogonality property}
\be
\sum_\Delta \frac{\psi_R(\Delta)\psi_{R'}(\Delta)}{z_\Delta} = \delta_{R,R'}
\ \ \  \ \ \ \Longleftrightarrow \ \ \ \ \ \
\sum_\Delta z_\Delta \varphi_R(\Delta) \varphi_{R'}(\Delta) = \frac{\delta_{R,R'}}{d_R^2}
\label{ortoR}
\ee
and the relation $\frac{z_\Delta}{m!} = \frac{1}{||\Delta||}$ imply that
\be
\sum_\Delta \frac{1}{||\Delta||}
\cdot {\cal N}^H(\Delta_1,\ldots,\Delta_r,\Delta) \cdot
{\cal N}^H(\Delta_{r+1},\ldots,\Delta_{r+r'},\Delta) =
{\cal N}^H(\Delta_{1},\ldots,\Delta_{r+r'})
\ee
The orthogonality relation (\ref{ortoR}) also implies that
\be
u_R(\Delta)=d_R\sqrt{z_\Delta}\varphi_R(\Delta) = \frac{\psi_R(\Delta)}{\sqrt{z_\Delta}}
\ee
is an orthogonal matrix, in particular, there is an orthogonality relation dual to (\ref{ortoR}):
\be
\sum_{R\,\vdash m} \psi_R(\Delta_1)\psi_R(\Delta_2) = z_\Delta\cdot \delta_{\Delta,\Delta'}
\ \ \ \ \Longleftrightarrow \ \ \ \
\sum_{R\,\vdash m} d_R^2 \varphi_R(\Delta)\varphi_R(\Delta')
= \frac{\delta_{\Delta,\Delta'}}{z_\Delta}
\label{ortoD}
\ee
Using the Cauchy formulas
\be
\sum_R d_R\cdot {\rm Schur}_R\{p\} = e^{p_1}
\ \ \ \ \ \Longleftarrow \ \ \ \ \
\sum_R  {\rm Schur}_R\{p\}\cdot {\rm Schur}_R\{p'\} = \exp\left(\sum_k \frac{p_kp'_k}{k}\right)
\label{CauchySchur}
\ee
where sums go over Young diagrams $R$ of all sizes,
one obtains from the Burnside-Frobenius formula (\ref{Fro})
\be
{\cal N}^H(\Delta) = |\Delta|!\sum_{R\,\vdash |\Delta|} d_R^2\varphi_R(\Delta)
 \stackrel{(\ref{Schurvarphi})}{=}
|\Delta|!\cdot {\rm coeff}_{p_\Delta} \sum_R d_R\cdot{\rm Schur}\{p\}
 \stackrel{(\ref{CauchySchur})}{=}
|\Delta|!\cdot {\rm coeff}_{p_\Delta} e^{p_1} = \delta(\Delta,[1^{|\Delta|}])
\label{1Hur}
\ee
and
\be
{\cal N}^H(\Delta,\Delta') = m! \sum_{R\,\vdash m} d_R^2\varphi_R(\Delta)\varphi_R(\Delta')
= m!\cdot {\rm coeff}_{p_\Delta p_\Delta'} \sum_R {\rm Schur}\{p\}\cdot{\rm Schur}\{p'\}
= \nn \\
\ \stackrel{(\ref{CauchySchur})}{=}\
m!\cdot {\rm coeff}_{p_\Delta p_\Delta'} e^{\sum p_kp'_k/k} =
||\Delta||\cdot \delta(\Delta,\Delta')
\label{2Hur}
\ee

\paragraph{Commutative ring in the group algebra.} The multi-point Hurwitz numbers involve $R$-independent structure  constants that describe the commutative ring, the center
of the group algebra \cite{MMN1}
\be
\varphi_R(\Delta_1)\varphi_R(\Delta_2) = \sum_{\Delta_3} C_{\Delta_1,\Delta_2}^{\Delta_3}
\varphi_R(\Delta_3)
\ee
for example,
\be
{\cal N}^H(\Delta_1,\Delta_2,\Delta_3) = m!
\sum_{R\,\vdash m} d_R^2\varphi_R(\Delta_1)\varphi_R(\Delta_2)\varphi_R(\Delta_3)
\ \stackrel{(\ref{2Hur})}{=}
 ||\Delta_3||\cdot C_{\Delta_1,\Delta_2}^{\Delta_3}
\label{3Hur}
\ee
and the product at the r.h.s. is symmetric in all the three diagrams.
Triple of the Young diagrams is called {\it admissible},
if the corresponding three-point Hurwitz number does not vanish, i.e. a Belyi function should exist.

Higher Hurwitz numbers are expressed through  powers of the matrix $||\Delta||\cdot C$.
The simplest structure constants $C_{\Delta_1,\Delta_2}^{\Delta_3}$
are listed in the multiplication tables in Appendix B2.
Note the obvious sum rule
\be
||\Delta_1||\cdot||\Delta_2|| = \sum_{\Delta_3} ||\Delta_3||\cdot
C_{\Delta_1,\Delta_2}^{\Delta_3}
\ee
in each box in these tables,
what implies that
\be
\sum_{\Delta_3}{\cal N}^H(\Delta_1,\Delta_2,\Delta_3) = ||\Delta_1||\cdot||\Delta_2||
\ee
The orthogonality relation (\ref{ortoD}) implies that
\be\label{Cddd}
C_{\Delta_1,\Delta_2}^{\Delta_3} =
{z_{{\Delta_3}}}\cdot
\sum_{R\,\vdash m} d_R^2 \varphi_R(\Delta_1)\varphi_R(\Delta_2)\varphi_R(\Delta_3)
= \frac{z_{\Delta_3}}{m!}\cdot {\cal N}^H(\Delta_1,\Delta_2,\Delta_3) =
\frac{1}{||\Delta_3||}\cdot {\cal N}^H(\Delta_1,\Delta_2,\Delta_3)
\ee
in accordance with (\ref{3Hur}).

\paragraph{Clebsch-Gordon coefficients.} From the same (\ref{ortoD}) it follows that
\be
\check C_{R_1,R_2,R_3} = d_{R_1}d_{R_2}d_{R_3} \sum_{\Delta\,\vdash m}
 z_\Delta^2 \varphi_{R_1}(\Delta)\varphi_{R_2}(\Delta)\varphi_{R_3}(\Delta)
=\sum_{\Delta\,\vdash m}
\frac{\psi_{R_1}(\Delta)\psi_{R_2}(\Delta)\psi_{R_3}(\Delta)}{z_\Delta}
\label{cleco}
\ee
serve as the structure constants of the dual algebra
\be
\sum_{R_3\,\vdash m} \check C_{R_1,R_2,R_3}\cdot \psi_{R_3}(\Delta)
= \psi_{R_1}(\Delta)\psi_{R_2}(\Delta)
\ee
Hence, they are nothing but the Clebsh-Gordan coefficients for decomposition
of representations $R_1\otimes R_2\otimes R_3\longrightarrow [1^m]$ in symmetric group $S_m$ (since the permutation $[1^m]$ plays the role of unity: its composition with any permutation gives that permutation).

\subsubsection{Number of operators in the Hurwitz gauge}

In \cite{KGT}, it was demonstrated, using the same lemma that is not Burnside's, that the number of the gauge-invariant operators,
is expressed through the Clebsch-Gordan coefficients (\ref{cleco}).
In particular, for $r=3$,
 \be
||{\cal S}_m^3|| =\sum_{R_1,R_2,R_3}\check C_{R_1,R_2,R_3}^2
\ee
Indeed, substituting expression (\ref{cleco}) and applying the orthogonality relation (\ref{ortoD})
three times, we obtain for the r.h.s.
\be
\sum_{R_1,R_2,R_3} \left(\sum_{\Delta\,\vdash m}
\frac{\psi_{R_1}(\Delta)\psi_{R_2}(\Delta)\psi_{R_3}(\Delta)}{z_\Delta}\right)^2
= \sum_{\Delta,\Delta'\,\vdash m} \frac{z_\Delta^3\cdot\delta_{\Delta,\Delta'}}
{z_\Delta z_{\Delta'}}
= \sum_\Delta z_\Delta
\ee
in accordance with (\ref{cosize'}) at $r=3$.

In practice, $\check C_{R_1,R_2,R_3}$ are all equal to 1 at the lowest levels,
only at level 5 there are a few appearances of 2.
At the same time, at high levels the number $||{\cal S}_m^3||$ grows with $m$ as $m!$:
the ratio $||{\cal S}_m^3||/m!$ is approximately
$$1,2,1.833, 1.792, 1.342, 1.251, 1.107,1.072,
1.043, 1.031, 1.023, 1.019, 1.015,
1.013, 1.011, \ldots $$
for $m=1,\ldots, 15$.

Note that the same number of gauge-invariant operators can be expressed through the Hurwitz numbers:
\be\label{Hurcount}
||{\cal S}_m^r|| =\sum_{\Delta_1,\ldots,\Delta_r} {1\over ||\Delta_1||^{r-1}}{\cal N}^H(\Delta_1,\ldots,\Delta_r)
\ee
Indeed, using the orthogonality relation (\ref{ortoR}) and its dual (\ref{ortoD}) and the fact that $\psi_{[r]}(\Delta)=1$, one easily proves that the r.h.s. of (\ref{Hurcount}) is equal to the sum $\sum_\Delta z_\Delta^{r-2}$.

\subsection{Calculations in the true Hurwitz gauge
}

As we explained in s.\ref{HuGa}, the Hurwitz gauge is not too convenient. In contrast, the true Hurwitz gauge looks much simpler. In this section, we enumerate the gauge-invariant operators in the Aristotelian $r=3$ model in the true Hurwitz gauge and describe the structure of operators at first low levels in detail.

\paragraph{The number of gauge-invariant operators.} In this gauge, we need to count the number of common conjugation orbits
of the permutation triple $({\cre\sigma_1},{\cg \sigma_2},{\cb \sigma_3})$
constrained by the condition
\be
{\cre \sigma_1}\circ{\cg \sigma_2}^{-1}\circ {\cb \sigma_3} = id \ \ \ \
\Longleftrightarrow \ \ \ {\cg \sigma_2} = {\cb \sigma_3} \circ{\cre \sigma_1}
\ee
Conjugation freedom allows us to fix a canonical $\sigma_1 = \sigma_1^{can}$
in the class $[{\cre \sigma_1}]$, and it remains to enumerate the conjugacy classes of ${\cb \sigma_3}$
under the condition that the conjugations of ${\cb \sigma_3}$ (and thus, simultaneously, of ${\cg \sigma_2}$)
preserve the selected ${\cre \sigma_1}^{can}$.
According to (\ref{cosizegen}), the number of such orbits,
which depends only on the conjugacy class of ${\cb\sigma_3}$, is
\be
\frac{1}{z_{[{\cre \sigma_1}]}}
\sum_{\sigma_3,h} \delta(h\circ {\cb \sigma_3}\circ h^{-1},{\cb \sigma_3})\cdot
\delta(h\circ {\cre \sigma_1}\circ h^{-1},{\cre \sigma_1})
= \frac{1}{z_{[{\cre \sigma_1}]}}
\sum_{h} z_h \cdot
\delta(h\circ {\cre \sigma_1}^{can}\circ h^{-1},{\cre \sigma_1}^{can})
\ee
Note that we already calculated this sum in (\ref{55}), and we can again take an average over all ${\cre \sigma_1}$ of the given type
$[{\cre \sigma_1}]$, which provides a factor
$z_h^{[{\cre \sigma_1}]}$:
\be
{\cal N}^{[{\cre \sigma_1}]} = \frac{1}{||{\cre \sigma_1}||\cdot z_{[{\cre \sigma_1}]}}
\sum_{h} z_h\cdot z_h^{[{\cre \sigma_1}]}
= \frac{1}{||{\cre \sigma_1}||\cdot z_{[{\cre \sigma_1}]}}
\sum_\Delta ||\Delta||\cdot z_\Delta\cdot z_\Delta^{[{\cre \sigma_1}]}
= \sum_\Delta z_\Delta^{[{\cre \sigma_1}]}
\ee
Similarly for arbitrary $r$ in the Hurwitz gauge we would get, as in (\ref{55p}),
\be
{\cal N}^{[{\cre \sigma_1}]}_r = \sum_\Delta z_\Delta^{[{\cre \sigma_1}}z_h\Delta^{r-3}
\ee
In other words, the counting is exactly the same as in the {\bf RG}-gauge,
though the pictorial interpretation in terms of the red-green cycles is now that immediate.

\paragraph{Structure of the operators at low levels.}

We explicitly describe the gauge-invariant operators at the first five levels $m=1,2,3,4,5$, the tables for the latter two being placed in Appendix B3. The notation for all operators up to level $m=5$ can be found in Appendix A.

\begin{itemize}
\item[${\bf m=1}$] The single triple is admissible (see the definition after formula (\ref{3Hur})), and it is associated with the
single gauge-invariant operator
\be
{\cal N}^H([\,],[\,],[\,]) = 1 \ \ \ \ \ \ \ \ \
{\cal K}_{{\cre [\,]},{\cg [\,]},{\cb[\,]}} = {\cal K}_1
\ee

\item[${\bf m=2}$] There are four admissible triples, and they are in one-to-one correspondence
with the gauge-invariant operators:
\be
\begin{array}{ccc|c|c}
{\cre [\sigma_1]} & {\cg [\sigma_2]} & {\cb [\sigma_3]}
& {\cal N}^H([\sigma_1],[\sigma_2],[\sigma_3])
& {\cal K}_{{\cre [\sigma_1]},{\cg [\sigma_2]},{\cb[\sigma_3]}}\\
&&&&\\
\hline
&&&&\\
\ph [\,]&[\,]&[\,]& 1 & {\cal K}_1^2 \\
&&&&\\
\hline
&&&&\\
\ph [2]&[2]&[\,]& 1 & {\cal K}_{\cb 2} \\
\ph [2]&[\,]&[2]& 1 & {\cal K}_{\cg 2} \\
\ph [\,]&[2]&[2]& 1 & {\cal K}_{\cre 2}
\end{array}
\label{Hur2}
\ee

\item[${\bf m=3}$] A similar table in this case is
\be
\begin{array}{ccc|c|c|c}
{\cre [\sigma_1]} & {\cg [\sigma_2]} & {\cb [\sigma_3]}
& {\cal N}^H([\sigma_1],[\sigma_2],[\sigma_3])
& {\cal K}_{{\cre [\sigma_1]},{\cg [\sigma_2]},{\cb[\sigma_3]}}&\text{{\bf RG}-gauge}\\
&&&&\\
\hline
&&&&\\
\ph [\,]&[\,]&[\,]& 1 & {\cal K}_1^2&{\cal K}_{{\cg [\,]},{\cb [\,]}} \\
&&&&\\
\ph [2]&[2]&[\,]& 3 & {\cal K}_{\cb 2}{\cal K}_1 &{\cal K}_{{\cg (12)},{\cb (12)}} \\
\ph [2]&[\,]&[2]& 3 & {\cal K}_{\cg 2}{\cal K}_1 &{\cal K}_{{\cg [2]},{\cb [\,]}} \\
\ph [\,]&[2]&[2]& 3 & {\cal K}_{\cre 2}{\cal K}_1 &{\cal K}_{{\cg [2]},{\cb [2]}}\\
&&&&\\
\hline
&&&&\\
\ph [3]&[3]&[\,]& 2 & {\cal K}_{\cb 3} &{\cal K}_{{\cg [\,]},{\cb [3]}} \\
\ph [3]&[\,]&[3]& 2 & {\cal K}_{3W} &{\cal K}_{{\cg (123)},{\cb (132)}}\\
\ph [\,]&[3]&[3]& 2 & {\cal K}_{\cre 3} &{\cal K}_{{\cg (123)},{\cb (123)}}\\
&&&&\\
\ph [2]&[2]&[3]& 6 & {\cal K}_{{\cre 2},{\cg 2}}&{\cal K}_{{\cg [3]},{\cb [2]}} \\
\ph [2]&[3]&[2]& 6 & {\cal K}_{{\cre 2},{\cb 2}}&{\cal K}_{{\cg [2]},{\cb [3]}} \\
\ph [3]&[2]&[2]& 6 & {\cal K}_{{\cg 2},{\cb 2}}&{\cal K}_{{\cg (12)},{\cb (13)}} \\
&&&&\\
\ph [3]&[3]&[3]& 2 & {\cal K}_{\cg 3} &{\cal K}_{{\cg [3]},{\cb [\,]} }  \\
\end{array}
\label{Hur3}
\ee
All other Hurwitz numbers are vanishing.
Again we have a one-to-one correspondence between operators and admissible triples,
with each operator counted exactly once irrelatively to the value of non-vanishing
Hurwitz number.

At level $m=3$, we already need to distinguish between inappropriate (\ref{HuGn}) and
appropriate (\ref{HuG}), when constructing the operators:
with $(123)\circ(132)\circ id=id$, one associates
${\cal K}_{{\cre (123)},{\cg (123)},{\cb [\,]}}= {\cal K}_{\cb 3}$
rather than ${\cal K}_{{\cre (123)},{\cg (132)},{\cb [\,]}}=K_{3W}$,
and, with
$(123)\circ(123)\circ(123)=id$, one associates ${\cal K}_{{\cre (123)},{\cg (132)},{\cb (123)}}
= K_{\cg 3}$ rather than ${\cal K}_{{\cre (123)},{\cg (123)},{\cb (123)}}= K_1^3$.
At the same time, for $(123)\circ id\circ (132)=id$ the inversion of $\sigma_2$ plays no role,
and the associated operator is
${\cal K}_{{\cre (123)},{\cg []},{\cb (132)}} = K_{3W}$.
Still, an apparent asymmetry between $\sigma_2$ and other two permutations leads
to a non-naive placing of ${\cal K}_{\cg 3}$ and ${\cal K}_{3W}$ in the table,
which makes the action of the global symmetry $S_3^{coloring}$ less straightforward
and signals about the problems with this labeling at higher levels.

Still, if $\sigma_2$ was not inverted, i.e. if we used (\ref{HuGn}) instead of (\ref{HuG}),
the troubles would be much stronger:
as already mentioned in sec.\ref{HuGa},
the two triples $[],[],[]$ and $[3],[3],[3]$ would
describe the same operator ${\cal K}_{[],[],[]} = {\cal K}_{(123),(123),(123)} = {\cal K}_1^3$.

In the last column, we have added correspondence with operators in the {\bf RG}-gauge, and one can see
how the description in terms of ${\cg [\sigma_2]},{\cb [\sigma_3]}$ gets degenerate:
at least, in two cases there are different operators associated with different permutations
${\cb \sigma_3}$ from the same conjugacy class ${\cb [\sigma_3]}$.

\item[${\bf m=4}$] Permutation triples from the subgroup $S_3\subset S_4$,
listed in table (\ref{Hur3}), are associated with the $11$ disconnected operators,
differing from those in (\ref{Hur3}) by an extra factor ${\cal K}_1$.
The  new $26$ connected operators are (see Appendix B3):
\be
3\times([4],[4],[3]), 3\times([4],[4],[22]), 3\times ([4],[4],[]), \
6\times ([4],[3],[2]), 6\times([4],[22],[2]), \nn \\
3\times ([3],[3],[22]), ([22],[22],[22]),  ([3],[3],[3])
\ee
The triple $([3],[3],[3])$ appears twice, one being connected, one, disconnected.
Also, there are $6$ disconnected operators ${\cal K}_2{\cal K}_2$
from the square of table (\ref{Hur2}) associated with the subgroup
$S_2\otimes S_2\subset S_4$:
\be
3\times ([22],[2],[2]), 3\times ([22],[22],[])
\ee
The number of new unordered connected operators is $8$.

Thus, at level $m=4$, there are $42$ ordered admissible triples
and $43$ gauge-invariant operators, the
degenerate triple being $([3],[3],[3])$.
The number of connected operators is $26$, and, out of 17 disconnected operators, there are $11$ operators
with ${\cal K}_1$ factors and $6$ operators of the type ${\cal K}_2{\cal K}_2$.

The number of unordered admissible triples is $14$, and the number
of gauge-invariant operators modulo $S_3^{coloring}$ symmetry is $15$,
the degenerate triple being again $([3],[3],[3])$.
Connected are $8$ of these operators, among the disconnected ones there are $5$ operators with the
${\cal K}_1$ factor and $2$ operators of the type ${\cal K}_2{\cal K}_2$, one with
coincident and one with distinct colors.

\item[${\bf m=5}$] At this level, one admissible triple, $([22],[4],[4])$ is triply degenerate.

\begin{itemize}
\item[$\surd$]
There are $28$ operators with $\sigma_1\in[5]$ and $19$ ordered admissible triples (see table \ref{t51}).
Degenerate are admissible triples $([5],[5],[5])$ (four times),
$([5],[4],[4])$ (three times),
$([5],[5],[32])$ and $([5],[32],[5])$ (two times each),
$([5],[4],[32])$ and $([5],[32],[4])$ (two times each).
\item[$\surd$]
There are $34$ operators with $\sigma_1\in[4]$ and $20$ ordered admissible triples (see table \ref{t52}).
Degenerate admissible triples are:
the already familiar
$([4],[5],[4])$ and $([4],[4],[5])$ (three times each), and
$([4],[5],[32])$ and $([4],[32],[5])$ (two times each);
the new degenerate triples are:
$([4],[4],[22])$ and $([4],[22],[4])$ (three times each),
$([4],[4],[3])$ and $([4],[3],[4])$ (two times each),
$([4],[32],[3])$ and $([4],[3],[32])$ (two times each).
\item[$\surd$]
There are $26$ operators with $\sigma_1\in[32]$ and $20$ ordered admissible triples (see table \ref{t53}).
Degenerate admissible triples are:
the already familiar
$([4],[5],[32])$ and $([4],[32],[5])$ (two times each)
and $([32],[4],[3])$ and $([32],[3],[4])$ (two times each);
only one pair of  degenerate triples is new,
$([32],[32],[3])$ and $([32],[3],[32])$ (two times each).
\item[$\surd$]
There are $26$ operators with $\sigma_1\in[3]$ and $20$ ordered admissible triples (see table \ref{t54}).
Degenerate admissible triples are:
the already familiar
$([3],[5],[5])$, $([3],[4],[4])$, $([3],[32],[32])$
$([3],[4],[32])$ and $([3],[32],[4])$ (two times each);
only one double-degenerate triple is new,
$([3],[3],[3])$.
\item [$\surd$]
There are $22$ operators and $20$ ordered admissible triples with $\sigma_1\in[22]$ (see table \ref{t55}).
This time there is only one triply-degenerate admissible triple,
the already familiar $([22],[4],[4])$.
\item [$\surd$]
For $\sigma_1\in[2]$, there are $18$ admissible triples and $18$ operators,
no triples are degenerate (see table \ref{t56}).
\item [$\surd$]
For $\sigma_1\in[]$, there are $7$ admissible triples and $7$ operators,
no triples are degenerate (see table \ref{t57}).
\end{itemize}
To summarize, at level $m=5$, we have $28+34+26+26+22+18+7=161$ gauge-invariant operator
(97 of them connected),
while there are only $19+20+20+20+20+18+7=124$ admissible ordered triples.
There are $34$ unordered admissible triples and $44$ $S_3^{coloring}$-symmetric operators,
$24$ of them being connected.

\end{itemize}

\section{Level-by-level analysis of operator ring and its CJ structure
\label{exalevels}}
\setcounter{equation}{0}

In this section, we illustrate our considerations by the diagram technique
for {\it operators}, introduced and explained in detail in \cite{IMMten2}.
Once again, the gauge-invariant operators are defined without any reference
to dynamics of the theory, thus these are {\it not} the Feynman diagrams
in the tensor model.
The Feynman diagrams arise when one introduces the bilinear action (\ref{quadact}),
perturbed by some {\it keystone} operators and their RG descendants.
Then the diagrams which we use in this section turn into various ``points",
vertices of the Feynman diagrams, and they are connected by Feynman propagators,
which are $r$-colored {\it tubes} (cables) made from $r$ thin lines of different colors (as is shown in figure below in s.7.1).
One can replace calculations with the rule (\ref{Wth}) by those with the
help of such Feynman diagrams, but this is of less practical use.
The Feynman rules and pictures can be more useful for the study of recursion
relations, but this is not the subject of the present section.

In this section, we will move up step by step in the ``level" $m$ of the gauge-invariant operators,
with the symmetric group $S_m$ underlying the description of the level $m$.
We attempt to distinguish operators with five descriptions:

\begin{itemize}
\item by diagrams (see Appendix A);
\item as elements of the coset ${\cal S}^r_m$ from (\ref{coset}), which
implies a labeling like ${\cal K}_{{\cre \sigma_1},{\cg \sigma_2},{\cb \sigma_3}}$
or just ${\cal K}_{ {\cg [\sigma_2]},{\cb \sigma_3}}$ in the
{\bf RG}-gauge ${\cre\sigma_1}=id$,
${\cg \sigma_2=[\sigma_2]}$;
\item by Gaussian averages (for which the general answer is known from \cite{MMten});
\item by their place among the trees formed by the triple of keystone operators
${\cal K}_{\cre 2}, {\cal K}_{\cg 2}, {\cal K}_{\cb 2}$,
which implies the labeling by bracket words like
$$
{\cal K}_{\left[[{\cre 2},{\cg 2}],[\cb 2,[\cre 2,\cre 2]]\right]}
= \Big\{\{{\cal K}_{\cre 2},{\cal K}_{\cg 2}\},\
\left\{{\cal K}_{\cb 2},\,\{{\cal K}_{\cre 2},{\cal K}_{\cre 2}\}\right\}\Big\}  \in {\cal R}_6
\ \ \ \ \ \ \ \ \ \ \ \ \ \ \ \ \ \ \ \ \ \ \ \ \ \ \ \
\ \ \ \ \ \ \ \ \ \ \ \ \ \ \ \ \ \ \ \ \ \ \ \ \ \ \ \
\ \ \ \ \ \ \ \ \ \ \ \ \ \ \ \ \
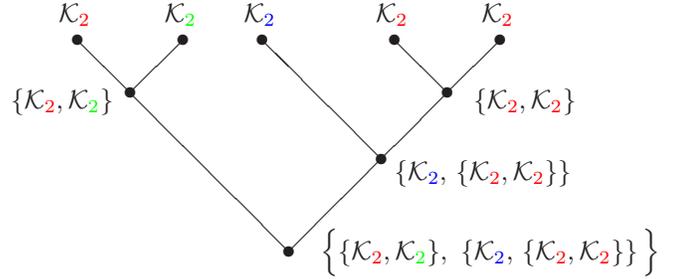
$$
$$
\ \ \ \ \phantom{R_6}
\ \ \ \ \ \ \ \ \ \ \ \ \ \ \ \ \ \ \ \ \ \ \ \ \ \ \ \
\ \ \ \ \ \ \ \ \ \ \ \ \ \ \ \ \ \ \ \ \ \ \ \ \ \ \ \
\ \ \ \ \ \ \ \ \ \ \ \ \ \ \ \ \
$$
\begin{picture}(100,44)(-360,-10)
\put(0,0){\line(-1,1){80}}
\put(0,0){\line(1,1){80}}
\put(-60,60){\line(1,1){20}}
\put(35,35){\line(-1,1){45}}
\put(60,60){\line(-1,1){20}}
\put(0,0){\circle*{4}}
\put(-60,60){\circle*{4}}
\put(60,60){\circle*{4}}
\put(35,35){\circle*{4}}
\put(-80,80){\circle*{4}}
\put(-40,80){\circle*{4}}
\put(-10,80){\circle*{4}}
\put(40,80){\circle*{4}}
\put(80,80){\circle*{4}}
\put(-87,87){\mbox{${\cal K}_{\cre 2}$}}
\put(-47,87){\mbox{${\cal K}_{\cg 2}$}}
\put(-17,87){\mbox{${\cal K}_{\cb 2}$}}
\put(33,87){\mbox{${\cal K}_{\cre 2}$}}
\put(73,87){\mbox{${\cal K}_{\cre 2}$}}
\put(70,54){\mbox{$\{{\cal K}_{\cre 2},{\cal K}_{\cre 2}\}$}}
\put(-105,54){\mbox{$\{{\cal K}_{\cre 2},{\cal K}_{\cg 2}\}$}}
\put(40,27){\mbox{$\left\{{\cal K}_{\cb 2},\, \{{\cal K}_{\cre 2},{\cal K}_{\cre 2}\}\right\}$}}
\put(12,-3){\mbox{$\Big\{\{{\cal K}_{\cre 2},{\cal K}_{\cg 2}\},\
\left\{{\cal K}_{\cb 2},\, \{{\cal K}_{\cre 2},{\cal K}_{\cre 2}\}\right\}\Big\}$}}
\end{picture}

\noindent
(in fact there are plenty of operators, which {\it can not}
be described this way: the join operation has a huge cokernel in the space
$({\cal S}_m^r)^{conn}$),
\item whether or not they have the same image under the cut operation $\Delta$.
\end{itemize}

\noindent
Immediate results from these five characterizations are often different. They receive more structure
by separation of connected and disconnected
operators and by the orbits of the global $S_r^{coloring}$ group.
We demonstrate how these organize themselves for the first five levels
$m\leq 5$ of the Aristotelian $r=3$ model.
In fact, the first three characterizations of operators can be found in Appendices A and B, and, in this section, we mostly concentrate on the CJ structure of operators, which is essential for the two latter characterizations.

\subsection{Level $S_1$}

At this level, there are no permutations and there is a single operator
${\cal K}_{[1],[1],[1]}=
{\cal K}_1 = M_{{\cre i_1}{\cg i_2}{\cb i_3}}\bar M^{{\cre i_1}{\cg i_2}{\cb i_3}}$.
This operator has no distinguished coloring: permutations of colorings leave it intact.
Pictorially, it is just

\begin{picture}(100,40)(-180,-20)

\put(0,0){{\cre
\put(0,0){\line(1,0){50}}\put(25,0){\vector(1,0){2}}
}}

\put(0,0){{\cg
\qbezier(0,0)(25,20)(50,0)\put(25,10){\vector(1,0){2}}
}}

\put(0,0){{\cb
\qbezier(0,0)(25,-20)(50,0)\put(25,-10){\vector(1,0){2}}
}}

\put(50,0){\circle*{4}}
\put(0,0){\circle{4}}
\put(-12,-2){\mbox{$1$}}
\put(55,-2){\mbox{$\bar 1$}}
\put(-62,-2){ \mbox{ $  {\cal K}_1\ = \ $    }}

\end{picture}

\noindent
where $1$ and $\bar 1$ refer to the fields $M$ and $\bar M$,
each of which, in this example, appear once.
We changed directions of the green and blue arrows as compared to
\cite{IMMten2} to better suit the permutation group language,
which we use in the present paper.

The Gaussian average in this case is depicted by the Feynman diagram,
where the above picture turns into the 2-valent vertex
with two external legs labeled by $1$ and $\bar 1$
connected by a thick propagator, a tube which has three colored
lines inside:

\begin{picture}(100,75)(-260,-30)

\put(0,20){

\put(0,0){{\cre
\put(0,0){\line(1,0){30}}\put(15,0){\vector(1,0){2}}
\qbezier(0,0)(-35,-40)(15,-40)
\qbezier(30,0)(65,-40)(15,-40)
}}

\put(0,0){{\cg
\qbezier(-3,3)(15,23)(33,3)\put(15,13){\vector(1,0){2}}
\qbezier(-3,3)(-40,-43)(15,-43)
\qbezier(33,3)(70,-43)(15,-43)
}}

\put(0,0){{\cb
\qbezier(3,-3)(15,-23)(27,-3)\put(15,-13){\vector(1,0){2}}
\qbezier(3,-3)(-30,-37)(15,-37)
\qbezier(27,-3)(60,-37)(15,-37)
}}
\put(-18,-2){\mbox{$1$}}
\put(45,-2){\mbox{$\bar 1$}}
}

\linethickness{0.8mm}
\put(-180,20){
\put(0,0){\circle*{8}}
\qbezier(0,0)(20,0)(20,-20)
\qbezier(0,0)(-20,0)(-20,-20)
\qbezier(0,-40)(20,-40)(20,-20)
\qbezier(0,-40)(-20,-40)(-20,-20)
\put(-12,5){\mbox{$1$}}
\put(7,5){\mbox{$\bar 1$}}

}

\put(-130,-2){\mbox{$=$}}
\put(-113,-2){ \mbox{ $  \Big<\!\!\Big<{\cal K}_1\Big>\!\!\Big>$}}
\put(-60,-2){\mbox{$=$  }}
\put(90,-2){\mbox{$= \ {\cre N_1}{\cg N_2}{\cb N_3} \ \equiv \ \beta $}}

\end{picture}

The cut and join operations act on ${\cal K}_1$ in the simple way:
\be
\Delta {\cal K}_1 =  {\cre N_1}{\cg N_2}{\cb N_3}=\beta
\ee
and
\be
\{{\cal K}_1,{\cal K}_1\} = {\cal K}_1
\ee
Moreover, for any other gauge-invariant operator ${\cal K}$
of a definite degree,
\be
\{{\cal K}_1,{\cal K}\} = {\rm deg}_{\cal K}\cdot{\cal K}
\label{joinK1}
\ee
i.e. ${\cal K}_1$ acts as a dilatation (grading) operator in the ring.
The  subring generated by the dilatation
(grading) operator ${\cal K}_1$, which consists of operators ${\cal K}_1^m$
at all levels $m$,
is by itself closed under the cut and join operations:
\be
\Delta\Big({\cal K}_1^m\Big) = m(m-1+\beta)\,{\cal K}_1^{m-1} \nn \\
\{{\cal K}_1^m,{\cal K}_1^n\} = mn\,{\cal K}_1^{m+n-1}
\ee
Also (\ref{joinK1}) generalizes to
\be
\{{\cal K}_1^m,{\cal K}\} = m\cdot {\rm deg}_{\cal K}\cdot{\cal K}{\cal K}_1^{m-1}
\label{joinKm}
\ee

\subsection{Level $S_2$}

In this subsection, we analyze the structure of gauge-invariant operators in detail, partly repeating some issues discussed earlier for illustrative purposes.

\subsubsection{Gauge-invariant  operators}

In this case, one can draw four different pictures for the gauge-invariant operators: ${\cal K}_1^2$, ${\cal K}_{\cre 2}$, ${\cal K}_{\cg 2}$ and ${\cal K}_{\cb 2}$ (see Appendix A2). The question is what they have to do with the eight elements
${\cre \sigma_1}\otimes{\cg \sigma_2}\otimes {\cb \sigma_3}\in S_2^{\otimes 3}\,$.
Working in the {\bf RG}-gauge, i.e. if we fix $\sigma_1=id$, the dictionary looks simple:
\be
\begin{array}{ccc|c}
{\cre \sigma_1} & {\cg \sigma_2} & {\cb \sigma_3} & {\rm operator} \\
&&\\
\hline
&&\\
\phantom.{\cre (1)(2)} & {\cg (1)(2)} & {\cb (1)(2)} & {\cal K}_1^2 \\
&&\\
\phantom.{\cre (1)(2)} &{\cg (1)(2)} & {\cb (12)} & {\cal K}_{\cb 2} \\
\phantom.{\cre (1)(2)} &{\cg (12)} & {\cb (1)(2)} & {\cal K}_{\cg 2} \\
\phantom.{\cre (1)(2)} &{\cg (12)} & {\cb (12)} & {\cal K}_{\cre 2} \\
\end{array}
\ee
The rule is just that the cycle ${\cb (12)}$ in the blue column means that the blue arrows
go from vertex $1$ to $\bar 2$ and from $2$ to $\bar 1$, while the pair of cycles
${\cb (1)(2)}$ means that the blue arrows go from $1$ to $\bar 1$ and from $2$ to $\bar 2$.

One can unfix $\sigma_1$ and obtain the second half of the table:
\be
\begin{array}{ccc|c}
{\cre \sigma_1} & {\cg \sigma_2} & {\cb \sigma_3} & {\rm operator} \\
&&\\
\hline
&&\\
\phantom.{\cre (12)} & {\cg (12)} & {\cb (12)}  & {\cal K}_1^2 \\
&&\\
\phantom.{\cre (12)} &{\cg (1)(2)} & {\cb (12)} & {\cal K}_{\cg 2} \\
\phantom.{\cre (12)} &{\cg (12)} & {\cb (1)(2)} & {\cal K}_{\cb 2} \\
\phantom.{\cre (12)} &{\cg (1)(2)} & {\cb (1)(2)} & {\cal K}_{\cre 2} \\
\end{array}
\ee
This, certainly restores the symmetry between the three ${\cal K}_2$, at the price of
each operator appearing twice.

The emerging rule is that when all the three
permutations are the same, we obtain ${\cal K}_1^2$,
while when two are the same, while the third differs, we
get ${\cal K}_2$, colored by the third permutation.

At the same time, the orbits of diagonal $S_2$ are also alike:
there are four of the size $2$,
and they are, indeed, in one-to-one correspondence
with the four operators.

The coset ${\cal S}_2^3=S_2\backslash S_2^{\otimes 3}/S_2$ consists of four classes
labeled by pairs of permutations.
Since, for $S_2$, the permutations are in one-to-one correspondence with the Young diagrams,
$()\leftrightarrow [1,1]$ and $(1,2)\leftrightarrow [2]$, an
alternative enumeration of the elements is by pairs of the Young diagrams.

In this case, there are four ordered admissible triples, and the corresponding four operators are
\be
[11],[11],[11] & {\cal K}_{[11],[11],[11]} = {\cal K}_1^2 =
M_{{\cre i_1}^{1}{\cg i_2}^1{\cb i_3}^1}M_{{\cre i_1}^2{\cg i_2}^2{\cb i_3}^2}
\bar M^{{\cre i_1}^1{\cg i_2}^1{\cb i_3}^1}\bar M^{{\cre i_1}^2{\cg i_2}^2{\cb i_3}^2}
\nn \\ \nn \\
\phantom. [2],[11],[2] & {\cal K}_{[2],[11],[2]} = {\cal K}_{\cg 2} =
M_{{\cre i_1}^1{\cg i_2}^1{\cb i_3}^1}M_{{\cre i_1}^2{\cg i_2}^2{\cb i_3}^2}
\bar M^{{\cre i_1}^1{\cg i_2}^2{\cb i_3}^1}\bar M^{{\cre i_1}^2{\cg i_2}^1{\cb i_3}^2}
\nn \\
\phantom. [11],[2],[2] & {\cal K}_{[11],[2],[11]} = {\cal K}_{\cb 2} =
M_{{\cre i_1}^1{\cg i_2}^1{\cb i_1}^1}M_{{\cre i_1}^2{\cg i_2}^2{\cb i_3}^2}
\bar M^{{\cre i_1}^1{\cg i_2}^1{\cb i_3}^2}\bar M^{{\cre i_1}^2{\cg i_2}^2{\cb i_3}^1}
\nn \\
\phantom. [2],[2],[11] & {\cal K}_{[2],[2],[11]} = {\cal K}_{\cre 2} =
M_{{\cre i_1}^1{\cg i_2}^1{\cb i_3}^1}M_{{\cre i_1}^2{\cg i_2}^2{\cb i_3}^2}
\bar M^{{\cre i_1}^1{\cg i_2}^2{\cb i_3}^2}\bar M^{{\cre i_1}^2{\cg i_2}^1{\cb i_3}^1}
\ee

The corresponding Gaussian averages are
\be
\left<\!\left<{\cal K}_{\cre 2}\right>\!\right>=N_1N_2N_3({\cre N_1}+{\cg N_2}{\cb N_3})
= \beta ({\cre N_1}+{\cg N_2}{\cb N_3})\nn \\
\left<\!\left<{\cal K}_{\cg 2}\right>\!\right>=N_1N_2N_3({\cg N_2}+{\cre N_1}{\cb N_3})
=\beta ({\cg N_2}+{\cre N_1}{\cb N_3})\nn \\
\left<\!\left<{\cal K}_{\cb 2}\right>\!\right>=N_1N_2N_3({\cb N_3}+{\cre N_1}{\cg N_2})
= \beta ({\cb N_3}+{\cre N_1}{\cg N_2})\nn \\
\nn \\
\left<\!\left<{\cal K}_{1}\,{\cal K}_{1}\right>\!\right>=N_1N_2N_3(N_1N_2N_3+1)=\beta(\beta+1)
\ee

At last, one may not distinguish between the three operators $ {\cal K}_{\cre 2}$, $ {\cal K}_{\cg 2}$ and ${\cal K}_{\cb 2}$:
they are related by the action of the group $S_3^{coloring}$, which
permutes the three colorings ${\cre N_1},\ {\cg N_2}, \ {\cb N_3}$.
This group has just {\it two} orbits in the space of operators
and, in this case, these are in one-to-one correspondence with
{\it the admissible un-ordered triples} of Young diagrams.

From now on, we often omit the tale of 1's from the Young diagram,
i.e., in example of $S_2$, write $[]$ instead of $[11]$.

\subsubsection{CJ structure at level $m=2$}

The join operation lifts a pair of the level-two operators to level three,
see (\ref{j3}) below.
Joining with ${\cal K}_1$ just multiplies by 2, in accordance with
(\ref{joinK1}).

The cut operation at level $2$ takes all operators from level 2 to level $1$, where there
is only one operator, ${\cal K}_1$:
\be
\Delta \Big({\cal K}_1^2\Big) = 2(N_1N_2N_3+1){\cal K}_1 = 2(\beta+1){\cal K}_1
\ee
and
\be
\Delta{\cal K}_{\cre 2} =  2(\overbrace{{\cre N_1}+{\cg N_2}{\cb N_3}}^{\cre\alpha}) {\cal K}_1 \nn \\
\Delta{\cal K}_{\cg 2} =  2({\cg N_2}+{\cre N_1}{\cb N_3}) {\cal K}_1 \nn \\
\Delta{\cal K}_{\cb 2} =  2(\underbrace{{\cb N_3}+{\cre N_1}{\cg N_2}}_{\cb\alpha}) {\cal K}_1
\ee
Thus, already at this level $\Delta$ has a big kernel
(of codimension one):
\be
{\rm Ker}^{(2)}_{conn}(\Delta) = {\rm span}\Big\{
(\beta+1){\cal K}_{\cre 2} -  {\cre\alpha} {\cal K}_1^2,\
(\beta+1){\cal K}_{\cg 2} -  {\cg\alpha} {\cal K}_1^2, \
(\beta+1){\cal K}_{\cb 2} -  {\cb\alpha} {\cal K}_1^2,
 \Big\}
\ee
in accordance with (\ref{kerDelta}).
Even in the sector of connected operators:
\be
{\rm Ker}^{(2)}_{conn}(\Delta) = {\rm span}\Big\{
{\cre \alpha}{\cal K}_{\cg 2} - {\cg\alpha}{\cal K}_{\cre 2}, \
{\cre \alpha}{\cal K}_{\cb 2} - {\cb\alpha}{\cal K}_{\cre 2}
\Big\}
\ee

Note that $\Delta$ converts all the four operators at level two into a single one
at level one.
This is what does not happen at $r=2$ (for matrix model),
where there is just one connected operator $\Tr (M\bar M)^m$ at each level,
but is necessarily happening in tensor models with $r\geq 3$, where the number
of operators is growing with the level.
Thus, $\Delta$ has a huge kernel
(which we mentioned as CJ ``cohomology" in the scheme in section 2).
From the point of view of Virasoro-like constraints, this means that
the $\frac{\p^2}{\p t\p t}$ part of $\Delta$ is highly degenerate,
and there is a large sub-sector, where only the $t\frac{\p}{\p t}$ part
is operative, which is fully controlled by the rooted-tree algebra.

\subsection{Level $S_3$}

\subsubsection{Operators}

Operators at this level are described by the four disconnected diagrams
for ${\cal K}_1^2$ and
${\cal K}_{\cre 2}{\cal K}_1$,  ${\cal K}_{\cg 2}{\cal K}_1$,
${\cal K}_{\cb 2}{\cal K}_1$.
and by seven
connected diagrams (see Appendix A3) already considered in \cite{IMMten2}.
The three diagrams in the last line of s.A3 are actually the same
(topologically equivalent).
Note the numeration of vertices: according to the {\bf RG} ``gauge fixing"  $\sigma_1=id$,
all red arrows should connect vertices with the same numbers: $1$ with $\bar 1$,
$2$ with $\bar 2$ and $3$ with $\bar 3$.

This is in full agreement with the size of the
coset ${\cal S}_3^3 =S_3\backslash S_3^{\otimes 3}/S_3$, which contains exactly
$11$ elements (conjugacy classes),  in one-to-one correspondence with the pictures of operators:

\bigskip

\be
\!\!\!\!\!\!\!\!\!\!\!\!\!
\begin{array}{c|c|c|c|c|c}
{\cal K}_1^3 &[\ ]\otimes[\ ] &(\ )(\ ) &{\cal K}_{{\cg [\,]},{\cb [\,]}} & [\ ]& 1 \\
&&&&&\\
{\cal K}_{\cre 3} &[3]\otimes[3] & (123)\otimes (123),\ (132)\otimes (132)
&{\cal K}_{{\cg (123)},{\cb (123)}}& [\ ]& 2 \\
{\cal K}_{\cb 3} &[\ ]\otimes[3]& (\ )\otimes (123), \ (\ )\otimes (132)
&{\cal K}_{{\cg [\,]},{\cb [3]}}& [3] & 2 \\
{\cal K}_{\cg 3} &[3]\otimes[\ ]& (123)\otimes (\ ), \ (132)\otimes (\ )
&{\cal K}_{{\cg [3]},{\cb [\,]}}& [3]& 2 \\
&&&&&\\
{\cal K}_{3W} &[3]\otimes[3]& (123)\otimes (132),\ (132)\otimes (123)
&{\cal K}_{{\cg (123)},{\cb (132)}}& [3] & 2 \\
&&&&&\\
{\cal K}_{\cre 2}{\cal K}_1 &[2]\otimes[2]
& (12)\otimes (12), \ (13)\otimes (13), (23)\otimes (23)
&{\cal K}_{{\cg (12)},{\cb (12)}}& [\ ]& 3 \\
{\cal K}_{\cg 2}{\cal K}_1 &[\ ]\otimes[2]
& (\ )\otimes (12), \  (\ )\otimes (12),  \  (\ )\otimes (12)
&{\cal K}_{{\cg [\,]},{\cb [2]}}&[2]& 3 \\
{\cal K}_{\cb 2}{\cal K}_1 &[2]\otimes[\ ]
& (12)\otimes (\ ), \ (13)\otimes (\ ), \ (23)\otimes(\ )
&{\cal K}_{{\cg [\,]},{\cb [2]}}& [2]& 3 \\
&&&&&\\
{\cal K}_{{\cg 2},{\cb 2}} &[2]\otimes[2]& (12)\otimes (13), \ (12)\otimes (23),
\ (13)\otimes(12),\ (13)\otimes(23),\
(23)\otimes (12),\ (23)\otimes (13)
&{\cal K}_{{\cg (12)},{\cb (13)}}& [3]&  6 \\
{\cal K}_{{\cb 2},{\cre 2}}&[2]\otimes[3]& (12)\otimes (123),\ (12)\otimes (132),
\ (13)\otimes (123),\ (13)\otimes (132),\
(23)\otimes (123),\ (23)\otimes (132)
&{\cal K}_{{\cg [2]},{\cb [3]}}&[2] & 6 \\
{\cal K}_{{\cre 2},{\cg 2}} &[3]\otimes[2]& (123)\otimes (12),\ (132)\otimes (12),
\ (123)\otimes (13),\ (132)\otimes (13),\
(123)\otimes (23),\ (132)\otimes (23)
&{\cal K}_{{\cg [3]},{\cb [2]}}& [2]& 6 \\
&&&&&\\
\hline
&&&&& 36
\end{array}\nn\\
\label{ops3}
\ee
The $m!=3! = 6$ permutations from $S_3$ are classified according to $3$ conjugacy classes $[111]=[\ ],[21]=[2]$, and $[3]$.
In the penultimate column, we show the conjugacy class of the composition
(product) $\sigma_3^{-1}\circ\sigma_2$, which is an invariant of the conjugation (\ref{conjAr}).
The last column is the size of the class, the sum of sizes is $(m!)^2 = 36$.

The first column lists the operators from diagrams,
in attempt to reflect their features seen pictorially, such as the ``wheel" for ${\cal K}_{3W}$.
This will become less and less systematic with increase of the level.
The fourth column lists operators mostly referring the pair of the conjugacy classes for the green-blue
permutations in the gauge. Sometimes, however, more close information on the relative permutation $\sigma_3^{-1}\circ\sigma_2$ is required,
and, for these cases, we provide representative $(\sigma_2,\sigma_3)$, in the lexicographic ordering such as $(12345)(67)(89)$, which is one-to-one correspondence
with an element of the coset ${\cal S}_m^3$. In the present case of level 3, the third label $[\sigma_3^{-1}\circ\sigma_2]$ is adequate to the classification. We will see, however, that this is not the case at higher levels.

\bigskip

The action of the group $S_r^{coloring}$, which permutes $r=3$ colorings, is non-transparent
in the gauge ${\cre \sigma_1}=id$.
Its $5$ orbits are, however, easily seen in the above table.
For example,
$$
{\cal K}_{\cre 3} = {\cal K}_{{\cre id},{\cg (123)},{\cb (123)}}
\ \stackrel{{\cre r}\longleftrightarrow {\cg g}}{\longrightarrow} \
{\cal K}_{{\cre (123)},{\cg id},{\cb (123)}}
\cong {\cal K}_{{\cre id},{\cg (123)^{-1}},{\cb (123)^{-1}\circ(123)}}
\cong {\cal K}_{{\cre id},{\cg (132)},{\cb id}} \cong {\cal K}_{{\cre id},{\cg(123)},{\cb id}}
= {\cal K}_{\cg 3}
$$
while
$$
{\cal K}_{3W} = {\cal K}_{{\cre id},{\cg (123)},{\cb (132)}}
\ \stackrel{{\cre r}\longleftrightarrow {\cg g}}{\longrightarrow} \
{\cal K}_{{\cre (123)},{\cg id},{\cb (132)}}
\cong {\cal K}_{{\cre id},{\cg (123)^{-1}},{\cb (123)^{-1}\circ(132)}}
\cong {\cal K}_{{\cre id},{\cg (132)},{\cb (123)}} \cong {\cal K}_{{\cre id},{\cg(123)},{\cb (132)}}
= {\cal K}_{3W}
$$
All averages (not obligatory Gaussian) of any two elements belonging to the same orbit
differ just by the permutations of ${\cre N_1}$, ${\cg N_2}$, and ${\cb N_3}$.
In particular, this is the case for the Gaussian averages, and we need to list only five:
\be
\begin{array}{c|c|c|c}
\phantom. [\ ],[\ ],[\ ] & {\cal K}_1^3 & 1 &  \beta(\beta+1)(\beta+2) \\
\phantom. [2],[2],[\ ] & {\cal K}_{\cre 2}{\cal K}_1 & 3 &  \beta(\beta+2)(N_2N_3+N_1) \\
&&&  \\
\phantom. [3],[3],[\ ] & {\cal K}_{\cre 3} & 3& \beta\Big(N_2^2N_3^2+3\beta+N_1^2+1\Big)\\
\phantom. [2],[2],[3] & {\cal K}_{{\cg 2},{\cb 2}}   & 3 &
\beta\Big( N_1(\beta+N_2^2+N_3^2)+2N_2N_3+N_1\Big) \\
\phantom. [3],[3],[3] & {\cal K}_{3W} & 1 &   \beta\Big(3\beta+N_1^2+N_2^2+N_3^2\Big)
\end{array}
\label{ops3G}
\ee
We also show in these table that the five orbits are in one-to-one correspondence
with the $5$ admissible non-ordered triples.
Indeed, one can observe this already in the table (\ref{ops3}):
each of the five groups there is characterized by its own triple of Young diagrams,
and one check that these five triples are indeed the five admissible ones
(those for which the Hurwitz numbers are non-vanishing).
Note, however, that, for this identification to work, one needs to consider the class
of $\sigma_2\circ\sigma_3^{-1}$, not that of $\sigma_2\circ\sigma_3$: otherwise, the
entries of the penultimate column for the lines ${\cal K}_{\cre 3}$ and ${\cal K}_{3W}$
in (\ref{ops3}) would interchange places and break the rule.

\bigskip

{\bf Summary:} At level $m=3$, there are $5$ orbits of $S_3^{coloring}$
in ${\cal S}^3_3 = S_3\backslash S_3^{\otimes 3}/S_3$, the latter being of size $11$.
The $3$ independent connected operators, representing the orbits of the symmetrized coset
at level $m=3$ are
\be
\boxed{
{\cal K}_{\cre 3}, \ {\cal K}_{{\cg 2},{\cb 2}},  \ {\cal K}_{3W}
}
\ee
All the three can be made from a single {\bf RG}  circle.

\subsubsection{CJ structure at level $m=3$}

The join operation lifts the operators from level $2$ to level $3$
\be
\{{\cal K}_1^2,{\cal K}_1^2\} = 4{\cal K}_1^3 \nn \\
\{{\cal K}_1^2,{\cal K}_{\cre 2}\} = 4{\cal K}_{\cre 2}{\cal K}_1 \nn \\
\{{\cal K}_{\cre 2},{\cal K}_{\cre 2}\} = 4{\cal K}_{\cre 3} \nn \\
\{{\cal K}_{\cre 2},{\cal K}_{\cg 2}\} = 4{\cal K}_{{\cre 2},{\cg 2}}
\label{j3}
\ee
Note that the wheel operator ${\cal K}_{3W}$ is {\it not} produced in this way, namely, it does not belong to the image of join operation:
\be
{\rm CoKer}^{(3)}(\{\,\}) = {\rm span}({\cal K}_{3W})
\ee
In other words, it has {\it no} labeling in the tree (bracket word) systematics.
We obtain the following dictionary
for the seven connected operators at level $m=3$, translating into the
keystone-trees notation:
\be
\begin{array}{c|c|c}
{\rm graphs} & {\rm permutations} & {\rm keystone\ trees} \\
&&\\
{\cal K}_{\cre 3} & {\cal K}_{{\cg (123)}{\cb (123)}} & {\cal K}_{\{{\cre 2},{\cre 2}\}} \\
&&\\
{\cal K}_{\cg 3} & {\cal K}_{{\cg [3]},{\cb [\,]}} & {\cal K}_{\{{\cg 2},{\cg 2}\}} \\
&&\\
{\cal K}_{\cb 3} & {\cal K}_{{\cg [\,]},{\cb [3]}} & {\cal K}_{\{{\cb 2},{\cb 2}\}} \\
&&\\
{\cal K}_{{\cre 2},{\cg 2}} & {\cal K}_{{\cg [3]},{\cb [2]}} & {\cal K}_{\{{\cre 2},{\cg 2}\}} \\
&&\\
{\cal K}_{{\cre 2},{\cb 2}} &{\cal K}_{{\cg [2]},{\cb [3]}} & {\cal K}_{\{{\cre 2},{\cb 2}\}} \\
&&\\
{\cal K}_{{\cg 2},{\cb 2}} &{\cal K}_{{\cg [2]},{\cb [2]}} & {\cal K}_{\{{\cg 2},{\cb 2}\}} \\
&&\\
{\cal K}_{3W} & {\cal K}_{{\cg (123)},{\cb (132)}} & - \\
\end{array}
\ee
The action of cut operation takes operators from level $3$ to level $2$:
\be
\Delta \Big({\cal K}_1^3\Big) = 3(\beta+2){\cal K}_1^2 \nn \\
\Delta \Big({\cal K}_{\cre 2}{\cal K}_1\Big) = (\beta+4){\cal K}_{\cre 2}
+ 2({\cre N_1}+{\cg N_2}{\cb N_3}){\cal K}_1^2 \nn \\
\Delta {\cal K}_{\cre 3} = 3({\cre N_1}+{\cg N_2}{\cb N_3}){\cal K}_{\cre 2}
+ 3{\cal K}_1^2 \nn \\
\Delta {\cal K}_{{\cre 2},{\cg 2}} = (2{\cg N_2}+{\cre N_1}{\cb N_3}){\cal K}_{\cre 2}
+ (2{\cre N_1}+{\cg N_2}{\cb N_3}){\cal K}_{\cg 2}
+ 2{\cal K}_{\cb 2} + {\cb N_3}{\cal K}_1^2  \nn \\
\Delta {\cal K}_{3W} = 3\Big({\cre N_1}{\cal K}_{\cre 2} +
{\cg N_2}{\cal K}_{\cg 2} + {\cb N_3}{\cal K}_{\cb 2}\Big)
\label{Deltafrom3level}
\ee
Seven connected operators from level $3$ are mapped by $\Delta$ into just
a 4-dimensional space, consisting of three connected and one disconnected,
all what is there at the level 2, i.e. $\Delta$ has a (11-4=)7-dimensional kernel
in ${\cal S}_3^3$, spanned by the deformations of connected operators,
as explained around (\ref{kerDelta}):
\be
{\rm Ker}^{(3)}(\Delta) = {\rm span}
\left\{\begin{array}{c}
3(\beta+4)(\beta+2){\cal K}_{{\cre 2},{\cg 2}} -3(\beta+2)\Big(
({\cre N_1}+{\cre \alpha}){\cal K}_{\cre 2}{\cal K}_1 +
({\cg N_2}+{\cg \alpha}){\cal K}_{\cg 2}{\cal K}_1 + 2{\cal K}_{\cb 2}{\cal K}_1\Big)
+\\
+ \Big(\beta(12-5{\cb N_3})-4{\cre N_1}{\cg N_2}+2(2+{\cb N_3}^2)({\cre N_1}+{\cg N_2})^2 \Big)
{\cal K}_1^3 
\\ \\
3(\beta+4)(\beta+2){\cal K}_{{\cre 2},{\cb 2}} -3(\beta+2)\Big(
({\cre N_1}+{\cre \alpha}){\cal K}_{\cre 2}{\cal K}_1 +
({\cb N_2}+{\cb \alpha}){\cal K}_{\cb 2}{\cal K}_1 + 2{\cal K}_{\cg 2}{\cal K}_1\Big)
+\\
+ \Big(\beta(12-5{\cg N_3})-4{\cre N_1}{\cb N_2}+2(2+{\cg N_3}^2)({\cre N_1}+{\cb N_2})^2 \Big)
{\cal K}_1^3
\\ \\
3(\beta+4)(\beta+2){\cal K}_{{\cb 2},{\cg 2}} -3(\beta+2)\Big(
({\cb N_1}+{\cb \alpha}){\cal K}_{\cb 2}{\cal K}_1 +
({\cg N_2}+{\cg \alpha}){\cal K}_{\cg 2}{\cal K}_1 + 2{\cal K}_{\cre 2}{\cal K}_1\Big)
+\\
+ \Big(\beta(12-5{\cre N_3})-4{\cb N_1}{\cg N_2}+2(2+{\cre N_3}^2)({\cb N_1}+{\cg N_2})^2 \Big)
{\cal K}_1^3
\\ \\(\beta+4)(\beta+2){\cal K}_{\cre 3} -3{\cre \alpha}(\beta+2){\cal K}_{\cre 2}{\cal K}_1
+ (2{\cre\alpha}^2-\beta-4){\cal K}_1^3 \\
(\beta+4)(\beta+2){\cal K}_{\cg 3} -3{\cg \alpha}(\beta+2){\cal K}_{\cg 2}{\cal K}_1
+ (2{\cg\alpha}^2-\beta-4){\cal K}_1^3 \\
(\beta+4)(\beta+2){\cal K}_{\cb 3} -3{\cb \alpha}(\beta+2){\cal K}_{\cb 2}{\cal K}_1
+ (2{\cb\alpha}^2-\beta-4){\cal K}_1^3 \\ \\
(\beta+4)(\beta+2){\cal K}_{3W} - 3(\beta+2)({\cre N_1}{\cal K}_{\cre 2}{\cal K}_1
+{\cg N_2}{\cal K}_{\cg 2}{\cal K}_1 + {\cb N_3}{\cal K}_{\cb 2}{\cal K}_1)
+2({\cre N_1}{\cre\alpha}+{\cg N_2}{\cg\alpha}+{\cb N_3}{\cb\alpha}){\cal K}_1^3
\end{array}
\right\}
\ee
Already in this example, we see that not only the dilatation operator ${\cal K}_1$,
but also the entire set of ${\cal K}_{\cre m}$ operators of one
definite coloring forms a sub-ring closed under the cut and join operations:
\be
\!\!\!\Delta\left(\prod_{s} {\cal K}_{\cre m_s}\right) =
 \sum_{s'}  m_{s'}\!\left({\cre \alpha}{\cal K}_{\cre m_{s'}-1}
+ \sum_{k=1}^{m_{s'}-1} {\cal K}_{\cre k}{\cal K}_{\cre m_{s'}-k-1}\right)
\prod_{s\neq s'} {\cal K}_{\cre m_{s}}
+\! \sum_{s'<s''} m_{s'}m_{s''}{\cal K}_{\cre m_{s'}+m_{s''}-1}
\!\!\!\!\prod_{s\neq s',s''} \!\!{\cal K}_{\cre m_{s}}
\ee
and
\be
\left\{\prod_{s} {\cal K}_{\cre m_s},\ \prod_{t} {\cal K}_{\cre n_{t}}\right\} =
\sum_{s',t'} m_{s'}n_{t'}\,{\cal K}_{\cre m_{s'}+n_{t'}-1}
\prod_{s\neq s'} {\cal K}_{\cre m_s}  \prod_{t\neq t'} {\cal K}_{\cre n_{t}}
\ee
This is what allowed us to introduce an RG-closed ``red" tensor model in \cite{IMMten2},
which is actually the rectangular complex matrix model (RCM) with a rectangular {\it matrix} of size
${\cre N_1}\times {\cg N_2}{\cb N_3}$.

The general situation in the presence of all three keystone operators
${\cal K}_{\cre 2},{\cal K}_{\cg 2},{\cal K}_{\cb 2}$ is much more involved
and interesting.
What we see at the level $3$  is that ${\cal K}_3$ and ${\cal K}_{2,2}$ are
generated from the keystones with the help of the join operation $\{\ ,\ \}$
i.e. are the {\it tree} operators in the CJ-ring, while ${\cal K}_{3W}$
is not. As we will see at the consideration of the level 4,
it appears as the {\it loop} operator, i.e. as a result of the action of $\Delta$
on the tree-operators emerging at that level. Because of this, it has to be attached to the CJ pyramid in order to have the Ward identities (\ref{vir}) closed.

The very top of the join operation pyramid
looks as follows:\\

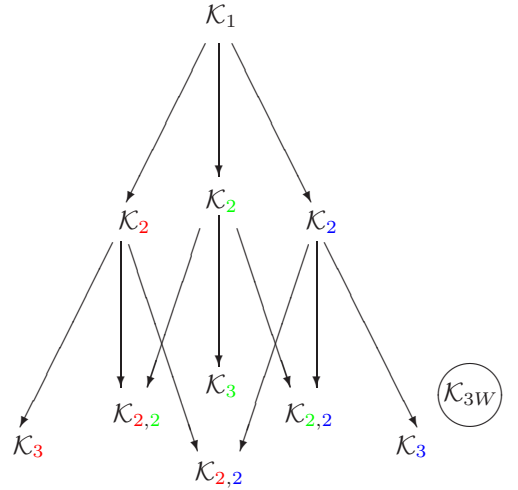
\vspace{3cm}

\parbox{10cm}{Note that, in principle, this pyramid does not include the operator ${\cal K}_{3W}$ at all. It is added here as an isolated point, and it should appear pleno jure only in the full CJ pyramid, which would contain also  up-going arrows describing the operation $\Delta$. Then, ${\cal K}_{3W}$ would be coming from higher levels of the join operation pyramid and further would go
to the three operators at level $2$.}

\begin{picture}(300,50)(-370,-210)

\put(-5,-2){\mbox{${\cal K}_1$}}
\put(-5,-10){\vector(-1,-2){30}}
\put(5,-10){\vector(1,-2){30}}
\put(0,-10){\vector(0,-1){50}}
\put(-38,-80){\mbox{${\cal K}_{\cre 2}$}}
\put(33,-80){\mbox{${\cal K}_{\cb 2}$}}
\put(-5,-72){\mbox{${\cal K}_{\cg 2}$}}
\put(-40,-85){\vector(-1,-2){35}}
\put(40,-85){\vector(1,-2){35}}
\put(0,-75){\vector(0,-1){57}}
\put(-78,-165){\mbox{${\cal K}_{\cre 3}$}}
\put(67,-165){\mbox{${\cal K}_{\cb 3}$}}
\put(-5,-142){\mbox{${\cal K}_{\cg 3}$}}
\put(-37,-85){\vector(0,-1){55}}
\put(-7,-80){\vector(-1,-3){20}}
\put(-40,-153){\mbox{${\cal K}_{{\cre 2},{\cg 2}}$}}
\put(37,-85){\vector(0,-1){55}}
\put(7,-80){\vector(1,-3){20}}
\put(25,-153){\mbox{${\cal K}_{{\cg 2},{\cb 2}}$}}
\put(-34,-86){\vector(1,-3){26}}
\put(34,-86){\vector(-1,-3){26}}
\put(-9,-175){\mbox{${\cal K}_{{\cre 2},{\cb 2}}$}}
\put(85,-145){\mbox{${\cal K}_{3W}$}}
\put(95,-143){\circle{22}}

\end{picture}

\subsection{Level $S_4$}

\subsubsection{Operators}

At this level,
in addition to disconnected and to already familiar connected ${\cal K}_4$ and ${\cal K}_{3,2}$,
we encounter six new types of connected operators (up to the permutations of colorings), not considered in \cite{IMMten2}, see Appendix A4.
Shown under the name of the operator in each diagram are ${\cg \sigma_2}\circ{\cb \sigma_3}$ and
$\sigma_3^{-1}\circ \sigma_2$.
The two first diagrams in the last line differ by colorings, say, of vertical lines:
there are two colors in the first case and three, in the second.

``C" in the last picture stands for ``cube", while an alternative notation can be ``WW" from double wheel.
Topologically, $K_{22W}$ is also a cube, but with a non-equivalent coloring:
each of the six faces in ${\cal K}_{4C}$ has edges of two different colorings,
while, in ${\cal K}_{22W}$, there are faces with edges of three colorings.
As to ${\cal K}_{222}$, it is symmetric under the permutations of three colorings.

There are $4!=24$ permutations and $5$ conjugacy classes $[4],[31]\longrightarrow [3],
[22],[211]\longrightarrow [2],[1111]\longrightarrow []$.

Coset ${\cal S}^3_4 = S_4\backslash S_4^{\otimes 3}/S_4$ consists of $43$ elements,
divided into $15$ orbits of coloring permutations $S_3^{coloring}$.
In particular, there are $15$ different Gaussian averages (modulo permutations of three
colorings). At the same time, the number of ordered admissible triples is $14$, which is smaller by one.
This degeneracy at level $m=4$ occurs between ${\cal K}_{3W}$ and ${\cal K}_{222}$ for $[3],[3],[3]$ and can be lifted by looking at $\sigma_3\circ\sigma_2$.
This gives the total number of operators $14+1=15$.

{\footnotesize
$$
\begin{array}{|cc|c|c|c|c|c|}
\sigma_2&\sigma_3&[\sigma_3^{-1}\circ\sigma_2]&[\sigma_3\circ\sigma_2]
&{\cal K} & S_3&\Big<\!\!\Big<{\cal K}\Big>\!\!\Big>\\
\hline
&&&&&&\\
\phantom. ()&()&[]&[]&  {\cal K}_1^4 & 1
& \beta(\beta+1)(\beta+2)(\beta+3)
\\
\phantom. (12)& (12)&[] & [] & {\cal K}_{\cre 2}{\cal K}_1^2 & 3
& {\cre\alpha}\beta (\beta+2)(\beta+3) \\
\phantom. (12)&(23)&[3]&[3]  & {\cal K}_{{\cg 2},{\cb 2}}{\cal K}_1 & 3
&\beta(\beta+3)\Big( N_1(\beta+N_2^2+N_3^2)+2N_2N_3+N_1\Big) \\
\phantom. (123)&(132)&[3]& [] & {\cal K}_{3W}{\cal K}_1 & 1
&\beta(\beta+3)\Big(3\beta+N_1^2+N_2^2+N_3^2
\Big)  \\
\phantom. (123)& (123)&[]& [3] & {\cal K}_{\cre 3}{\cal K}_1   & 3
 & \beta(\beta+3)\Big(N_2^2N_3^2+3\beta+N_1^2+1\Big)  \\
 &&&&&&\\
\phantom. (123)&(142)&[22]&[3] &{\cal K}_{{\cre 2},{\cg 2},{\cb 2}} & 3
& G_1\\
\phantom. (123)&(124)&[3]&[22] & {\cal K}_{222} & 1
&G_2 \\
&&&&&&\\
\phantom. (12)(34)&(12)(34)&[]&[] & {\cal K}_{\cre 2}^2  & 3
&\beta\Big(\underline{\underline{{\cre\alpha}^2(\beta}}+4)+2(\beta+1)
\Big)\\
\phantom. (12)(34)&(12)&[2]&[2]   &  {\cal K}_{\cre 2}{\cal K}_{\cg 2} & 3
&\beta\Big(
\underline{\underline{N_3\beta^2+N_3(N_1^2+N_2^2)(\beta}}+4)
+(\underline{\underline{N_1N_2}}+4N_3)\beta
+6N_1N_2+2N_3  \Big) \\
&&&&&&\\
\phantom. (1234)&(1432)&[22]&[] & {\cal K}_{\cre 22W} & 3
&G_3 \\
\phantom. (1234)&(12)(34)&[2]&[2] & {\cal K}_{{\cre 2},{\cg 2},{\cre 2}} & 6
&G_4 \\
\phantom. (13)&(124)&[4]&[4] & {\cal K}_{{\cre 3},{\cg 2}} & 6
&G_5\\
\phantom. (12)(34)&(13)(24)&[22]&[22] & {\cal K}_{4C} & 1
&G_6  \\
\phantom. (1234)&(1234)&[]&[22] &  {\cal K}_{\cre 4}  & 3
&\underline{{\cre\alpha}\beta\Big(N_2^2N_3^2+5\beta+N_1^2+5
\Big)}  \\
\phantom. (1234)&(1342)&[3]&[3] & {\cal K}_{\cre 31W} & 3
&G_7 \\
&&&&&&\\
\hline
&&&&&&\\
&&&&& 43 & \\
\end{array}
$$
}

\bigskip

\be
G_1=\beta\Big(\beta^2+(N_1^2+N_2^2+N_3^2) \beta
+ 2N_1^2N_2^2+2N_1^2N_3^2+N_2^2N_3^2+9\beta+N_1^2+2N_2^2+2N_3^2+1\Big)\nn\\
G_2=\beta\Big(\beta^2 + (N_1^2+N_2^2+N_3^2)(\beta+1)+2(N_1^2N_2^2+N_1^2N_3^2+N_2^2N_3^2)
+9\beta +2\Big)\nn\\
G_3=\beta\Big(2\beta(N_2N_3+3N_1)
+N_2^3N_3+N_2N_3^3+N_1^3+4N_1N_2^2+4N_1N_3^2+4N_2N_3+N_1
\Big)\nn\\
G_4=\beta\Big(
N_3^2N_2\beta+N_3^2N_2^3 +(2N_1N_3+5N_2)\beta
+N_1^3N_3+3N_3^2N_2+4N_2N_1^2+5N_1N_3+2N_2
\Big)\nn\\
G_5=\beta\Big( N_1^2N_2\beta+N_1^2N_2^3+(3N_1N_3+5N_2)\beta
+N_1N_3^3+2N_1^2N_2+3N_2N_3^2+5N_1N_3+3N_2
\Big) \nn\\
G_6=\beta\Big((N_1^2+N_2^2+N_3^2)(\beta+2)+2(N_1^2N_2^2+N_1^2N_3^2+N_2^2N_3^2)+9\beta
+2N_1^2+2N_2^2+2N_3^2
\Big)\nn\\
G_7=\beta\Big(3\beta(N_2N_3+2N_1)
+N_2^3N_3+N_2N_3^3+N_1^3+3N_1N_2^2+3N_1N_3^2+4N_2N_3+2N_1
\Big) \nn
\ee

\noindent
Note that the six operators ${\cal K}_2{\cal K}_2$  can be of two types: those
with coincident and those with different colorings. They are not related by coloring
permutations and have different Gaussian and any other averages.
Despite these operators are reducible (composite), their Gaussian averages do not
factorize. The terms, surviving the planar limit
are double-underlined: they are the terms with the highest possible power of {\it any}
of ${\cre N_1}$, ${\cg N_2}$ and ${\cb N_3}$ (see \cite{IMMten2}).
They from factorizable combinations.
The Gaussian average of $\left<\!\left<{\cal K}_{\cre 4}\right>\!\right>$
has an accidental factor $({\cre N_1} + {\cg N_2}{\cb N_3})$ (underlined once).

Thus, we have obtained $43$ operators with $43$ different averages.
Modulo permutations of colorings, there are just $15$ different operators with $15$ different averages,
i.e. $15$ different orbits of the coloring permutation group $S_3^{coloring}$.

\bigskip

{\bf Summary:}
At level $m=4$, there are $15$ orbits of $S_3^{coloring}$
in ${\cal S}^3_4 = S_4\backslash S_4^{\otimes 3}/S_4$ of total size $43$. This gives $15$ different operators, and
$8$ of them are independent connected operators:
\be
\boxed{
{\cal K}_{\cre 4}, \ {\cal K}_{{\cre 3},{\cg 2}}, \ {\cal K}_{{\cre 22W}}, \ {\cal K}_{\cre 31W},\
{\cal K}_{{\cre 2},{\cg 2},{\cre 2}}, \
{\cal K}_{222},\ {\cal K}_{{\cre 2},{\cb 2},{\cg 2}}, {\cal K}_{4C}
}
\ee
The first $5$ operators contain a single {\bf RG}  circle,
the last $3$   are made from two {\bf RG}  circles.
Note that  the operator ${\cal K}_{{\cre 2},{\cb 2},{\cg 2}} $ is red-green symmetric and
$S_3^{coloring}$ interchanges this one into two others of this kind to form an orbit,
while ${\cal K}_{{\cre 2},{\cg 2},{\cre 2}}$  and five others of this kind
get interchanged also by $S_3^{coloring}$.

For $\sigma_2\otimes \sigma_3\in [3]\otimes[3]$ with $\sigma_3^{-1}\circ\sigma_2\in [3]$,
for the first time, there is a new phenomenon:
multiplication can take place inside $S_3$ or can be essentially in $S_4$.
In the both cases, $(132)^{-1}\circ (123) = (132)\in [3]$ and
$(124)^{-1}\circ(123) = (413)\in [3]$
we obtain elements from the class $[31]$ in $S_4$, but in the former case it is also
the class $[3]$ in $S_3$, and it corresponds to the disconnected operator ${\cal K}_{3W}{\cal K}_1$,
while, in the latter case, we obtain the connected operator ${\cal K}_{222}$.
This is the origin of 15 operator classes instead of 14.

For $S_5$, the same phenomenon takes place:
$(1234)\circ(1234)=(13)(24)\in [22]\in S_4\subset S_5$
while $(1325)\circ(1234)=(14)(25)\in S_5$, etc.

\subsubsection{CJ structure}

The join operation can lift the operators from level 3 to level 4. We list here the join operations among the connected operator representatives which belong
to the symmetrized coset by the action of  the $S_3^{coloring}$:
\be
\{{\cal K}_{\cre 3}, {\cal K}_{\cre 2}  \} = 6 {\cal K}_{\cre 4}
\nn \\
\{{\cal K}_{\cre 3}, {\cal K}_{{\cg 2}} \} = 6 {\cal K}_{{\cre 3},{\cg 2}}
\nn \\
\{ {\cal K}_{{\cre 2},{\cg 2}}, {\cal K}_{\cre 2}  \}  =
2 {\cal K}_{{\cre 2},{\cg 2},{\cre 2}} + 4 {\cal K}_{{\cre 3},{\cg 2}}
\nn \\
\{ {\cal K}_{{\cre 2}, {\cg 2}}, {\cal K}_{\cb 2} \} =
4 {\cal K}_{{\cre 2},{\cb 2},{\cg 2}}  + 2 {\cal K}_{222 }
\nn \\
\{ {\cal K}_{3W} , {\cal K}_{\cre 2} \} = 6 {\cal K}_{{\cre 31W} }
\nn \\
\ldots
\ee
It follows, for representative of the symmetrized cosets, the following classification table:
\be
\begin{array}{c|c|c}
{\rm graphs} & {\rm permutations} & {\rm keystone\ trees} \\
&&\\
{\cal K}_{\cre 4} & {\cal K}_{{\cg (1234)}{\cb (1234)}} &
{\cal K}_{\{\{{\cre 2},{\cre 2}\},{\cre 2}\}} \\
&&\\
{\cal K}_{{\cre 3},{\cg 2}} & {\cal K}_{{\cg (1234)},{\cb (123)}} &
{\cal K}_{\{\{{\cre 2},{\cre 2}\},{\cg 2}\}} \\
&&\\
{\cal K}_{{\cre 31W} } &  {\cal K}_{{\cg (1234)},{\cb (1342)}} &  {\cal K}_{\{3W,{\cre 2}\}} \\
&&\\
{\cal K}_{{\cre 22W} } &  {\cal K}_{{\cg (1234)},{\cb (1432)}} & - \\
&&\\
{\cal K}_{{\cre 2},{\cg 2},{\cre 2}} &  {\cal K}_{{\cg (1234)},{\cb (12)(34)}} & 3{\cal K}_{\{\{{\cre 2},{\cg 2}\},{\cre 2}\}} 
-2{\cal K}_{\{\{{\cre 2},{\cre 2}\},{\cg 2}\}}\\
&&\\
{\cal K}_{{\cre 2},{\cb 2},{\cg 2}} &  {\cal K}_{{\cg (12)(34)},{\cb (123)}} & - \\
&&\\
{\cal K}_{222 } &  {\cal K}_{{\cg (123)},{\cb (124)}} & - \\
&&\\
{\cal K}_{4C } &  {\cal K}_{{\cg (12)(34)},{\cb (13)(24)}} & - \\
\end{array}
\ee
Here, in variance with level $m=3$, most of descendants of ${\cal K}_2$'s turn out to be linear combinations in the ring basis associated with elements of the coset ${\cal S}_m^3$. In particular, the four-dimensional sub-ring of ${\cal R}_4$ that consists of ${\cal K}_{{\cre 2},{\cb 2},{\cg 2}}$, ${\cal K}_{{\cre 2},{\cg 2},{\cb 2}}$, ${\cal K}_{{\cg 2},{\cre 2},{\cg 2}}$ (taking into account the symmetry of ${\cal K}_{{\cre 2},{\cb 2},{\cg 2}}$ w.r.t. permutations of red and green) and ${\cal K}_{222 }$ intersects with the join pyramid only at a subspace of dimension three (through $\{ {\cal K}_{{\cre 2}, {\cg 2}}, {\cal K}_{\cb 2} \}$, $\{ {\cal K}_{{\cre 2}, {\cb 2}}, {\cal K}_{\cg 2} \}$ and $\{ {\cal K}_{{\cg 2}, {\cre 2}}, {\cal K}_{\cb 2} \}$), i.e. the coker of the join operation in this sub-ring has dimension 1. At the same time, ${\cal K}_{{\cre 31W} } $ is obtained from the secondary operator of the first degree ${\cal K}_{3W}$:
\be
{\cal K}_{3W} \in \Delta\Big({\cal K}_{{\cre 2},{\cb 2},{\cg 2}}\Big)
\in \Delta\Big(\{ {\cal K}_{{\cre 2},{\cg 2}},{\cal K}_{\cb 2}\}\Big)
\in \Delta\left( \Big\{ \{{\cal K}_{\cre 2},{\cal K}_{\cg 2}\},{\cal K}_{\cb 2}\Big\}\right)
\ee
The operators ${\cal K}_{{\cre 22W} }$, ${\cal K}_{4C }$ are new secondary operators at level $m=4$.

The action of cut operation at level $m=4$ is
\be
\Delta({\cal K}_{{\cre 4}}^{{\cg (1234)},{\cb (1234)}}) =
4({\cre N_1}+{\cg N_2}{\cb N_3}){\cal K}_{\cre 3}^{{\cg (123)},{\cb (123)}}
+ 8{\cal K}_{\cre 2}^{{\cg (12)},{\cb (12)}}{\cal K}_1
\nn\\ \nn\\
\Delta({\cal K}_{{\cre 3},{\cg 2}}^{{\cg (1234)},{\cb (123)}}) =
(2{\cg N_2}+{\cre N_1}{\cb N_3}){\cal K}_{{\cre 3}} ^{{\cg (123)},{\cb (123)}}
+(3{\cre N_1}+2{\cg N_2}{\cb N_3}){\cal K}_{{\cre 2},{\cg 2}}^{{\cg (123)},{\cb (12)}}
+4{\cal K}_{{\cre 2},{\cb 2}}^{{\cg (12)},{\cb (123)}}
+{\cb N_3}{\cal K}_{{\cre 2}}^{{\cg (12)},{\cb (12)}}{\cal K}_1
+ 3{\cal K}_{{\cg 2}}^{{\cg (12)},{\cb ()}}{\cal K}_1
\nn\\ \nn\\
\!\!\!\!\!\!\!\!\!\!\!\!
\Delta({\cal K}_{{\cre 31W} }^{{\cg (1234)},{\cb (1342)}}) =
2{\cre N_1}{\cal K}_{{\cre 3}}^{{\cg (123)},{\cb (123)}}
+ (2{\cre N_1}+{\cg N_2}{\cb N_3}){\cal K}_{3W}^{{\cg (123)},{\cb (132)}}
+3{\cg N_2}{\cal K}_{{\cre 2},{\cg 2}}^{{\cg (123)},{\cb (12)}}
+3{\cb N_3}{\cal K}_{{\cre 2},{\cb 2}}^{{\cg (12)},{\cb (123)}}
+ 4{\cal K}_{{\cg 2},{\cb 2}}^{{\cg (12)},{\cb (23)}}
+ {\cal K}_{{\cre 2}}^{{\cg (12)},{\cb (12)}}{\cal K}_1
\nn \\ \nn\\
\Delta({\cal K}_{{\cre 22W} }^{{\cg (1234)},{\cb (1432)}}) = 
4{\cre N_1}{\cal K}_{3W}^{{\cg (123)},{\cb (132)}}
+ 4{\cg N_2}{\cal K}_{{\cre 2},{\cg 2}}^{{\cg (123)},{\cb (12)}}
+ 4{\cb N_3}{\cal K}_{{\cre 2},{\cb 2}}^{{\cg (12)},{\cb (123)}}
+ 4{\cal K}_{{\cg 2},{\cb 2}}^{{\cg (12)},{\cb (23)}}
\nn\\ \nn\\
\Delta({\cal K}_{{\cre 2},{\cg 2},{\cre 2}}^{{\cg (1234)},{\cb (12)(34)}}) = 
2(2{\cre N_1}+{\cg N_2}{\cb N_3}){\cal K}_{{\cre 2},{\cg 2}}^{{\cg (123)},{\cb (12)}}
+ 6{\cal K}_{{\cre 2},{\cb 2}}^{{\cg (12)},{\cb (123)}}
+ 2{\cg N_2}{\cal K}_{{\cre 3}}^{{\cg (123)},{\cb (123)}}
+ 2{\cb N_3}{\cal K}_{{\cre 2}}^{{\cg (12)},{\cb (12)}}{\cal K}_1
\nn\\ \nn\\
\Delta({\cal K}_{{\cre 2},{\cb 2},{\cg 2}}^{{\cg (12)(34)},{\cb (123)}}) = 
(2{\cre N_1}+{\cg N_2}{\cb N_3}){\cal K}_{{\cg 2},{\cb 2}}^{{\cg (12)},{\cb (23)}}
+ (2{\cg N_2}+{\cre N_1}{\cb N_3}){\cal K}_{{\cre 2},{\cb 2}}^{{\cg (12)},{\cb (123)}}
+2{\cb N_3}{\cal K}_{{\cre 2},{\cg 2}}^{{\cg (123)},{\cb (12)}}
+ 2{\cal K}_{{\cg 3}}^{{\cg (123)},{\cb ( )}}
+\nn\\
+ 2{\cal K}_{{\cre 3}}^{{\cg (123)},{\cb (123)}}
+ 2{\cal K}_{3W}^{{\cg (123)},{\cb (132)}} +
{\cre N_1} {\cal K}_{{\cre 2}}^{{\cg (12)},{\cb (12)}}{\cal K}_1
+ {\cg N_2}{\cal K}_{{\cg 2}}^{{\cg (12)},{\cb ()}}{\cal K}_1
\nn\\ \nn\\
\Delta({\cal K}_{222 }^{{\cg (123)},{\cb (124)}}) =
(2{\cre N_1}+{\cg N_2}{\cb N_3}){\cal K}_{{\cg 2},{\cb 2}}^{{\cg (12)},{\cb (23)}}
+(2{\cg N_2} + {\cre N_1}{\cb N_3})
{\cal K}_{{\cre 2},{\cb 2}}^{{\cg (12)},{\cb (123)}}
+(2{\cb N_3}+{\cre N_1}{\cg N_2}){\cal K}_{{\cre 2},{\cg 2}}^{{\cg (123)},{\cb (12)}}
+\nn\\
+2{\cal K}_{{\cre 3}}^{{\cg (123)},{\cb (123)}}
+2{\cal K}_{{\cg 3}}^{{\cg (123)},{\cb ( )}}
+ 2{\cal K}_{{\cb 3}}^{{\cg ( )},{\cb (123)}} +{\cal K}_1^3
\nn\\ \nn\\
\Delta({\cal K}_{4C }^{{\cg (12)(34)},{\cb (13)(24)}}) =
4\cdot \Big({\cre N_1}{\cal K}_{{\cg 2},{\cb 2}}^{{\cg (12)},{\cb (23)}}
+  {\cg N_2}{\cal K}_{{\cre 2},{\cb 2}}^{{\cg (12)},{\cb (123)}}
+  {\cb N_3}{\cal K}_{{\cre 2},{\cg 2}}^{{\cg (123)},{\cb (12)}}
+ {\cal K}_{3W}^{{\cg (123)},{\cb (132)}}\Big)
\nn\\
\ee
The kernel of $\Delta$ in this case is (26-11=)15-dimensional.

\subsection{Level $S_5$}

At this level, there are $5!=120$ permutations and $7$ conjugacy classes, which we abbreviate as $[5],[41]\longrightarrow [4], [32],
[311]\longrightarrow [3],[221]\longrightarrow [22],[2111]\longrightarrow[2],
[11111]\longrightarrow []$.

The coset ${\cal S}^3_5 = S_5\backslash S_5^{\otimes 3}/S_5$ consists of $161$ elements
divided into $44$ orbits of coloring permutations $S_3^{coloring}$.
The number of admissible unordered triples is $34$.

At this level, there are 97 connected operators, which we do not draw all and restrict ourselves only with an essential set, all remaining being easily restorable by permuting colors, see Appendix A5. The set of all pictures is divided into two sets: those with one red-green cycle (28 operators), and those with more red-green cycles (7 operators).
Note that the operators with a single reg-green cycle (the first group) are all connected.

Let us start from the first group of Appendix A5. The red-green symmetry from $S_3^{coloring}$ leaves $6$ of these $28$ operators
($\#\# 5,7,14,19,25,28$)
intact and interchanges the remaining operators within the $11$ remaining pairs
($9\leftrightarrow 4$, $11\leftrightarrow 3$, $15\leftrightarrow 2$,
$16\leftrightarrow 12$, $17\leftrightarrow 13$, $20\leftrightarrow 18$,
$21\leftrightarrow 1$, $22\leftrightarrow 8$,
$23\leftrightarrow 10$, $24\leftrightarrow 6$, $27\leftrightarrow 26$).

The red-blue and green-blue symmetries can change the number of
red-green cycles, but not always: in the above list, operators $6,8,10$
have also a single blue-green cycle, $22,23,24$, a single blue-red cycle,
and $25,27,28$ have a single cycle of each pair of colors.
They are, however, left intact by the corresponding {\bf GB} and {\bf RB} symmetries.

This gives $17$
independent connected operators with a single circle:
\be
{\cal K}_{I}, {\cal K}_{II}, {\cal K}_{III}, {\cal K}_{IV}, {\cal K}_{V}, {\cal K}_{VI},
{\cal K}_{VII}, {\cal K}_{VIII}, {\cal K}_{X}, {\cal K}_{XII}, {\cal K}_{XIII},
{\cal K}_{XIV}, {\cal K}_{XVIII}, {\cal K}_{XIX}, {\cal K}_{XXV}, {\cal K}_{XXVI},
{\cal K}_{XXVIII}\nn
\ee
\be
{\tiny
\begin{array}{ccccccccccccccccc}
I&II&III&IV&V&VI&VII&VIII&X&XII&XIII&XIV&XVIII&XIX&XXV&XXVI&XXIII \\ \\
\updownarrow & \updownarrow &\updownarrow &\updownarrow & - &\updownarrow & - &
\updownarrow &\updownarrow &\updownarrow &\updownarrow &-&\updownarrow & -&
- &\updownarrow & -  \\ \\
XXI&XV&XI&IX&&XXIV&&XXII&XXIII&XVI&XVII&&XX&&
&XXVII&
\end{array}
}\nn
\ee

The remaining $24-17=7$ orbits in $S_3^{coloring}$ (the second group in Appendix A5) have several cycles in every channel ({\bf RG}, {\bf GB} and {\bf RB}).
Actually, five are of the type $4+1$, and two of the type $3+2$ (or $3+1+1$, depending
on the choice of the channel, {\bf RG} , {\bf GB} or {\bf RB}).

All other operators with several {\bf GB} cycles are either disconnected or
related to I-XXXV by {\bf RGB} symmetry ($S_3^{coloring}$).

We now elaborate on these 24 connected operators from the point of view of the three conjugacy classes $[\sigma_2]$, $[\sigma_3]$ and $[\sigma_3^{-1} \circ \sigma_2]$, which are, in fact, conjugation invariants of the double cosets ${\cal S}_5^3$ in our {\bf RG} gauge, and can be used to label the operators. Among the $34$ unordered admissible triples designated by a set of three Young diagrams, which can be read off from the table of Appendix B2, there are 12 ones that contain $[5]$ and that are associated with connected operators: these are $([5],[],[5])$, $([5],[2],[32])$, $([5],[2],[4])$, $([5],[22],[22])$, $([5],[22],[3])$, $([5],[22],[5])$, $([5],[3],[3])$,  $([5],[3],[5])$, $([5],[32],[32])$, $([5],[32],[4])$, $([5],[4],[4])$, $([5],[5],[5])$.
Out of these, we have found that, in each of the cases $([5],[3],[5])$, $([5],[32],[4])$, $([5],[4],[4])$, there are two distinct connected operators associated: namely, they are doubly degenerate. In addition, we have found that there are three connected operators associated with $([5],[5],[5])$: they are triply degenerate. There are, in fact, 17 distinct connected operators of this kind.

Likewise, there are 6 unordered admissible triples that do  not contain $[5]$ and that are associated with connected operators: these are $([4],[22],[32])$, $([4],[22],[4])$, $([4],[3],[32])$, $([4],[3],[4])$, $([32],[22],[32])$, $([32],[3],[32])$. We have found that the case $([4],[3],[32])$ is doubly degenerate\footnote{Actually, the case $([32],[3],[32])$ is doubly degenerate as well, but one of them obviously gives a disconnected operator $K_{3W}K_2$.}. There are, in fact, 7 distinct connected operators of this kind. The figure for each of the first four cases contains one {\bf RG}-cycle of length 4 and  one {\bf RG}-cycle of length 1, while the figure for each of the last two cases contains of one {\bf RG}-cycle of length 3 and one {\bf RG}-cycle of length 2.

Let us explain briefly how we have managed to identify these degenerate operators. As we have already noted in the case of ${\cal S}_4^3$, one may consider $[\sigma_3 \circ \sigma_2]$ in contrary to $[\sigma_3^{-1} \circ \sigma_2]$. This one $[\sigma_3 \circ \sigma_2]$, unlike $[\sigma_3^{-1} \circ \sigma_2]$, is not a conjugation invariant of the double coset, but one can study the automorphism of the elements of the table in Appendix B2 (which encodes the elements of the center of the group algebra for the symmetric group $S_5$ with proper normalization) under $\sigma_3^{-1} \leftrightarrow \sigma_3$ to be able to tell which elements of the table correspond to degenerate operators.

Let us take an example to demonstrate this phenomenon. The element of the lowest right corner of the table for $S_5$ in Appendix B2 is read
\be
\label{55goesto}
[5][5] = 24[] + 8[22] + 12[3] + 8[5].
\ee
By studying the transformation of operators under $\sigma_3^{-1} \leftrightarrow \sigma_3$, we see that this is grouped into
\be
\label{55goestosplitting}
\left( 24[] + [5] \right) + \left( 8[22] + 6[3] \right) + \left( 6[3] + 5[5] \right) + 2[5].
\ee
Here, the first and the second terms inside each of the three parentheses have the same number of elements and get interchanged under $\sigma_3^{-1} \leftrightarrow \sigma_3$. Eq.(\ref{55goestosplitting}) tells us that $([5],[5],[3])$ splits into two pieces and corresponds to doubly degenerate operators, while $([5],[5],[5])$ splits into three pieces and corresponds to triply degenerate operators.

Let us note that the last four entries at level 5 with cycles of length 5 belong to four different conjugacy classes (are different elements of the coset ${\cal S}_5^3$), but have the same Gaussian average. They still coincide in the background of ${\cal K}_1$ and ${\cal K}_1^2$, but start getting separated already by ${\cal K}_{\cg 2}$.

Similarly the last two entries at level 5 with cycles of length 4 have the same Gaussian averages as $(1243)$, but belong to different conjugacy classes. They remain un-separated by insertions of ${\cal K}_{\cg 2}$ and ${\cal K}_{\cg 3}$, but $(1253)$ gets separated from $(1243)$ in the background of ${\cal K}_{{\cre 2},{\cg 2}}$, while $(1254)$, which differs from $(1243)$ just by the arrow inversion (and it is the first case when this changes the conjugacy class, i.e. the point in (\ref{coset})), only in that of ${\cal K}_{(12345),(1243)}={\cal K}_{XVIII}$.

The number of different Gaussian averages in different sectors
(the number of independent operators can be bigger) is given by the following table:
\be
N^G_{[\sigma_2]}= \ \ \ \
\begin{array}{lcccccccccc}
{\cg \sigma_2} &m=& 1&2&3&4&5&6&\ldots \\
\hline
() \ \underbrace{1+1+\ldots+1}_m &\longrightarrow& 1&2&3&5&7&11&\ldots &= \#(m)   \\
(12) &\longrightarrow&& 2&3&10&18&34&\ldots  \\
(123) &\longrightarrow &&&4&10&26&55&\ldots   \\
(1234) &\longrightarrow  &&&&10&28&88&\ldots   \\
(12345) &\longrightarrow &&&&&23&100&\ldots  \\
(123456) &\longrightarrow &&&&&&98&\ldots \\
(12)(34) &\longrightarrow &&&&8&21&62&  \ldots   \\
(123)(45) &\longrightarrow &&&&&26& 102& \ldots \\
(1234)(56) &\longrightarrow &&&&&&89&\ldots \\
(123)(456) &\longrightarrow &&&&&&55&\ldots \\
(12)(34)(56) &\longrightarrow &&&&&&34&\ldots \\
\hline
&&1&4&11&43& 149  & 728  & \ldots \\
needed &&&&&&161&901&\ldots \\
unseparated &&&&&&12& 173 &\ldots
\end{array}
\ee

In various backgrounds,
we get instead:
\be
\begin{array}{c|cccccc}
{\rm background}  & 1 &{\cal K}_1 & {\cal K}_1^2 & {\cal K}_{\cg 2}& {\cal K}_{\cg 3}\\
() & 7 & 7 & 7 & 7& 7 \\
(12) &18&&&18&  \\
(123) & 26 &&& 26& \\
(1234)& 28 &&& 29& 29 \\
(12345) &23&23&23& 25&25  \\
(12)(34)& 21 &&& 21&   \\
(123)(45) & 26 &&& 26&  \\
\hline
&149 &&& 152&  \\
\end{array}
\ee

\bigskip

{\bf Summary:}
At level $m=5$, there are $44$ orbits of $S_3^{coloring}$
in ${\cal S}^3_5 = S_5\backslash S_5^{\otimes 3}/S_5$ of size $161$.
The $24$ independent connected operators representing the orbits
at level $m=5$ are:
\be
\boxed{
\begin{array}{c}
{\cal K}_{I}, {\cal K}_{II}, {\cal K}_{III}, {\cal K}_{IV}, {\cal K}_{V}, {\cal K}_{VI},
{\cal K}_{VII}, {\cal K}_{VIII}, {\cal K}_{X}, {\cal K}_{XII}, {\cal K}_{XIII},
{\cal K}_{XIV},
{\cal K}_{XVIII}, {\cal K}_{XIX},
{\cal K}_{XXV}, {\cal K}_{XXVI},
{\cal K}_{XXVIII},\\ \ {\cal K}_{XXIX},{\cal K}_{XXX},{\cal K}_{XXXI},{\cal K}_{XXXII},{\cal K}_{XXXIII},{\cal K}_{XXIV},{\cal K}_{XXV}
\end{array}
}
\nn\ee
The first $17$ operators contain a single {\bf RG}  cycle,
the last $7$ operators are made from two {\bf RG}  cycles.

\section{Conclusion}
\setcounter{equation}{0}

Like eigenvalue matrix models, the rainbow tensor models have a good chance of being
a kind of superintegrable. Usually, superintegrability is understood as an explicit knowledge of more
than $N$ integrals of motion in phase space of dimension $2N$,
while the complete integrability requires exactly $N$.
Extra integrals often allow one to solve the system {\it explicitly}.
The best known example in ordinary mechanics is the existence of closed orbits.
The 2-dimensional harmonic oscillator has two commuting Hamiltonians,
$H_a=p_a^2+\omega_a^2q_a^2$, $a=1,2$, but it has closed orbits only when the
frequencies are rationally related, $m\omega_1=n\omega_2$ with some integer
$m$ and $n$.
In these cases, there are additional conservation laws like
\be
\omega_2=\omega_1: &  p_1q_2-p_2q_1 & \&\ \  p_1p_2+\omega^2q_1q_2\nn \\
\omega_2=2\omega_1: & \omega_1\omega_2 p_1q_1q_2+(p_1^2-\omega_1^2 q_1^2)p_2
& \&\ \  (p_1^2-\omega_1^2q_1^2)q_2-p_1p_2q_1   \nn \\
\ldots
\ee
Another example is the Runge-Lenz conservation law for motion in the Coulomb potential.

For matrix models, the superintegrability would mean that the partition functions
are not just generic $\tau$-functions, but some very special ones,
in some sense, ``better", ``simpler" and ``more explicit".
This feeling is, indeed, present in everybody who studied the problem,
and this is reflected in the peculiar notion of ``matrix-model $\tau$-functions"
widely used since \cite{UFN3}.
Technically, what is special about them is the large variety of Ward identities
(Virasoro constraints and alike), which, in the space of $\tau$-functions, usually
reduces to just one ``string equation".
A much stronger manifestation of hidden superintegrability should be the
existence of explicit formulas for all matrix model correlators in the Gaussian phase
recently discovered in \cite{AMMNhur,KR,MMmamo}.
This fact leaves not many doubts that superintegrability {\it is} the
pertinent feature of matrix models.
Still, it remains to understand this phenomenon in the more standard
terms of the Hamiltonian dynamics.

Remarkably, this understanding is not needed to discover
that the same explicit formulas for the Gaussian correlators are straightforwardly
lifted from matrix to tensor models \cite{IMMten2,Di,MMten,KGT},
where even the complete integrability is
far from being observed.
This observation gives a hope to bypass all the seeming difficulties in
developing the tensor models theory: they look like {\it super}integrable,
and this should stimulate further
investigations and guarantee a fast advance.

\bigskip

In the present paper, we considered different ways to describe the operator ring
in the simplest Aristotelian model with complex tensor of rank $r=3$ and with the
gauge symmetry $U({\cre N_1})\otimes U({\cg N_2})\otimes U({\cb N_3})$.
They are originally labeled by points of the double coset
${\cal S}^3_m = S_m\backslash S_m^{\otimes 3} / S_m$ made from the symmetric group $S_m$,
i.e. by the conjugacy classes ${\cre \sigma_1}\otimes {\cg \sigma_2}\otimes {\cb \sigma_3}$
w.r.t. the left and right multiplication by diagonal $S_m$ in $S_m^{\otimes 3}$.

(a) We considered various ``gauge choices" and pictorial descriptions of the operators,
found the number of connected operators and dimensions of the orbits of the colorings-permutation group $S_3$
for $m\leq 5$.

(b) We demonstrated that, starting from $m=5$, the Gaussian correlators do not distinguish
all independent operators, and complexity of the backgrounds
needed to lift the degeneracies strongly depends on the operator.

(c) We began analysis of the action of cut and join operations $\Delta$ and $\{\ ,\ \}$
on the operator ring, which is necessary to efficiently formulate the Ward identities
(Virasoro-like constraints).

One of the issues to address in the close future is the dependence of operator classification and of the
Gaussian calculus on the choice of rainbow model.
Two directions are most important from this point of view:
the single-tensor model of arbitrary rank $r$ and the starfish models with $|I|=r+1$
different tensors.

\section*{Acknowledgements}

Our work was partly supported by the grant of the Foundation for the Advancement of Theoretical Physics ``BASIS" (A.Mor.), by RFBR grants 16-01-00291 (A.Mir.) and 16-02-01021 (A.Mor.), by joint grants 15-51-52031-NSC-a, 16-51-53034-GFEN, 16-51-45029-IND-a (A.M.'s). The work of H.I. was supported by JSPS KAKENHI Grant Number JP15K05059. Support from RFBR/JSPS bilateral collaboration program ``Topological field theory and string theory: from topological recursion to quantum toroidal algebra" (17-51-50051-YaF) is appreciated.

\section*{Appendix A}

In this Appendix, we draw all connected operators emerging at the given level. Only at level $m=5$ we list only part of operators, all remaining being easily obtained by permuting colors. The notation of operator pictures throughout the paper refers to this Appendix.

\subsection*{A1. $m=1$}
\setcounter{equation}{0}
\def\theequation{B.\arabic{equation}}



\subsection*{A2. $m=2$}

%

\subsection*{A3. $m=3$}

%

\subsection*{A4. $m=4$}

%

\subsection*{A5. $m=5$}

\subsubsection*{Connected operators with one red-green cycle}

%

\subsubsection*{Connected operators with two red-green cycles}

%

\section*{Appendix B}

This Appendix contains various tables discussed in the main body of the text.

\subsection*{B1. Tables of lowest $z_h^R$}

{\footnotesize
\be
\!\!\!\!\!\!\!\!\!\!\!\!\!\!\!
\begin{array}{c|c||c|cc|ccc|ccccc|ccccccc|}
&& \ \ R: &[2]&[11] & [3] & [21] &[111]  & [4] & [31]& [22] & [211] & [1111]
&[5]&[41]&[32]&[311]&[221]&[2111]&[11111] \\
h & ||h||& z_h &&&&&&&&&& &&&&&&&\\
\hline
(12) &1 & 2 & 1& 1& &&  &&&&& &&&&&&&\\
() & 1& 2 & 1& 1& &&    &&&&& &&&&&&&\\
\hline
(123) &2&3&&&2&0&1  &&&&& &&&&&&&\\
(12) &3&2&&&0&1&1   &&&&& &&&&&&&\\
() &1& 6&&&2&3&1     &&&&& &&&&&&&\\
\hline
(1234) &6&4&&&&&   &2&0&1&0&1 &&&&&&&\\
(123) &8&3&&&&&    &0&2&0&0&1 &&&&&&&\\
(12)(34) &3&8&&&&& &2&0&3&2&1 &&&&&&&\\
(12) &6&4&&&&&     &0&0&1&2&1 &&&&&&&\\
() &1&24&&&&&       &6&8&3&6&1 &&&&&&&\\
\hline
(12345)&24&5&&&&&&&&&& &\boxed{4}&0&0&0&0&0&1\\
(1234)&30&4&&&&&&&&&& &0&2&0&0&1&0&1\\
(123)(45)&20&6&&&&&&&&&& &0&0&2&2&0&1&1\\
(123)&20&6&&&&&&&&&& &0&0&2&2&0&1&1\\
(12)(34)&15&8&&&&&&&&&& &0&2&0&0&3&2&1\\
(12)&10&12&&&&&&&&&& &0&0&2&2&3&4&1\\
()&1&120&&&&&&&&&& &24&30&20&20&15&10&1\\
\hline
&&&&&&&&&&&& &&&&&&&\\
\sum_\Delta z_\Delta^R &&&2&2& 4&4&3 &10&10&8&10&5
&28&34&26&26&22&18&7\\
&&&&&&&&&&&& &&&&&&&\\
&&&&\!\!\!\!\!\!\!\!\!\!\!\!\!\! 4&&11&&&&43\!\!\!\!&& &&&&161\!\!\!\!\!&&&\\
\end{array}
\label{tabzhR}
\ee}
It can be made symmetric by multiplication of each line with $||h||=\frac{|h|!}{z_h}$,
because the both sides  of
\be
||\Delta||\cdot z^R_\Delta   = ||R||\cdot z^\Delta_R
\ee
count the number of commuting permutations,
one of the type $R$, the other one of the type $\Delta$.
The entries in the row sum into $z_h=\sum_{R\,\vdash |h|} z_h^R$,
the entries in the columns sum into the numbers in the bottom of the table.

Comment on the boxed entry of the table:
the four permutations of the type $[5]$, which commute with $(12345)$ are:
$(12345)$, its inverse $(15432)$ and $(13524)$ with its inverse $(14253)$.

\subsection*{B2. Multiplication tables for the structure constants in the center of the group algebra, (\ref{Cddd})}

\be
\begin{array}{|c|c|}
\hline
1 & 1 \\
\hline\hline
[11] & [2] \\
\hline
[2] & [11] \\
\hline
\end{array}
\ee

\be
\begin{array}{|c|c|c|}
\hline
1 & 3 & 2 \\
\hline\hline
[111] & [21] & [3] \\
\hline
[21] & 3[111]+3[3] & 2[21] \\
\hline
[3] & 2[21] & 2[111]+[3] \\
\hline
\end{array}
\ee

\be
\begin{array}{|c|c|c|c|c|}
\hline
1 & 6 & 3 & 8 & 6 \\
\hline\hline
[1111] & [211] & [22] & [31] & [4] \\
\hline
[211] & 6[1111]+2[22]+3[31] & [211]+2[4]& 4[211]+4[4] & 4[22]+3[31] \\
\hline
[22] & [211]+2[4] & 3[1111]+2[22] & 3[31] & 2[211]+[4] \\
\hline
[31] & 4[211]+4[4] & 3[31] & 8[1111]+8[22]+4[31] & 4[211]+4[4] \\
\hline
[4] & 4[22]+3[31] & 2[211]+[4] & 4[211]+4[4] & 6[1111]+2[22]+3[31]\\
\hline
\end{array}
\ee

{\tiny
\be
\!\!\!\!\!\!\!\!\!\!\!\!\!\!\!\!\!\!\!
\begin{array}{|c|c|c|c|c|c|c|}
\hline
1&10&15&20&20&30&24 \\
\hline\hline
[11111] & [2111] & [221] & [311] & [32] & [41] & [5] \\
\hline
[2111] & 10[ ]+2[22 ]+3[3 ] & 3[2 ]+3[32]+2[4 ]& 6[2 ]+[32]+4[4 ]
& 4[22 ]+[3 ]+5[5] & 4[22 ]+6[3 ]+5[5] & 6[32]+4[4 ] \\
\hline
[221] & 3[2 ]+3[32]+2[4 ] & 15[ ]+2[22 ]+3[3 ]+5[5] &4[22 ]+6[3 ]+5[5]
&6[2 ]+6[32]+4[4 ]&6[2 ]+6[32]+9[4 ]&8[22 ]+6[3 ]+5[5] \\
\hline
[311] & 6[2 ]+[32] +4[4 ] & 4[22 ]+6[3 ]+5[5]  & 20[ ]+8[22 ]+7[3 ]+5[5]
& 2[2 ]+7[32]+8[4 ]&12[2]+12[32]+8[4 ]&8[22 ]+6[3 ]+10[5] \\
\hline
[32] & 4[22 ]+[3 ]+5[5] &6[2 ]+6[32]+4[4 ] &2[2 ]+7[32]+8[4 ]
&20[ ]+8[22 ]+7[3 ]+5[5]&8[22 ]+12[3 ]+10[5]&12[2 ]+6[32]+8[4 ]\\
\hline
[41] & 4[22 ]+6[3 ]+5[5] &6[2 ]+6[32]+9[4 ]&12[2]+12[32]+8[4 ]
&8[22 ]+12[3 ]+10[5]&30[ ]+18[22 ]+12[3 ]+15[5]&12[22 ]+12[32]+12[4 ] \\
\hline
[5] & 6[32]+4[4 ] &8[22 ]+6[3 ]+5[5]&8[22 ]+6[3 ]+10[5]
&12[2 ]+6[32]+8[4 ]&12[22 ]+12[32]+12[4 ]&24[ ]+8[22 ]+12[3 ]+8[5]\\
\hline
\end{array}
\nn
\ee
}

\bigskip

\noindent
We used the abbreviated notation inside the table, omitting all 1's from the Young diagram,
i.e. substituting $[11111]\longrightarrow[],[2111]\longrightarrow[2],[221]\longrightarrow[22],
[311]\longrightarrow[3],[41]\longrightarrow[4]$.
The first columns list both the diagrams $\Delta_1$ and their products with $\Delta=[1^m]=[\,]$,
likewise the second lines contain both $\Delta_2$ and their products with $\Delta=[1^m]=[\,]$.
The top lines list the multiplicities $||\Delta_2||$ (the number of permutations of the given type)
for the diagrams in the second line.

\subsection*{B3. The Hurwitz gauge: operators at levels $m=4,5$}

These tables describe computations in the Hurwitz gauge, see s.6. Here $\sigma_3$ and thus $\sigma_2$ are chosen randomly
among the representatives of the conjugacy class.

\paragraph{Operators at level $m=4$.}

{\footnotesize
\be
\begin{array}{c|ccc|ccc|c|}
&{\cre [\sigma_1]} & {\cg [\sigma_2]} & {\cb [\sigma_3]}
&{\cre \sigma_1^{can}} & {\cg \sigma_2}=\sigma_3\circ\sigma_1 & {\cb \sigma_3}
& {\cal N}^H([\sigma_1,\sigma_2,\sigma_3])
\\
&&&&&&&\\
\hline
&&&&&&&\\
\ph  &[4] &[4]&[3]&(1234)&(1324)&(123)&24\\
\ph   &[4]&[3]&[4]&(1234)&(243)&(1423)&24\\
\ph   &[4]&[4]&[22]&(1234)&(1432)&(13)(24)&6\\
\ph   &[4]&[22]&[4]&(1234)&(13)(24)&(1234)&6\\
\ph   &[4]&[4]&[]&(1234)&(1234)&()&6\\
\ph  10 &[4]&[]&[4]&(1234)&()&(1432)&6\\
\ph   &[4]&[3]&[2]&(1234)&(124)&(23)&24\\
\ph   &[4]&[2]&[3]&(1234)&(14)&(132)&24\\
\ph   &[4]&[22]&[2]&(1234)&(14)(23)&(13)&12\\
\ph   &[4]&[2]&[22]&(1234)&(24)&(14)(23)&12\\
&&&&&&&\\
\hline
&&&&&&&\\
\ph   &[3]&[4]&[4]&(123)&(1432)&(1423)&24\\
\ph   &[3]&[4]&[2]&(123)&(1423)&(14)&24\\
\ph   &[3]&[2]&[4]&(123)&(14)&(1432)&24\\
\ph   &[3]&[3]&[22]&(123)&(142)&(14)(23)&24\\
\ph   &[3]&[22]&[3]&(123)&(12)(34)&(234)&24\\
\ph   10&[3]&[3]&[3]&(123)&(132)&(123)&32\\
\ph   &[3]&[3]&[3]&(123)&(124)&(243)&32\\
\ph   &[3]&[2]&[2]&(123)&(23)&(13)&24\\
\ph   &[3]&[3]&[]&(123)&(123)&()&8\\
\ph   &[3]&[]&[3]&(123)&()&(132)&8\\
&&&&&&&\\
\hline
&&&&&&&\\
\ph   &[22]&[4]&[4]&(12)(34)&(1324)&(1423)&6\\
\ph   &[22]&[4]&[2]&(12)(34)&(1432)&(13)&12\\
\ph   &[22]&[2]&[4]&(12)(34)&(13)&(1432)&12\\
\ph   &[22]&[3]&[3]&(12)(34)&(143)&(132)&24\\
\ph   8&[22]&[22]&[22]&(12)(34)&(13)(24)&(14)(23)&6\\
\ph   &[22]&[2]&[2]&(12)(34)&(34)&(12)&6\\
\ph   &[22]&[22]&[]&(12)(34)&(12)(34)&()&3\\
\ph   &[22]&[]&[22]&(12)(34)&()&(12)(34)&3\\
&&&&&&&\\
\hline
&&&&&&&\\
\ph   &[2]&[4]&[3]&(12)&(1234)&(234)&24\\
\ph   &[2]&[3]&[4]&(12)&(143)&(1432)&24\\
\ph   &[2]&[4]&[22]&(12)&(1423)&(14)(23)&12\\
\ph   &[2]&[22]&[4]&(12)&(14)(23)&(1423)&12\\
\ph   &[2]&[3]&[2]&(12)&(132)&(13)&24\\
\ph   10&[2]&[2]&[3]&(12)&(13)&(132)&24\\
\ph   &[2]&[22]&[2]&(12)&(12)(34)&(34)&6\\
\ph   &[2]&[2]&[22]&(12)&(34)&(12)(34)&6\\
\ph   &[2]&[2]&[]&(12)&(12)&()&6\\
\ph   &[2]&[]&[2]&(12)&()&(12)&6\\
&&&&&&&\\
\hline
&&&&&&&\\
\ph   &[]&[4]&[4]&()&(1423)&(1423)&6\\
\ph   &[]&[3]&[3]&()&(132)&(132)&8\\
\ph  5 &[]&[22]&[22]&()&(14)(23)&(14)(23)&3\\
\ph   &[]&[2]&[2]&()&(13)&(13)&6\\
\ph   &[]&[]&[]&()&()&()&1\\
\end{array}
\ee}

\paragraph{Operators at level $m=5$.}

{\tiny
\be\label{t51}
\begin{array}{ccc|ccc|c|}
{\cre [\sigma_1]} & {\cg [\sigma_2]} & {\cb [\sigma_3]}
&{\cre \sigma_1^{can}} & {\cg \sigma_2}=\sigma_3\circ\sigma_1 & {\cb \sigma_3}
& {\cal N}^H([\sigma_1,\sigma_2,\sigma_3])\\
&&&&&&\\
\hline
&&&&&&\\
\ph [5]&[5]&[5]&(12345)&(15243)&(14235)&192\\
\ph [5]&[5]&[5]&(12345)&(15432)&(14253)&192\\
\ph [5]&[5]&[5]&(12345)&(14253)&(13524)&192\\
\ph [5]&[5]&[5]&(12345)&(13524)&(12345)&192\\
\hline
\ph [5]&[5]&[3]&(12345)&(13245)&(123)&240\\
\ph [5]&[5]&[3]&(12345)&(12453)&(235)&240\\
\ph [5]&[3]&[5]&(12345)&(354)&(15342)&240\\
\ph [5]&[3]&[5]&(12345)&(253)&(15243)&240\\
\hline
\ph [5]&[5]&[22]&(12345)&(15342)&(14)(25)&120\\
\ph [5]&[22]&[5]&(12345)&(24)(35)&(15234)&120\\
\hline
\ph [5]&[4]&[4]&(12345)&(1253)&(2435)&360\\
\ph [5]&[4]&[4]&(12345)&(1254)&(2453)&360\\
\ph [5]&[4]&[4]&(12345)&(2354)&(1534)&360\\
\hline
\ph [5]&[4]&[32]&(12345)&(1432)&(13)(254)&240\\
\ph [5]&[4]&[32]&(12345)&(1354)&(12)(345)&240\\
\ph [5]&[32]&[4]&(12345)&(124)(35)&(2345)&240\\
\ph [5]&[32]&[4]&(12345)&(12)(354)&(2534)&240\\
\hline
\ph [5]&[4]&[2]&(12345)&(1245)&(23)&120\\
\ph [5]&[2]&[4]&(12345)&(12)&(2543)&120\\
\hline
\ph [5]&[3]&[3]&(12345)&(145)&(132)&120\\
\hline
\ph [5]&[32]&[32]&(12345)&(153)(24)&(14)(235)&120\\
\hline
\ph [5]&[32]&[2]&(12345)&(145)(23)&(13)&120\\
\ph [5]&[2]&[32]&(12345)&(13)&(12)(354)&120\\
\hline
\ph [5]&[3]&[22]&(12345)&(134)&(12)(45)&120\\
\ph [5]&[22]&[3]&(12345)&(12)(34)&(254)&120\\
\hline
\ph [5]&[22]&[22]&(12345)&(15)(24)&(14)(23)&120\\
\hline
\ph [5]&[5]&[]&(12345)&(15432)&()&24\\
\ph [5]&[]&[5]&(12345)&()&(12345)&24\\
\end{array}
\ee

\be\label{t52}
\begin{array}{ccc|ccc|c|}
{\cre [\sigma_1]} & {\cg [\sigma_2]} & {\cb [\sigma_3]}
&{\cre \sigma_1^{can}} & {\cg \sigma_2}=\sigma_3\circ\sigma_1 & {\cb \sigma_3}
& {\cal N}^H([\sigma_1,\sigma_2,\sigma_3])
\\
&&&&&&\\
\hline
&&&&&&\\
\ph [4]&[5]&[4]&(1234)&(12543)&(2534)&360\\
\ph [4]&[5]&[4]&(1234)&(15423)&(1534)&360\\
\ph [4]&[5]&[4]&(1234)&(12435)&(2354)&360\\
\ph [4]&[4]&[5]&(1234)&(1543)&(15342)&360\\
\ph [4]&[4]&[5]&(1234)&(1532)&(15243)&360\\
\ph [4]&[4]&[5]&(1234)&(1542)&(15324)&360\\
\hline
\ph [4]&[5]&[32]&(1234)&(14532)&(13)(245)&240\\
\ph [4]&[5]&[32]&(1234)&(14253)&(134)(25)&240\\
\ph [4]&[32]&[5]&(1234)&(153)(24)&(15234)&240\\
\ph [4]&[32]&[5]&(1234)&(15)(243)&(15423)&240\\
\hline
\ph [4]&[5]&[2]&(1234)&(15234)&(15)&120\\
\ph [4]&[2]&[5]&(1234)&(15)&(15432)&120\\
\hline
\ph [4]&[4]&[3]&(1234)&(1324)&(123)&240\\
\ph [4]&[4]&[3]&(1234)&(1235)&(354)&240\\
\ph [4]&[3]&[4]&(1234)&(125)&(2543)&240\\
\ph [4]&[3]&[4]&(1234)&(243 )&(1432 )&240\\
\hline
\ph [4]&[4]&[22]&(1234)&(2534)&(14)(25)&270\\
\ph [4]&[4]&[22]&(1234)&(2354)&(14)(35)&270\\
\ph [4]&[4]&[22]&(1234)&(1432)&(13)(24)&270\\
\ph [4]&[22]&[4]&(1234)&(12)(35)&(2435)&270\\
\ph [4]&[22]&[4]&(1234)&(12)(45)&(2453)&270\\
\ph [4]&[22]&[4]&(1234)&(13)(24)&(1234)&270\\
\hline
\ph [4]&[32]&[3]&(1234)&(125)(34)&(254)&240\\
\ph [4]&[32]&[3]&(1234)&(125)(45)&(345)&240\\
 \ph [4]&[3]&[32]&(1234)&(135)&(12)(354)&240\\
\ph [4]&[3]&[32]&(1234)&(354)&(142)(35)&240\\
\hline
\ph [4]&[32]&[22]&(1234)&(14)(253)&(13)(25)&120\\
\ph [4]&[22]&[32]&(1234)&(24)(35)&(14)(235)&120\\
\hline
\ph [4]&[3]&[2]&(1234)&(124)&(23)&120\\
\ph [4]&[2]&[3]&(1234)&(14)&(132)&120\\
\hline
\ph [4]&[22]&[2]&(1234)&(14)(23)&(13)&60\\
\ph [4]&[2]&[22]&(1234)&(24)&(14)(23)&60\\
\hline
\ph [4]&[4]&[]&(1234)&(1234)&()&30\\
\ph [4]&[]&[4]&(1234)&()&(1432)&30\\
\end{array}
\ee

\be\label{t53}
\begin{array}{ccc|ccc|c|}
{\cre [\sigma_1]} & {\cg [\sigma_2]} & {\cb [\sigma_3]}
&{\cre \sigma_1^{can}} & {\cg \sigma_2}=\sigma_3\circ\sigma_1 & {\cb \sigma_3}
& {\cal N}^H([\sigma_1,\sigma_2,\sigma_3])
\\
&&&&&&\\
\hline
&&&&&&\\
\ph [32]&[5]&[4]&(123)(45)&(12534)&(2435)&240\\
\ph [32]&[5]&[4]&(123)(45)&(14532)&(1523)&240\\
\ph [32]&[4]&[5]&(123)(45)&(1435)&(15342)&240\\
\ph [32]&[4]&[5]&(123)(45)&(1432)&(15423)&240\\
\hline
\ph [32]&[5]&[32]&(123)(45)&(15342)&(14)(235)&120\\
\ph [32]&[32]&[5]&(123)(45)&(142)(35)&(15234)&120\\
\ph [32]&[5]&[2]&(123)(45)&(15423)&(14)&120\\
\ph [32]&[2]&[5]&(123)(45)&(14)&(15432)&120\\
\hline
\ph [32]&[4]&[3]&(123)(45)&(1243)&(254)&240\\
\ph [32]&[4]&[3]&(123)(45)&(1254)&(243)&240\\
\ph [32]&[3]&[4]&(123)(45)&(125)&(2453)&240\\
\ph [32]&[3]&[4]&(123)(45)&(145)&(2532)&240\\
\hline
\ph [32]&[4]&[22]&(123)(45)&(1542)&(14)(23)&120\\
\ph [32]&[22]&[4]&(123)(45)&(14)(23)&(1543)&120\\
\hline
\ph [32]&[32]&[3]&(123)(45)&(132)(45)&(123)&140\\
\ph [32]&[32]&[3]&(123)(45)&(12)(354)&(234)&140\\
\ph [32]&[3]&[32]&(123)(45)&(132)&(123)(45)&140\\
\ph [32]&[3]&[32]&(123)(45)&(253)&(13)(245)&140\\
\hline
\ph [32]&[32]&[22]&(123)(45)&(153)(24)&(14)(25)&120\\
\ph [32]&[22]&[32]&(123)(45)&(14)(35)&(152)(34)&120\\
\hline
\ph [32]&[3]&[2]&(123)(45)&(123)&(45)&20\\
\ph [32]&[2]&[3]&(123)(45)&(45)&(132)&20\\
\hline
\ph [32]&[22]&[2]&(123)(45)&(23)(45)&(13)&60\\
\ph [32]&[2]&[22]&(123)(45)&(23)&(13)(45)&60\\
\hline
\ph [32]&[32]&[]&(123)(45)&(123)(45)&()&20\\
\ph [32]&[]&[32]&(123)(45)&()&(132)(45)&20\\
\end{array}
\ee

\be\label{t54}
\begin{array}{ccc|ccc|c|}
{\cre [\sigma_1]} & {\cg [\sigma_2]} & {\cb [\sigma_3]}
&{\cre \sigma_1^{can}} & {\cg \sigma_2}=\sigma_3\circ\sigma_1 & {\cb \sigma_3}
& {\cal N}^H([\sigma_1,\sigma_2,\sigma_3])
\\
&&&&&&\\
\hline
&&&&&&\\
\ph [3]&[5]&[5]&(123)&(15432)&(15423)&240\\
\ph [3]&[5]&[5]&(123)&(15342)&(15234)&240\\
\hline
\ph [3]&[4]&[4]&(123)&(1245)&(2453)&240\\
\ph [3]&[4]&[4]&(123)&(1532)&(1523)&240\\
\hline
\ph [3]&[32]&[32]&(123)&(142)(35)&(14)(235)&140\\
\ph [3]&[32]&[32]&(123)&(132)(45)&(123)(45)&140\\
\hline
\ph [3]&[3]&[3]&(123)&(132)&(123)&140\\
\ph [3]&[3]&[3]&(123)&(124)&(243)&140\\
\hline
\ph [3]&[22]&[22]&(123)&(23)(45)&(13)(45)&60\\
\hline
\ph [3]&[2]&[2]&(123)&(23)&(13)&60\\
\hline
\ph [3]&[5]&[3]&(123)&(12543)&(254)&120\\
\ph [3]&[3]&[5]&(123)&(154)&(15432)&120\\
\hline
\ph [3]&[5]&[22]&(123)&(14253)&(14)(25)&120\\
\ph [3]&[22]&[5]&(123)&(15)(34)&(15342)&120\\
\hline
\ph [3]&[4]&[32]&(123)&(2453)&(13)(245)&240\\
\ph [3]&[4]&[32]&(123)&(1534)&(152)(34)&240\\
\ph [3]&[32]&[4]&(123)&(154)(23)&(1543)&240\\
\ph [3]&[32]&[4]&(123)&(124)(35)&(2435)&240\\
\hline
\ph [3]&[4]&[2]&(123)&(1423)&(14)&120\\
\ph [3]&[2]&[4]&(123)&(15)&(1532)&120\\
\hline
\ph [3]&[32]&[2]&(123)&(123)(45)&(45)&20\\
\ph [3]&[2]&[32]&(123)&(45)&(132)(45)&20\\
\hline
\ph [3]&[3]&[22]&(123)&(142)&(14)(23)&120\\
\ph [3]&[22]&[3]&(123)&(12)(34)&(234)&120\\
\hline
\ph [3]&[3]&[]&(123)&(123)&()&20\\
\ph [3]&[]&[3]&(123)&()&(132)&20\\
\end{array}
\ee

\be\label{t55}
\begin{array}{ccc|ccc|c|}
{\cre [\sigma_1]} & {\cg [\sigma_2]} & {\cb [\sigma_3]}
&{\cre \sigma_1^{can}} & {\cg \sigma_2}=\sigma_3\circ\sigma_1 & {\cb \sigma_3}
& {\cal N}^H([\sigma_1,\sigma_2,\sigma_3])
\\
&&&&&&\\
\hline
&&&&&&\\
\ph [22]&[5]&[5]&(12)(34)&(15324)&(15423)&120\\
\hline
\ph [22]&[4]&[4]&(12)(34)&(1235)&(2435)&270\\
\ph [22]&[4]&[4]&(12)(34)&(1254)&(2534)&270\\
\ph [22]&[4]&[4]&(12)(34)&(1324)&(1423)&270\\
\hline
\ph [22]&[32]&[32]&(12)(34)&(135)(24)&(14)(235)&120\\
\hline
\ph [22]&[3]&[3]&(12)(34)&(143)&(132)&120\\
\hline
\ph [22]&[22]&[22]&(12)(34)&(13)(24)&(14)(23)&30\\
\hline
\ph [22]&[2]&[2]&(12)(34)&(34)&(12)&30\\
\hline
\ph [22]&[5]&[3]&(12)(34)&(12534)&(254)&120\\
\ph [22]&[3]&[5]&(12)(34)&(154)&(15342)&120\\
\hline
\ph [22]&[5]&[22]&(12)(34)&(13425)&(14)(25)&120\\
\ph [22]&[22]&[5]&(12)(34)&(13)(45)&(14532)&120\\
\hline
\ph [22]&[4]&[32]&(12)(34)&(1425)&(134)(25)&120\\
\ph [22]&[32]&[4]&(12)(34)&(123)(45)&(2453)&120\\
\hline
\ph [22]&[4]&[2]&(12)(34)&(1432)&(13)&60\\
\ph [22]&[2]&[4]&(12)(34)&(13)&(1432)&60\\
\hline
\ph [22]&[32]&[2]&(12)(34)&(152)(34)&(15)&60\\
\ph [22]&[2]&[32]&(12)(34)&(45)&(12)(345)&60\\
\hline
\ph [22]&[3]&[22]&(12)(34)&(345)&(12)(45)&60\\
\ph [22]&[22]&[3]&(12)(34)&(12)(45)&(345)&60\\
\hline
\ph [22]&[22]&[]&(12)(34)&(12)(34)&()&15\\
\ph [22]&[]&[22]&(12)(34)&()&(12)(34)&15\\

\end{array}
\ee

\be\label{t56}
\begin{array}{ccc|ccc|c|}
{\cre [\sigma_1]} & {\cg [\sigma_2]} & {\cb [\sigma_3]}
&{\cre \sigma_1^{can}} & {\cg \sigma_2}=\sigma_3\circ\sigma_1 & {\cb \sigma_3}
& {\cal N}^H([\sigma_1,\sigma_2,\sigma_3])
\\
&&&&&&\\
\hline
&&&&&&\\
\ph [2]&[5]&[4]&(12)&(12435)&(2435)&120\\
\ph [2]&[4]&[5]&(12)&(1534)&(15342)&120\\
\hline
\ph [2]&[5]&[32]&(12)&(14235)&(14)(235)&120\\
\ph [2]&[32]&[5]&(12)&(154)(23)&(15423)&120\\
\hline
\ph [2]&[4]&[3]&(12)&(1254)&(254)&120\\
\ph [2]&[3]&[4]&(12)&(154)&(1542)&120\\
\hline
\ph [2]&[4]&[22]&(12)&(1423)&(14)(23)&60\\
\ph [2]&[22]&[4]&(12)&(15)(23)&(1523)&60\\
\hline
\ph [2]&[32]&[3]&(12)&(12)(345)&(345)&20\\
\ph [2]&[3]&[32]&(12)&(345)&(12)(345)&20\\
\hline
\ph [2]&[32]&[22]&(12)&(142)(35)&(14)(35)&60\\
\ph [2]&[22]&[32]&(12)&(15)(34)&(152)(34)&60\\
\hline
\ph [2]&[3]&[2]&(12)&(132)&(13)&60\\
\ph [2]&[2]&[3]&(12)&(13)&(132)&60\\
\hline
\ph [2]&[22]&[2]&(12)&(12)(34)&(34)&30\\
\ph [2]&[2]&[22]&(12)&(45)&(12)(45)&30\\
\hline
\ph [2]&[2]&[]&(12)&(12)&()&10\\
\ph [2]&[]&[2]&(12)&()&(12)&10\\
\end{array}
\ee

\be\label{t57}
\begin{array}{ccc|ccc|c|}
{\cre [\sigma_1]} & {\cg [\sigma_2]} & {\cb [\sigma_3]}
&{\cre \sigma_1^{can}} & {\cg \sigma_2}=\sigma_3\circ\sigma_1 & {\cb \sigma_3}
& {\cal N}^H([\sigma_1,\sigma_2,\sigma_3])
\\
&&&&&&\\
\hline
&&&&&&\\
\ph []&[5]&[5]&()&(12345)&(12345)&24\\
\hline
\ph []&[4]&[4]&()&(1234)&(1234)&30\\
\hline
\ph []&[32]&[32]&()&(123)(45)&(123)(45)&20\\
\hline
\ph []&[3]&[3]&()&(123)&(123)&20\\
\hline
\ph []&[22]&[22]&()&(12)(34)&(12)(34)&15\\
\hline
\ph []&[2]&[2]&()&(12)&(12)&10\\
\hline
\ph []&[]&[]&()&()&()&1\\
\end{array}
\ee
}

\subsection*{B4. The RG-gauge: operators with fixed conjugacy class of $\sigma_2$}

In this Appendix, we list the operators with fixed numbers of red-green cycles, see ss.\ref{rgcycles} and \ref{RGcycles}.

\subsubsection*{Operators with one red-green cycle}

Here the first column labels the level, in the second column there are lengths of the red-green cycle, in the third column there is total number of operators at given $\sigma_2$, and the 9th column describes the symmetry under permutations of the red and green colorings. The numbers $N^G$ count numbers of the distinct Gaussian correlators.

\bigskip

{\footnotesize
\be
\!\!\!\!\!\!\!\!\!\!\!\!\!\!\!
\begin{array}{c|c|c|c|c|c|c|c|c|c}
m&{\rm Length} & {\rm number} & {\cg \sigma_2} & {\cb \sigma_3}
&{\rm equivalent}\ {\cb \sigma_3}
&||\sigma_3||=\frac{m!}{z_{[\sigma_3]}}&N_{[\sigma_2],[\sigma_3]}
&{\cre r}\leftrightarrow{\cg g}  & {\rm operator}\\
\hline\hline\hline
1&1 & 1 & () & () & &1&N_{[1],[1]}=1&self&{\cal K}_1\\
\hline\hline\hline
2&2 & N_{[2]}=2 & (12) & () & &1&N_{[2],[11]}=1&self& {\cal K}_{\cg 2} \\
&&&& (12) & &1&N_{[2],[2]}=1&self &{\cal K}_{\cre 2}  \\
\hline\hline\hline
3&3 & N_{[3]}=4 & (123) & () & &1&N_{[3],[111]}=1&(123)& {\cal K}_{\cg 3} \\
&&&&(12)&(13),(23) &3&N_{[3],[21]}=1&self& {\cal K}_{{\cre 2},{\cg 2}}\\
&&&&(123)& &2&N_{[3],[3]}=2&()& {\cal K}_{\cre 3}\\
&&&&(132)& &&&self & {\cal K}_{3W}\\
\hline\hline\hline
4&4 & N_{[4]}=10  & (1234) & () & &1&N_{[4],[1^4]}=1&(1234)&  {\cal K}_{\cg 4}\\
&&&(1234) & (12) &(23),(34),(14) &6&N_{[4],[211]}=2& (123) & {\cal K}_{{\cre 2},{\cg 3}}  \\
&&&(1234) & (13) &(24) &&& (12)(34) & {\cal K}_{{\cg 2},{\cre 2},{\cg 2}}  \\
&&&(1234) & (12)(34) & (14)(23)&3&N_{[4],[22]}=2& (13)& {\cal K}_{{\cre 2},{\cg 2},{\cre 2}}  \\
&&&(1234) & (13)(24) &&&&(1432) & {\cal K}_{\cg 22W}  \\
&&&(1234) & (123) &(234),(134),(124)&8&N_{[4],[31]}=2& (12)&  {\cal K}_{{\cg 2},{\cre 3}}  \\
&&&(1234) & (132) &(243),(143),(142)&&& (1243) & {\cal K}_{\cg 31W}  \\
&&&(1234) & (1234)& &6&N_{[4],[4]}=3 &() &{\cal K}_{\cre 4}  \\
&&&(1234) & (1243)&(1324),(1342),(1423)&&&(132) &{\cal K}_{\cre 31W}  \\
&&&(1234) & (1432) &&&&(13)(24)&{\cal K}_{\cre 22W}  \\
\hline\hline\hline
5&5 & N_{[5]}=28 & (12345) & () &1&1&N_{[5],[1^5]}=1&(12345)& {\cal K}_{\cg 5}={\cal K}_I \\
\hline\hline
&& &(12345) &(12)&5&10&N_{[5],[2111]}=2&&{\cal K}_{II}\\
&&N^G_{[5]}=23&(12345) &(13)&5&&&&{\cal K}_{III}\\
\hline\hline
&&&(12345) &(12)(34)&5&15&N_{[5],[221]}=3&&{\cal K}_{IV}\\
&&&(12345) &(12)(35)&5&&&&{\cal K}_{V}\\
&&&(12345) &(13)(25)&5&&&&{\cal K}_{VI}\\
\hline\hline
&&&(12345) &(123)&5&20&N_{[5],[311]}=4&&{\cal K}_{VII}\\
&&&(12345) &(132)&5&&&&{\cal K}_{VIII}\\
&&&(12345) &(124)&5&&&&{\cal K}_{IX}\\
&&&(12345) &(142)&5&&&&{\cal K}_{X}\\
\hline\hline
&&&(12345) &(12)(345)&5&20&N_{[5],[32]}=4&&{\cal K}_{XI}\\
&&&(12345) &(12)(354)&5&&&&{\cal K}_{XII}\\
&&&(12345) &(13)(245)&5&&&&{\cal K}_{XIII}\\
&&&(12345) &(13)(254)&5&&&&{\cal K}_{XIV}\\
\hline\hline
&&&(12345) &(1234)&(1235),(1245),(1345),(2345)&30&N^G_{[5],[41]}=4&&{\cal K}_{XV}\\
&&&(12345) &(1324)&(1352),(1354),(1524),(2435)&&&&{\cal K}_{XVI}\\
&&&(12345) &(1432)&(1532),(1542),(1543),(2543)&&&&{\cal K}_{XVII}\\
\hline
&&&(12345) &(1243)&(1325),(1452),(1534),(2354)&&&&{\cal K}_{XVIII}\\
&&&(12345) & {(1253)}&(1423),(1425),(1453),(2534)&&N_{[5],[41]}=6&&{\cal K}_{XIX}\\
&&&(12345) & {(1254)}&(1342),(1435),(1523),(2453)&&&&{\cal K}_{XX}\\
\hline\hline
&&&(12345)&(12345) &1&24&N^G_{[5],[5]}=5&()& {\cal K}_{\cre 5}={\cal K}_{XXI}\\
&&&(12345) &(12354)&5&&&&{\cal K}_{XXII}\\
&&&(12345) &(12453)&5&&&&{\cal K}_{XXIII}\\
&&&(12345) &(15423)&5&&&&{\cal K}_{XXIV}\\
\hline
&&&(12345) &(13254)&5&&&&{\cal K}_{XXV}\\
&&&(12345) & {(15432)}&1&&N_{[5],[5]}=8&&{\cal K}_{XXVI}\\
&&&(12345) & {(13524)}&1&&&&{\cal K}_{XXVII}\\
&&&(12345) & {(14253)}&1&&&&{\cal K}_{XXVIII}\\
\hline\hline\hline
6&6 & N^G_{[6]}=98 & (123456) & () &&&&(123456)& {\cal K}_{\cg 6} \\
&&&&(123456) &&&&()& {\cal K}_{\cre 6}\\
\hline
\ldots &&&&&&&&&
\end{array}
\!\!\!\!\!\!
\label{tabsym}
\ee}

\subsubsection*{Operators with several red-green cycles}

Here the second column describes the lengths of cycles.

\be
\begin{array}{c|c|c|c|c|c}
m& {\rm Length} & {\rm number} & {\cg \sigma_2} & {\cb \sigma_3} & {\rm operator} \\
\hline\hline
2 & 1+1 &N_{[11]}=2& ()& () & {\cal K}_1^2 \\
&1+1 & &()& (12) & {\cal K}_{\cb 2} \\
\hline\hline
3 & 1+1+1 & N_{[111]}=3 & () & ( ) & {\cal K}_{1}^3 \\
&1+1+1 &   & () & (12) & {\cal K}_{\cb 2} {\cal K}_1 \\
&1+1+1 &   & () & (123) & {\cal K}_{\cb 3} \\
\hline
&2+1 & N_{[21]}=4&  (12) & ( ) & {\cal K}_{\cg 2} {\cal K}_1  \\
&2+1 &  &  (12) & (12) & {\cal K}_{\cre 2} {\cal K}_1  \\
&2+1 &  &  (12) & (23) & {\cal K}_{{\cg 2},{\cb 2}}  \\
&2+1 & &  (12) & (123) & {\cal K}_{{\cre 2},{\cb 2}}\\
\hline\hline
4 & 1+1+1+1 & N_{[1^4]}=5 & () & () & {\cal K}_1^4 \\
& 1 + 1 + 1 + 1 &&() & (12) & {\cal K}_{\cb 2}{\cal K}_1^2 \\
& 1 + 1 + 1 + 1 &&() & (12)(34) & {\cal K}_{\cb 2}^2 \\
& 1 + 1 + 1 + 1 &&() & (123) & {\cal K}_{\cb 3}{\cal K}_1 \\
& 1 + 1 + 1 + 1 &&() & (1234) & {\cal K}_{\cb 4} \\
\hline
& 2+1+1 &N_{[211]}=10 & (12) & () & {\cal K}_{\cre 2}{\cal K}_1^2 \\
& 2+1+1 & & (12) & (12) & {\cal K}_{\cg 2}{\cal K}_1^2 \\
& 2+1+1 & & (12) & (23) & {\cal K}_{{\cg 2},{\cb 2}}{\cal K}_1 \\
& 2+1+1 & & (12) & (34) & {\cal K}_{\cre 2}{\cal K}_{\cb 2} \\
& 2+1+1 & & (12) & (12)(34) & {\cal K}_{\cg 2}{\cal K}_{\cb 2} \\
& 2+1+1 & & (12) & (13)(24) & {\cal K}_{{\cb 2},{\cg 2},{\cb 2}} \\
& 2+1+1 & & (12) & (123) & {\cal K}_{{\cre 2},{\cb 2}}{\cal K}_1 \\
& 2+1+1 & & (12) & (134) & {\cal K}_{{\cg 2},{\cb 3}} \\
& 2+1+1 & & (12) & (1234) & {\cal K}_{{\cre 2},{\cb 3}} \\
& 2+1+1 & & (12) & (1324) & {\cal K}_{{\cb 2},{\cre 2},{\cb 2}} \\
\hline
& 2+2 & N_{[22]}=8 & (12)(34) & () & {\cal K}_{\cg 2}^2 \\
& 2+2 &  & (12)(34) & (12) & {\cal K}_{\cre 2}{\cal K}_{\cg 2} \\
& 2+2 &  & (12)(34) & (13) & {\cal K}_{{\cg 2},{\cb 2},{\cg 2}} \\
& 2+2 &  & (12)(34) & (12)(34) & {\cal K}_{\cre 2}^2 \\
& 2+2 &  & (12)(34) & (13)(24) & {\cal K}_{4C} \\
& 2+2 &  & (12)(34) & (123) & {\cal K}_{{\cre 2},{\cb 2},{\cg 2}} \\
& 2+2 &  & (12)(34) & (1234) & {\cal K}_{{\cre 2},{\cb 2},{\cre 2}} \\
& 2+2 &  & (12)(34) & (1423) & {\cal K}_{\cb 22W}  \\
\hline
& 3+1 &N_{[31]}=10 & (123) & () & {\cal K}_{\cg 3}{\cal K}_1 \\
& 3+1 &  & (123) & (12) & {\cal K}_{{\cre 2},{\cg 2}}{\cal K}_1 \\
& 3+1 &  & (123) & (14) & {\cal K}_{{\cb 2},{\cg 3}}\\
& 3+1 &  & (123) & (12)(34) & {\cal K}_{{\cre 2},{\cg 2},{\cb 2}} \\
& 3+1 &  & (123) & (123) & {\cal K}_{\cre 3}{\cal K}_1 \\
& 3+1 &  & (123) & (132) & {\cal K}_{3W}{\cal K}_1 \\
& 3+1 &  & (123) & (124) & {\cal K}_{222} \\
& 3+1 &  & (123) & (1234) & {\cal K}_{{\cb 2},{\cre 3}} \\
& 3+1 &  & (123) & (143) & {\cal K}_{{\cg 2},{\cre 2},{\cb 2}} \\
& 3+1 &  & (123) & (1324) & {\cal K}_{\cb 31W} \\
\hline
\end{array}
\label{tabmanyrgcycles}
\ee

\bigskip

In the table for $m=5$, we do not list the operators at the rightmost column, since, at this level, part of operators is not drawn in Appendix A, and is obtained from those drawn by permutations of colorings.

\be
\begin{array}{c|c|c|c|c|c|c}
m& {\rm Length} & {\rm number} & {\cg \sigma_2} & {\cb \sigma_3}
&{\rm equivalent\ } {\cb \sigma_3}& {\rm operator} \\
\hline\hline
5 & 1+1+1+1+1 &N_{[1^5]}=7& ()& () &1& {\cal K}_1^5 \\
 & 1+1+1+1+1 & & ()& (12) &10& {\cal K}_{\cb 2}{\cal K}_1^3   \\
 & 1+1+1+1+1 &N^G_{[1^5]}=7 & ()& (12)(34) &15& {\cal K}_{\cb 2}^2{\cal K}_1   \\
 & 1+1+1+1+1 & & ()& (123) &20& {\cal K}_{\cb 3}{\cal K}_1^2  \\
 & 1+1+1+1+1 & & ()& (123)(45) &20& {\cal K}_{\cb 3}{\cal K}_{\cb 2} \\
 & 1+1+1+1+1 & & ()& (1234) &30& {\cal K}_{\cb 4}{\cal K}_1   \\
 & 1+1+1+1+1 & & ()& (12345) &24&  {\cal K}_{\cb 5}  \\
 &   & &  &   &  & \\
 \hline\hline
  &   & &  &   &  & \\
 & 2+1+1+1 &N_{[2111]}=18 & (12)& () &1&   \\
 \hline
 & 2+1+1+1 & & (12)& (12) &1&   \\
 & 2+1+1+1 &N^G_{[2111]}=18 & (12)& (13) &6 &   \\
 & 2+1+1+1 & & (12)& (34) &3&   \\
 \hline
 & 2+1+1+1 & & (12)& (12)(34) & 3&  \\
 & 2+1+1+1 & & (12)& (13)(24) &6&   \\
 & 2+1+1+1 & & (12)& (13)(45) &6&   \\
 \hline
 & 2+1+1+1 & & (12)& (123) &6&   \\
 & 2+1+1+1 & & (12)& (134) &12&   \\
 & 2+1+1+1 & & (12)& (345) &2&   \\
 \hline
 & 2+1+1+1 & & (12)& (123)(45) &6&   \\
 & 2+1+1+1 & & (12)& (135)(24) &12&   \\
 & 2+1+1+1 & & (12)& (345)(12) &2&   \\
 \hline
 & 2+1+1+1 & & (12)& (1234) &12&   \\
 & 2+1+1+1 & & (12)& (1324) &6&   \\
 & 2+1+1+1 & & (12)& (1324) &12&   \\
 \hline
 & 2+1+1+1 & & (12)& (12345) &12&   \\
 & 2+1+1+1 & & (12)& (13425) &12   \\
  &   & &  &   &   \\
  \hline\hline
   &   & &  &   &  & \\
 & 2+2+1 &N_{[221]}=22 & (12)(34)& ( ) &1&   \\
 \hline
 & 2+2+1 & & (12)(34)& (12) &2&   \\
 & 2+2+1 &N^G_{[221]}=21 & (12)(34)& (13) &4&   \\
  & 2+2+1 & & (12)(34)& (15) & 4&   \\
  \hline
   & 2+2+1 & & (12)(34)& (12)(34) &1&   \\
    & 2+2+1 & & (12)(34)& (13)(24) &2&   \\
     & 2+2+1 & & (12)(34)& (12)(35) &4&   \\
      & 2+2+1 & & (12)(34)& (13)(25) &8&   \\
      \hline
      & 2+2+1 & & (12)(34)& (123) &8&   \\
      & 2+2+1 & & (12)(34)& (125) &4&   \\
      & 2+2+1 & & (12)(34)& (135) &8&   \\
      \hline
      & 2+2+1 & & (12)(34)& (123)(45) &8&   \\
      & 2+2+1 & & (12)(34)& (125)(34) &4&   \\
      & 2+2+1 & & (12)(34)& (135)(24) &8&   \\
      \hline
      & 2+2+1 & & (12)(34)& (1234) &4&   \\
      & 2+2+1 & & (12)(34)& (1324) &2&   \\
      & 2+2+1 & & (12)(34)& (1235) &8&   \\
      & 2+2+1 & & (12)(34)& (1325) &8&   \\
      & 2+2+1 & & (12)(34)& (1532) &8&   \\
      \hline
      & 2+2+1 & & (12)(34)& (12345) & 8&  \\
      & 2+2+1 & & (12)(34)& (12354) & 8&  \\
      & 2+2+1 & & (12)(34)& (13245) & 8&  \\
  &   & &  &   &   \\
\end{array}\nn
\ee
\be
\begin{array}{c|c|c|c|c|c|c}
m& {\rm Length} & {\rm number} & {\cg \sigma_2} & {\cb \sigma_3}
&{\rm equivalent\ } {\cb \sigma_3}& {\rm operator} \\
\hline
5
 & 3+1+1 &N_{[311]}= 26& (123)& () & 1 &  \\
 \hline
 & 3+1+1 & & (123)& (12) & 3&   \\
  & 3+1+1 & N^G_{[311]}=26& (123)& (14) & 6&  \\
 & 3+1+1 & & (123)& (45) & 1 &   \\
 \hline
 & 3+1+1 & & (123)& (12)(34) &6&   \\
 & 3+1+1 & & (123)& (12)(45) &3&   \\
 & 3+1+1 & & (123)& (14)(25) &6&   \\
 \hline
 & 3+1+1 & & (123)& (123) & 1&  \\
 & 3+1+1 & & (123)& (132) &1&   \\
 & 3+1+1 & & (123)& (124) &6&   \\
 & 3+1+1 & & (123)& (142) &6&   \\
 & 3+1+1 & & (123)& (145) &6&   \\
 \hline
 & 3+1+1 & & (123)& (123)(45) & 1&  \\
 & 3+1+1 & & (123)& (132)(45) &1&   \\
 & 3+1+1 & & (123)& (124)(35) & 6&  \\
 & 3+1+1 & & (123)& (142)(35) & 6&  \\
 & 3+1+1 & & (123)& (145)(23) & 6&  \\
 \hline
 & 3+1+1 & & (123)& (1234) & 6&   \\
 & 3+1+1 & & (123)& (1324) & 6&  \\
 & 3+1+1 & & (123)& (1245) & 6&   \\
 & 3+1+1 & & (123)& (1425) & 6&  \\
 & 3+1+1 & & (123)& (1452) & 6&  \\
 \hline
 & 3+1+1 & & (123)& (12345) &6&   \\
 & 3+1+1 & & (123)& (13245) &6&   \\
 & 3+1+1 & & (123)& (12435) &6&   \\
 & 3+1+1 & & (123)& (13425) &6&   \\
  &   & &  &   &   \\
  \hline\hline
  &   & &  &   &   \\
 & 3+2 &N_{[32]}=26 & (123)(45)& () &1&   \\
 \hline
 & 3+2 & & (123)(45)& (12) &3&   \\
 & 3+2 &N^G_{[32]}=26 & (123)(45)& (14) &6&   \\
 & 3+2 & & (123)(45)& (45) &1&   \\
 \hline
 & 3+2 & & (123)(45)& (12)(34) & 6&  \\
 & 3+2 & & (123)(45)& (12)(45) & 3&  \\
 & 3+2 & & (123)(45)& (14)(25) & 6&  \\
 \hline
 & 3+2 & & (123)(45)& (123) & 1&  \\
 & 3+2 & & (123)(45)& (132) & 1&  \\
 & 3+2 & & (123)(45)& (124) & 6&  \\
 & 3+2 & & (123)(45)& (142) & 6&  \\
 & 3+2 & & (123)(45)& (145) & 6&  \\
 \hline
 & 3+2 & & (123)(45)& (123)(45) & 1&  \\
 & 3+2 & & (123)(45)& (132)(45) & 1&  \\
 & 3+2 & & (123)(45)& (124)(35) & 6&  \\
 & 3+2 & & (123)(45)& (142)(35) & 6&  \\
 & 3+2 & & (123)(45)& (145)(23) & 6&  \\
 \hline
 & 3+2 & & (123)(45)& (1234) &6&   \\
 & 3+2 & & (123)(45)& (1324) &6&   \\
 & 3+2 & & (123)(45)& (1245) &6&   \\
 & 3+2 & & (123)(45)& (1425) &6&   \\
 & 3+2 & & (123)(45)& (1452) &6&   \\
 \hline
 & 3+2 & & (123)(45)& (12345) &6&   \\
 & 3+2 & & (123)(45)& (13245) &6&   \\
 & 3+2 & & (123)(45)& (12435) &6&   \\
 & 3+2 & & (123)(45)& (13425) &6&   \\
  &   & &  &   &   \\
  \end{array}
\ee
\be
\begin{array}{c|c|c|c|c|c|c}
m& {\rm Length} & {\rm number} & {\cg \sigma_2} & {\cb \sigma_3}
&{\rm equivalent\ } {\cb \sigma_3}& {\rm operator} \\
\hline
5
 & 4+1 &N_{[41]}=34 & (1234)& () &1&   \\
 \hline
 & 4+1 & & (1234)& (12) &4&   \\
 & 4+1 &N^G_{[41]}=28 & (1234)& (13) &2&   \\
 & 4+1 & & (1234)& (15) &4&   \\
 \hline
 & 4+1 & & (1234)& (12)(34) &2&   \\
 & 4+1 & & (1234)& (13)(24) &1&   \\
 & 4+1 & & (1234)& (12)(45) &4&   \\
 & 4+1 & & (1234)& (13)(45) &4&   \\
 & 4+1 & & (1234)& (14)(25) &4&   \\
 \hline
 & 4+1 & & (1234)& (123) & 4&  \\
 & 4+1 & & (1234)& (132) & 4&  \\
 & 4+1 & & (1234)& (345) & 4&  \\
 & 4+1 & & (1234)& (245) & 4&  \\
 & 4+1 & & (1234)& (354) & 4&  \\
 \hline
 & 4+1 & & (1234)& (123)(45) & 4&  \\
 & 4+1 & & (1234)& (132)(45) & 4&  \\
 & 4+1 & & (1234)& (345)(12) & 4&  \\
 & 4+1 & & (1234)& (245)(13) & 4&  \\
 & 4+1 & & (1234)& (354)(12) & 4&  \\
 \hline
 & 4+1 & & (1234)& (1234) &1&   \\
 & 4+1 & & (1234)& (1432) &1&   \\
 & 4+1 & & (1234)& (1243) &4&   \\
 & 4+1 & & (1234)& (1245) &4&   \\
 & 4+1 & & (1234)& (1235) &4&   \\
 & 4+1 & & (1234)& (1253) &4&   \\
 & 4+1 & & (1234)& (1325) &4&   \\
 & 4+1 & & (1234)& (1352) &4&   \\
 & 4+1 & & (1234)& (1452) &4&   \\
 \hline
 & 4+1 & & (1234)& (12345) &4&   \\
 & 4+1 & & (1234)& (12435) &4&   \\
 & 4+1 & & (1234)& (12453) &4&   \\
 & 4+1 & & (1234)& (12543) &4&   \\
 & 4+1 & & (1234)& (13245) &4&   \\
 & 4+1 & & (1234)& (14325) &4&   \\
\hline
\end{array}\nn
\ee

\end{document}